\documentclass[12pt,a4paper]{article}
\usepackage[utf8]{inputenc}
\usepackage{epsf,amsmath,amssymb,graphicx,dcolumn}
\usepackage{caption}
\usepackage{subcaption}
\usepackage{scalefnt,ulem,pstricks}
\usepackage{booktabs,multirow,tabularx}
\usepackage{cite,hyperref}
\usepackage{cleveref}
\usepackage{color}
\usepackage{rotating}
\usepackage{microtype}
\usepackage[titletoc,title]{appendix}
 
\newcommand{\sphid}[1]{}

\providecommand{\href}[2]{#2}

\def\ltap{\raisebox{-.6ex}{\rlap{$\,\sim\,$}} \raisebox{.4ex}{$\,<\,$}}

\newcommand\as{\alpha_{\mathrm{S}}}

\def\to{\rightarrow}

\def\qT{q_T}
\def\pT{p_T}

\newcommand\Matrix{{\sc Matrix}}
\newcommand\Munich{{\sc Munich}}
\newcommand\OpenLoops{{\sc OpenLoops}}
\newcommand\Collier{{\sc Collier}}
\newcommand\GINAC{{\sc Ginac}}
\newcommand{\CutTools}{{\sc CutTools}}
\newcommand{\OneLOop}{{\sc OneLOop}}
\newcommand{\qt}{\ensuremath{q_T}}
\newcommand{\pt}{\ensuremath{p_T}}

\newcommand{\eqn}[1]{Eq.\,(\ref{#1})}

\newcommand{\fig}[1]{Figure\,\ref{#1}}
\newcommand{\figs}[1]{Figures\,\ref{#1}}
\newcommand{\tab}[1]{Table\,\ref{#1}}
\newcommand{\sct}[1]{Section~\ref{#1}}

\newcommand{\app}[1]{Appendix~\ref{#1}}

\def\refeq#1{\mbox{Eq.~\eqref{#1}}}

\def\reffi#1{\mbox{Figure~\ref{#1}}}

\def\reffis#1#2{\mbox{Figures~\ref{#1}--\ref{#2}}}

\def\refse#1{\mbox{Section~\ref{#1}}}
\def\refsetwo#1#2{\mbox{Sections~\ref{#1} and \ref{#2}}}

\def\citere#1{\mbox{Ref.~\cite{#1}}}
\def\citeres#1{\mbox{Refs.~\cite{#1}}}

\newcommand{\rcut}{\ensuremath{r_{\mathrm{cut}}}}
\newcommand{\zz}{\ensuremath{ZZ}}
\newcommand{\ww}{\ensuremath{W^+W^-}}
\newcommand{\wz}{\ensuremath{W^\pm Z}}

\newcommand{\z}{\ensuremath{Z}}
\newcommand{\w}{\ensuremath{W}}

\newcommand{\abbrev}{}

\newcommand{\nnll}{\text{\abbrev NNLL}}

\newcommand{\nlo}{\text{\abbrev NLO}}
\newcommand{\nnlo}{\text{\abbrev NNLO}}

\newcommand{\qcd}{{\abbrev QCD}}
\newcommand{\D}{\mathrm{d}}

\newcommand\Bstrut{\rule[-1.5ex]{0pt}{0pt}}   

\newcommand{\elle}{\ensuremath{\ell}}
\newcommand{\genllln}{\ensuremath{\elle\elle\elle\nu}}
\newcommand{\llln}{\elle'^\pm{\nu}_{\elle^\prime} \elle^-\elle^+}
\newcommand{\mllln}{\ensuremath{m_{\llln}}}
\newcommand{\ptllln}{\ensuremath{p_{T,\llln}}}

\begin{document} 
\begin{titlepage}
\renewcommand{\thefootnote}{\fnsymbol{footnote}}
\begin{flushright}
ZU-TH 06/17\\
CERN-TH-2017-065\\
\end{flushright}
\vspace*{1cm}

\begin{center}
{\Large \bf \wz{} production at the LHC:\\[0.4cm]
fiducial cross sections and distributions in NNLO QCD
}
\end{center}

\par \vspace{2mm}
\begin{center}
{\bf Massimiliano Grazzini$^{(a)}$},
{\bf Stefan Kallweit$^{(b)}$}\\[0.3cm]
{\bf Dirk Rathlev$^{(a)}$} and {\bf Marius Wiesemann$^{(a,b)}$}
\vspace{5mm}

$^{(a)}$Physik-Institut, Universit\"at Z\"urich, CH-8057 Z\"urich, Switzerland 

$^{(b)}$TH Division, Physics Department, CERN, CH-1211 Geneva 23, Switzerland

\vspace{5mm}

\end{center}

\par \vspace{2mm}
\begin{center} {\large \bf Abstract} \end{center}
\begin{quote}
\pretolerance 10000

We report on the first fully differential calculation for \wz{} production in hadron collisions up to next-to-next-to-leading order (NNLO) in QCD perturbation theory.
Leptonic decays of the $W$ and $Z$ bosons are consistently taken into account, i.e.\ we
include all resonant and non-resonant diagrams that contribute to the process
$pp\to \ell^{'\pm} \nu_{\ell^{'}} \ell^+\ell^-+X$ 
both in the same-flavour ($\ell'=\ell$) and the different-flavour ($\ell'\neq \ell$) channel.
Fiducial cross sections and distributions are presented in the presence of standard selection cuts applied
in the experimental \wz{} analyses by ATLAS and CMS at centre-of-mass energies of 8 and 13\,TeV.
As previously shown for the inclusive cross section, \nnlo{} corrections increase the \nlo{} result by about $10\%$, thereby leading to an improved agreement with experimental data.
The importance of NNLO accurate predictions is also shown in the case of new-physics scenarios,
where, especially in high-$p_T$ categories, their impact can reach ${\cal O}(20\%)$. 
The availability of differential NNLO predictions will play a crucial role in the rich physics
programme that is based on precision studies of \wz{} signatures at the LHC.

\end{quote}

\vspace*{\fill}
\begin{flushleft}
March 2017

\end{flushleft}
\end{titlepage}

\setcounter{footnote}{1}
\renewcommand{\thefootnote}{\fnsymbol{footnote}}

\section{Introduction}

The production of a pair of vector bosons is among the most relevant physics processes 
at the Large Hadron Collider~(LHC).
Besides playing a central role in precision tests of the gauge structure
of electroweak (EW) interactions and in studies of the mechanism of EW symmetry breaking,
vector-boson pair production constitutes an irreducible background in most of the Higgs-boson measurements and in
many searches for physics beyond the Standard Model (SM).

The production of \wz{} pairs, in particular, offers a valuable test of the triple gauge-boson couplings,
and is an important SM background in many SUSY searches (see e.g. \citere{Morrissey:2009tf}).
The \wz{} cross section has been measured at the
Tevatron~\cite{Aaltonen:2012vu,Abazov:2012cj} and at the LHC for centre-of-mass energies of
7\,TeV~\cite{Aad:2012twa,Khachatryan:2016poo}, 8\,TeV~\cite{Aad:2016ett,Khachatryan:2016poo} 
and 13\,TeV~\cite{Aaboud:2016yus,Khachatryan:2016tgp}. 
Thanks to the increasing reach in energy of LHC Run 2, more statistics --- the above-cited 13\,TeV results
are only based on the early 2015 data --- will make \wz{} measurements a powerful tool to extend the 
current bounds on the corresponding anomalous couplings.
To this purpose, a good control over the SM predictions in
the tails of some kinematic distributions is particularly important. 
As a SM background, \wz{} production is especially relevant in searches based on
final states with three leptons and missing transverse energy, which feature a clean
experimental signature, but miss a full reconstruction of the $W$ boson. 
As a result, the irreducible \wz{} background is not easily extracted from data with a 
simple side-band approach. For the above reasons, the availability of accurate theoretical predictions 
of the differential \wz{} cross section is necessary in order to ensure a high sensitivity
to anomalous couplings and to control the SM background
in searches based on the trilepton plus missing transverse energy signature.

Accurate theoretical predictions for the \wz{} cross section were obtained 
at \nlo{} in perturbative QCD a long time ago~\cite{Ohnemus:1991gb}. Leptonic 
decays of the $W$ and $Z$ bosons were added only a few years later 
\cite{Ohnemus:1994ff}, while initially omitting
spin correlations in the virtual matrix elements. The first complete off-shell \nlo{} 
computations, including leptonic decays and 
spin correlations, were performed \cite{Campbell:1999ah,Dixon:1999di,Campbell:2011bn} after
the relevant one-loop helicity amplitudes~\cite{Dixon:1998py} became available.
The corresponding computation of the off-shell $\wz{}+{\rm jet}$ cross section in \nlo{} \qcd{} 
was presented in \citere{Campanario:2010hp}. EW corrections to \wz{} production 
are known only in an on-shell approach \cite{Bierweiler:2013dja,Baglio:2013toa} so far.
Recently, the first NNLO QCD accurate prediction of the inclusive \wz{} cross section 
became available in~\citere{Grazzini:2016swo}. Due to the difference of the \w{}- and \z{}-boson masses, 
this computation already used the off-shell two-loop helicity amplitudes of 
\citere{Gehrmann:2015ora} (another calculation of these amplitudes was described in \citere{Caola:2014iua}), 
which allow for the computation of all
vector-boson pair production processes, including leptonic decays, spin correlations 
and off-shell effects.

\wz{} production is the only remaining di-boson process for which a fully exclusive \nnlo{} calculation was not available so far.
In this paper, we finally fill this gap 
by presenting, for the first time, \nnlo{}-accurate fully differential predictions for the \wz{} cross section.
More precisely, our off-shell calculation includes the leptonic decays of the vector bosons by considering 
the full process that leads to three leptons and one neutrino~(\genllln),
$pp\to \ell^{'\pm} \nu_{\ell^{'}} \ell^+\ell^-+X$, in both the same-flavour ($\ell' = \ell$) and 
the different-flavour ($\ell'\neq\ell$) channel. Thereby, we take into account all non-resonant, single-resonant and 
double-resonant components, including intermediate $W^\pm\gamma^*$ contributions and all interference effects as well as spin correlations and off-shell effects, 
consistently in the complex-mass scheme~\cite{Denner:2005fg}.

Our calculation is performed in the \Matrix\footnote{\Matrix{} is the abbreviation of 
``\Munich{} Automates qT subtraction and Resummation
to Integrate X-sections'', by M.~Grazzini, S.~Kallweit, D.~Rathlev, M.~Wiesemann. 
In preparation.} framework, 
which applies the $\qT$-subtraction~\cite{Catani:2007vq} and 
-resummation~\cite{Bozzi:2005wk} formalisms in their process-independent implementation within the 
Monte Carlo program \Munich{}\footnote{\Munich{} is the abbreviation of 
``MUlti-chaNnel Integrator at Swiss~(CH) precision'' --- an automated parton-level NLO 
generator by S.~Kallweit. In preparation.}. 
\Munich{} facilitates the fully automated computation of NLO corrections to any SM process
by using the Catani--Seymour dipole subtraction method~\cite{Catani:1996jh,Catani:1996vz},
an efficient phase-space integration, as well as an interface to the one-loop
generator \OpenLoops{}~\cite{Cascioli:2011va} to obtain all required (spin- and colour-correlated) 
tree-level and one-loop amplitudes. For the numerical stability in the tensor reductions of the one-loop amplitudes, \OpenLoops{} relies on 
the \Collier{} library~\cite{Denner:2014gla,Denner:2016kdg}.
Our implementation of $\qT$ subtraction and resummation\footnote{The first application 
of the transverse-momentum resummation framework implemented in \Matrix{} at 
\nnll{}+\nnlo{} to on-shell \ww{} and \zz{} production was presented 
in \citere{Grazzini:2015wpa} (see also \citere{Wiesemann:2016tae} for more details).}
for the production of colourless final states is fully general, 
and it is based on the universality of the hard-collinear 
coefficients~\cite{Catani:2013tia} appearing in 
the transverse-momentum resummation formalism.
These coefficients were explicitly computed for quark-initiated processes
in \citeres{Catani:2012qa,Gehrmann:2012ze,Gehrmann:2014yya}.
For the two-loop helicity amplitudes we use the results of \citere{Gehrmann:2015ora},
and of \citere{Matsuura:1988sm} for Drell-Yan like topologies.
Their implementation in \Matrix{}
is applicable to any \genllln{} final state.
This widely automated framework has already been used, in combination with
the two-loop scattering amplitudes of~\citeres{Gehrmann:2011ab,Gehrmann:2015ora}, for the
calculations of 
$Z\gamma$~\cite{Grazzini:2013bna,Grazzini:2015nwa},
$ZZ$~\cite{Cascioli:2014yka,Grazzini:2015hta}, 
\ww{}~\cite{Gehrmann:2014fva,Grazzini:2016ctr}, 
$W^\pm\gamma$~\cite{Grazzini:2015nwa} and  
$W^\pm Z$~\cite{Grazzini:2016swo} 
production at NNLO QCD as well as in 
the resummed computations of the $ZZ$ and \ww{} transverse-momentum 
spectra~\cite{Grazzini:2015wpa} at NNLL+NNLO.

NNLO corrections to the \wz{} process have been shown to be sizeable already in the case of the 
total inclusive cross section \cite{Grazzini:2016swo}. This is explained by the existence of an approximate radiation zero \cite{Baur:1994ia} at LO, which is broken only by real corrections starting at NLO.
In this paper we will show that NNLO corrections to \wz{} production are equally relevant
to provide reliable QCD predictions for fiducial cross sections and distributions, and
to obtain agreement with the LHC data. At the same time, the inclusion of NNLO corrections
will be shown to be essential to obtain a good control of SM backgrounds in SUSY searches based on the trilepton + missing energy signature \cite{CMS:2016gvu}.

The manuscript is organized as follows.
In \sct{sec:calculation} we give details on the technical implementation of our computation, 
including a brief introduction of 
the \Matrix{} framework (\sct{subsec:matrix})
and a discussion of the stability of the \wz{} cross section at (N)NLO 
based on \qt{} subtraction (\sct{subsec:stability}).
\sct{sec:results} gives an extensive collection of numerical results for $pp\to\ell^{(')\pm} \nu_{\ell^{(')}} \ell^+\ell^-+X$:
We present 
cross sections (\sct{sec:fiducial}) and distributions (\sct{sec:distributions}) in the fiducial volume
for \wz{} measurements, including their comparison to experimental data, where available, and with cuts corresponding to new-physics searches (\sct{sec:results-np}).
The main results are summarized in \sct{sec:summary}.

\section{Description of the calculation}
\label{sec:calculation}

We study the process 
\begin{equation}
\label{eq:process}
pp\to \ell'^\pm{\nu}_{\ell^\prime} \ell^+\ell^-+X,\quad \ell,\ell'\in\{e,\mu\}, 
\end{equation}
including all Feynman
diagrams that contribute to the production of three charged leptons --- one opposite-sign, same-flavour~(OSSF) lepton pair, and another charged lepton of either the same ($\ell'=\ell$) 
or a different ($\ell'\neq\ell$) flavour, later referred to as same-flavour~(SF) and different-flavour~(DF) channel --- and one corresponding neutrino.

Our calculation is performed in the
complex-mass scheme~\cite{Denner:2005fg}, and
besides resonances, it includes also 
contributions from off-shell EW bosons
and all relevant interferences; no resonance approximation is applied.
Our implementation can deal with any combination of leptonic flavours,
$\ell,\ell^\prime\in \{e,\mu,\tau\}$. For the sake of brevity,
we will often denote this process as \wz{} production though.

\begin{figure}
\begin{center}
\begin{tabular}{cccc}
\includegraphics[width=.23\textwidth]{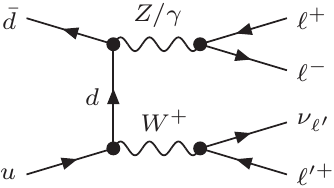} & 
\includegraphics[width=.23\textwidth]{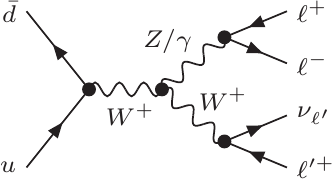} &
\includegraphics[width=.23\textwidth]{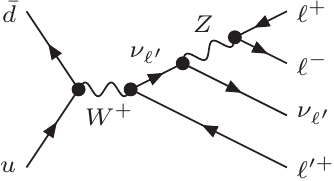} &
\includegraphics[width=.23\textwidth]{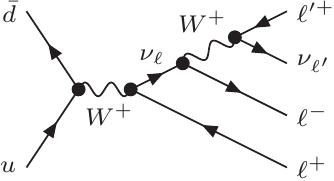} \\[2ex]
(a) & (b) & (c) & (d)
\end{tabular}
\caption[]{\label{fig:Borndiagrams}{Sample of Born diagrams contributing to $W^+Z$ production both in the different-flavour channel ($\ell\neq \ell^\prime$) and in the same-flavour channel ($\ell=\ell^\prime$). The analogous diagrams for $W^-Z$ production are achieved by charge conjugation.}}
\end{center}
\vspace*{3ex}
\end{figure}
The \genllln{} final states are generated, as shown in \fig{fig:Borndiagrams} for the $u\bar d\to \elle^{\prime+}\nu_{\elle^{\prime}}\elle^-\elle^+$ process at LO,

\setlength\parskip{0em}
\begin{itemize}\setlength\itemsep{0em}
\item[(a)] via resonant $t$-channel \wz{} production with subsequent \mbox{$W^\pm\to\elle^{\prime\pm}\nu_{\elle^\prime}$}
and \mbox{$Z\to \elle^-\elle^+$} decays, where the intermediate $Z$ boson can be replaced by an off-shell photon $\gamma^\ast$; 
\item[(b)] via $s$-channel production in \mbox{$W^\pm\to \wz{}/W^\pm\gamma^\ast$} topologies through a triple-gauge-boson vertex $WWZ$ or $WW\gamma$
with subsequent \mbox{$W^\pm\to\elle^{\prime\pm}\nu_{\elle^\prime}$}
and \mbox{$Z/\gamma^\ast\to \elle^-\elle^+$} decays;  
\item[(c)] via $W^{\pm(\ast)}$ production with a subsequent decay 
\mbox{$W^{\pm(\ast)}\to \elle^{\prime\pm}\nu_{\elle^\prime}Z^{(\ast)}/\gamma^\ast \to\elle^{\prime\pm}\nu_{\elle^{\prime}}\elle^-\elle^+$};
\item[(d)] via $W^{\pm(\ast)}$ production with a subsequent decay 
\mbox{$W^{\pm(\ast)}\to \elle^-\elle^+W^{\pm(\ast)} \to\elle^{\prime\pm}\nu_{\elle^{\prime}}\elle^-\elle^+$}.
\end{itemize}
\setlength\parskip{2ex}
In the SF channel, each diagram has to be supplemented with the analogous diagram obtained by exchanging the momenta of the identical charged leptons, 
but the generic resonance structure is not modified as compared to the DF channel.
Note that in both SF and DF channels the appearance of infrared (IR) divergent $\gamma^\ast\to \elle^-\elle^+$ splittings 
prevents a fully inclusive phase-space integration for massless leptons. In the DF channel, the 
usual experimental requirement of a mass window around the $Z$-boson mass for the OSSF lepton pair 
is already sufficient to avoid such divergences and render the cross section finite, while in the SF channel a lepton separation must be applied
on both possible combinations of OSSF lepton pairs.

The NNLO computation requires the following scattering amplitudes at $\mathcal{O}(\as^2)$:
\setlength\parskip{0em}
\begin{itemize}\setlength\itemsep{0em}
\item tree amplitudes for 
$q\bar q^{\prime} \to \llln\, gg$,\; $q\bar q^{\prime} \to \llln\,q^{\prime\prime}\bar q^{\prime\prime}$, 
and crossing-related processes;
\item one-loop amplitudes for $q\bar q^{\prime} \to \llln\, g$, and crossing-related processes; 
\item squared one-loop and two-loop amplitudes for $q\bar q^{\prime} \to \llln$.
\end{itemize}
\setlength\parskip{2ex}
The $q\bar q^{\prime}$ pair is of type $u\bar d$ and $d\bar u$ for $W^+Z$ and $W^-Z$ production, respectively, and $q^{\prime\prime}=q$ or $q^{\prime\prime}=q^{\prime}$ are explicitly allowed. Note that there is no loop-induced $gg$ channel in \wz{} production due to the electric charge of the final state.

All required tree-level and one-loop amplitudes are
obtained from the {\sc OpenLoops} generator~\cite{Cascioli:2011va,hepforge:OpenLoops}, which
implements a fast numerical recursion for the calculation of NLO scattering
amplitudes within the SM.  For the numerically stable evaluation of tensor
integrals we employ the {\sc Collier} library~\cite{Denner:2014gla,Denner:2016kdg,hepforge:COLLIER}, which
is based on the Denner--Dittmaier reduction
techniques~\cite{Denner:2002ii,Denner:2005nn} and the scalar integrals
of~\citere{Denner:2010tr}. 
To guarantee numerical stability in exceptional phase-space regions --- more precisely for phase-space points where the two independent tensor-reduction implementations of {\sc Collier} disagree by more than a certain threshold --- \OpenLoops{} provides a rescue system based on the quadruple-precision implementation of \CutTools{}~\cite{Ossola:2007ax}, which applies scalar integrals from \OneLOop{}~\cite{vanHameren:2010cp}.

For the two-loop helicity
amplitudes we rely on a public C++ library~\cite{hepforge:VVamp} that implements 
the results of~\citere{Gehrmann:2015ora},
and for the numerical evaluation of the relevant 
multiple polylogarithms we use the implementation \cite{Vollinga:2004sn} in
the \GINAC\cite{Bauer:2000cp} library.
The contribution of the massive-quark loops
is neglected in the two-loop amplitudes, but accounted for everywhere else.

\subsection{Organization of the calculation in M{\small ATRIX}}
\label{subsec:matrix}

The widely automated framework \Matrix{} is used for our NNLO calculation of the \wz{} cross section.
\Matrix{} entails a fully automated implementation of the \qt{}-subtraction formalism to compute NNLO corrections, 
and is thus applicable to any production process of an arbitrary set of colourless final-state particles in hadronic collisions, 
as long as the respective two-loop virtual amplitudes of the Born-level process are known. On the same basis
\Matrix{} automates also the small-\qt{} resummation of logarithmically enhanced terms at NNLL
accuracy (see \citere{Grazzini:2015wpa}, and \citere{Wiesemann:2016tae} for more details).

The core of the \Matrix{} framework is the Monte Carlo program \Munich{}, which includes 
a fully automated implementation of the Catani--Seymour dipole-subtraction method for massless~\cite{Catani:1996jh,Catani:1996vz}
and massive~\cite{Catani:2002hc} partons,
an efficient phase-space integration,
as well as an interface to the one-loop generator \OpenLoops{}~\cite{Cascioli:2011va,hepforge:OpenLoops} 
to obtain all required (spin- and colour-correlated) 
tree-level and one-loop amplitudes. 
The extension of \Munich{} and \OpenLoops{} to deal with EW 
corrections~\cite{Kallweit:2014xda,Kallweit:2015dum} allows for the fully automated computation 
of EW and QCD corrections to arbitrary SM processes at NLO accuracy.

Through an extension of \Munich{} by a generic implementation of the \qt{}-subtraction and -resummation 
techniques, \Matrix{} achieves NNLL+NNLO accuracy in QCD for the production of colourless final states
at a level of automation that is limited only 
by the process dependence of the two-loop amplitudes that enter the hard-collinear coefficient
${\cal H}^{\mathrm{F}}_{\mathrm{NNLO}}$. 
Any other process-dependent constituents of the calculation are formally (N)LO quantities
and can thus be automatically computed by \Munich{}+\OpenLoops{}.

In order to give some technical details on its practical implementation, we 
recall the master formula for \qt{}-subtraction for the calculation of
the $pp\to {\mathrm F}+X$ cross section at (N)NLO accuracy:
\begin{equation}
\label{eq:main}
\D{\sigma}^{\mathrm{F}}_{\mathrm{(N)NLO}}={\cal H}^{\mathrm{F}}_{\mathrm{(N)NLO}}\otimes \D{\sigma}^{\mathrm{F}}_{\mathrm{LO}}
+\left[ \D{\sigma}^{\mathrm{F + jet}}_{\mathrm{(N)LO}}-
\D{\sigma}^{\mathrm{CT}}_{\mathrm{(N)NLO}}\right].
\end{equation}
In \refeq{eq:main} the label $F$ denotes an arbitrary combination of colourless particles and $\D{\sigma}^{\mathrm{F + jet}}_{\mathrm{(N)LO}}$ is the 
(N)LO cross section for $\mathrm{F + jet}$ production. The explicit expression of the process-independent counterterm 
$\D{\sigma}^{\mathrm{CT}}_{\mathrm{(N)NLO}}$ is provided in \citere{Bozzi:2005wk}.
The general structure of the hard-collinear coefficient ${\cal H}^{\mathrm{F}}_{\mathrm{NLO}}$
is known from \citere{deFlorian:2001zd}, and that of ${\cal H}^{\mathrm{F}}_{\mathrm{NNLO}}$
from \citere{Catani:2013tia}. The latter exploits the explicit results for Higgs
\cite{Catani:2011kr} and vector-boson~\cite{Catani:2012qa} production.
More details on the implementation of \refeq{eq:main} in \Matrix{} can be found in \citere{Grazzini:2016ctr}.

The subtraction in the square brackets of \refeq{eq:main} is not local, but the cross section 
is formally finite in the limit $\qt\to 0$. In practice, a technical cut on $\qt$
is introduced to render $\D{\sigma}^{\mathrm{F + jet}}_{\mathrm{(N)LO}}$ and 
$\D{\sigma}^{\mathrm{CT}}_{\mathrm{(N)NLO}}$ separately finite.
In this respect, the \qt{}-subtraction method is 
very similar to a phase-space slicing method. 
It turns out that a cut, $r_{\mathrm{cut}}$, on the 
dimensionless quantity $r=\qt/M$, where $M$ 
denotes the invariant mass of $\mathrm{F}$, 
is more convenient from a practical point of view. 
The absence of any residual logarithmic
dependence on $r_{\mathrm{cut}}$ is a strong evidence of the correctness of the 
computation as any mismatch between the contributions would result in a divergence 
of the cross section when $r_{\mathrm{cut}}\to0$.
The remaining power-suppressed contributions vanish in that limit, and can be controlled by monitoring the $r_{\mathrm{cut}}$ dependence of the cross section.

\subsection{Stability of $\boldsymbol{q_T}$ subtraction for \wz{} production}
\label{subsec:stability}

\begin{figure}[t]
\begin{center}
\includegraphics[width=0.48\textwidth]{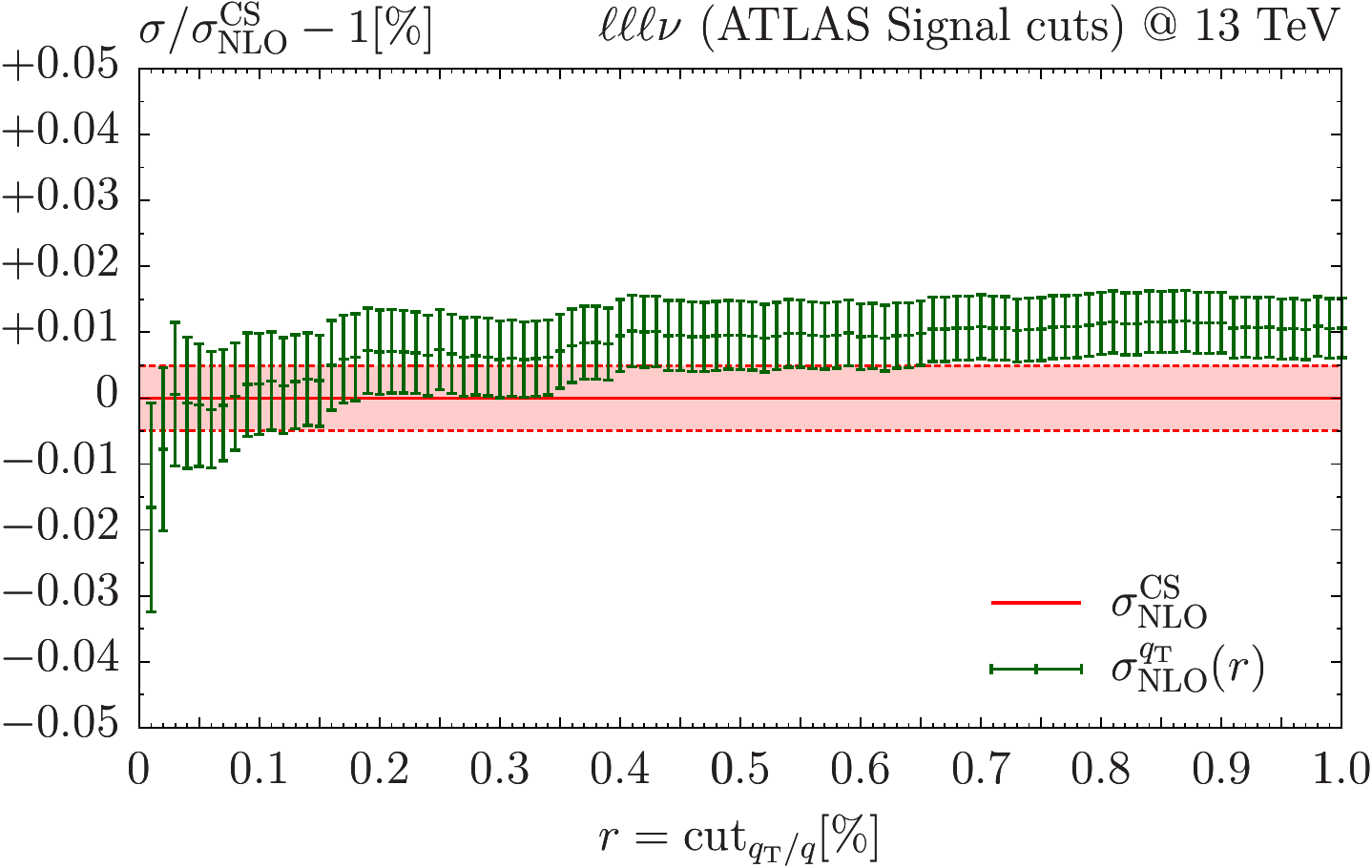}\hfill
\includegraphics[width=0.48\textwidth]{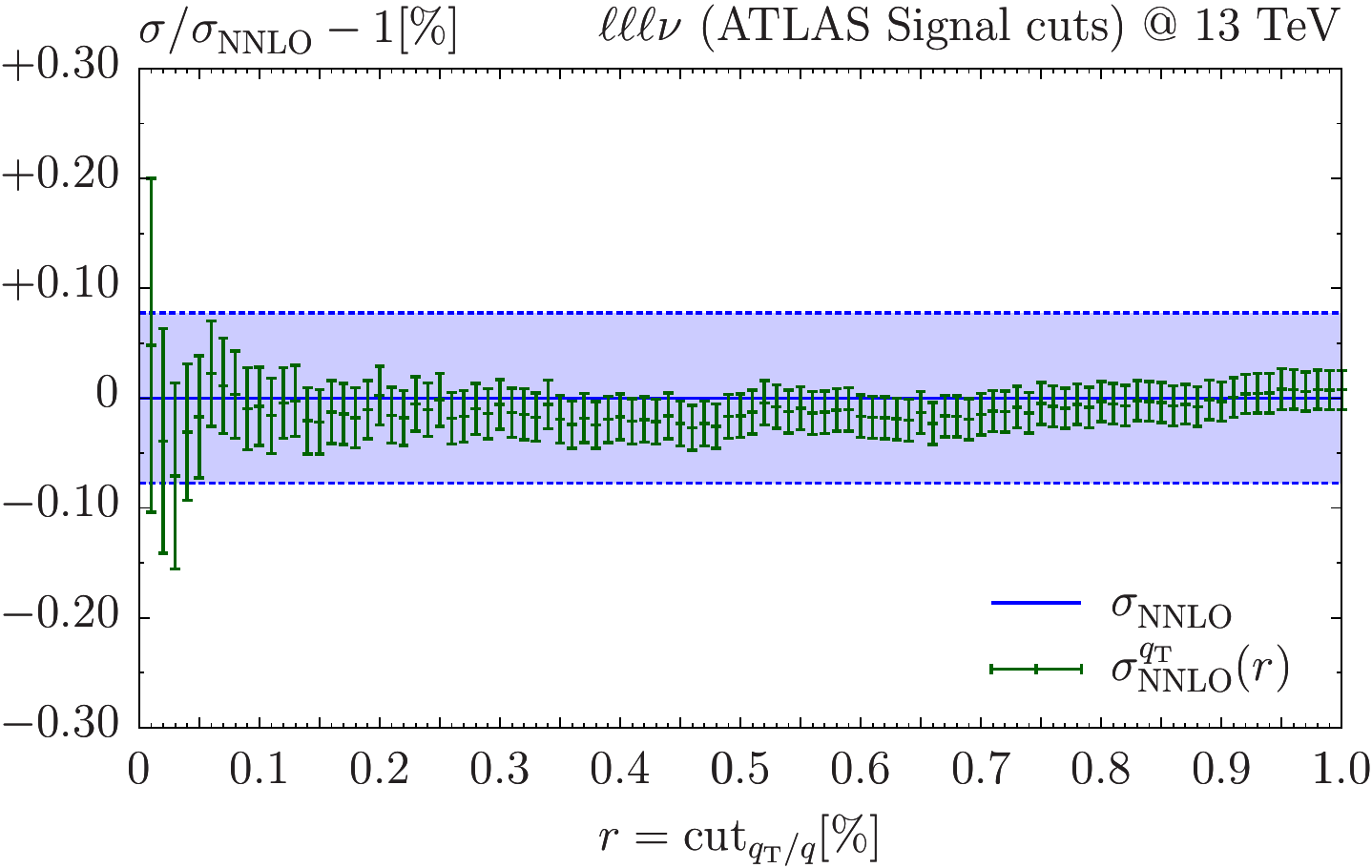}\\[1ex]
\includegraphics[width=0.48\textwidth]{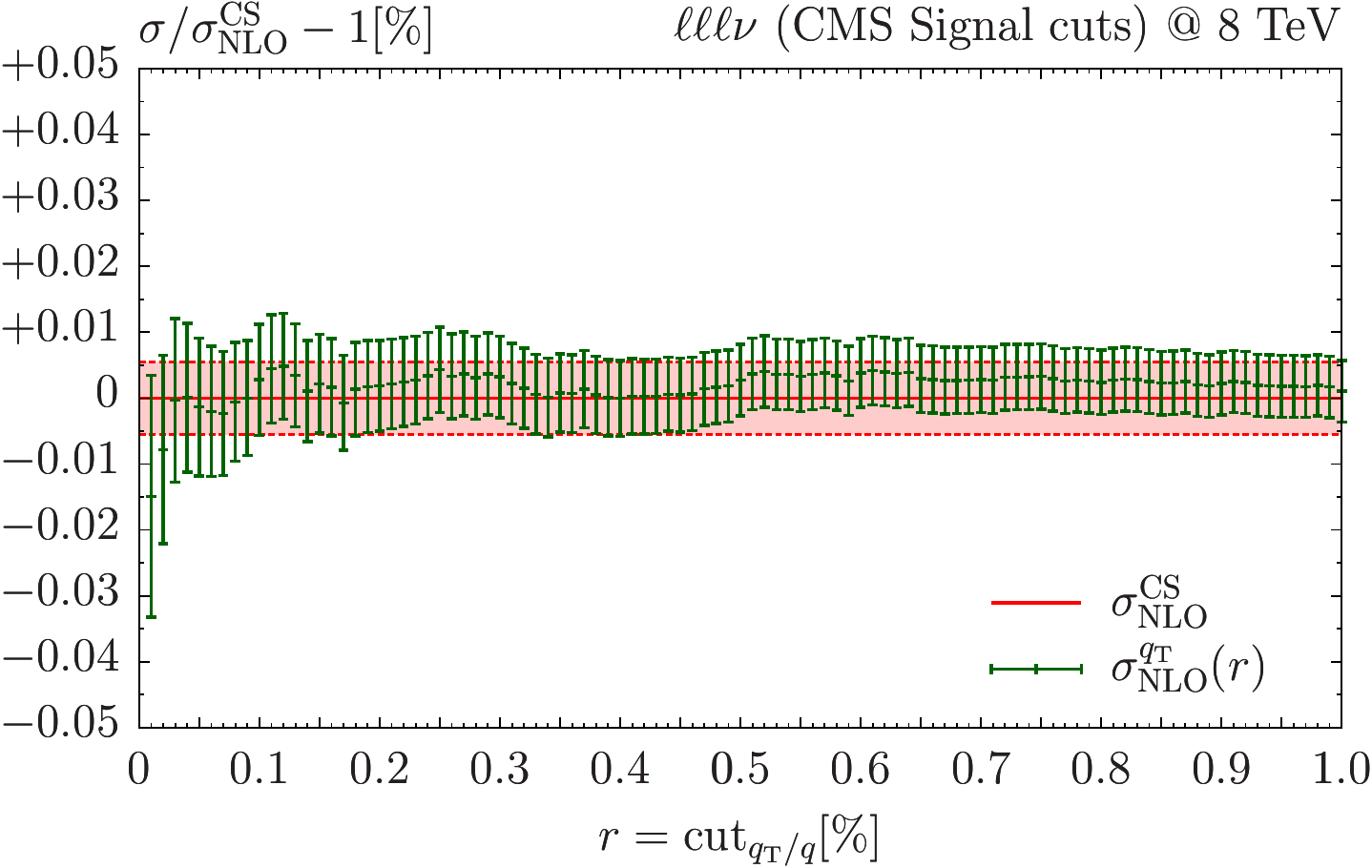}\hfill
\includegraphics[width=0.48\textwidth]{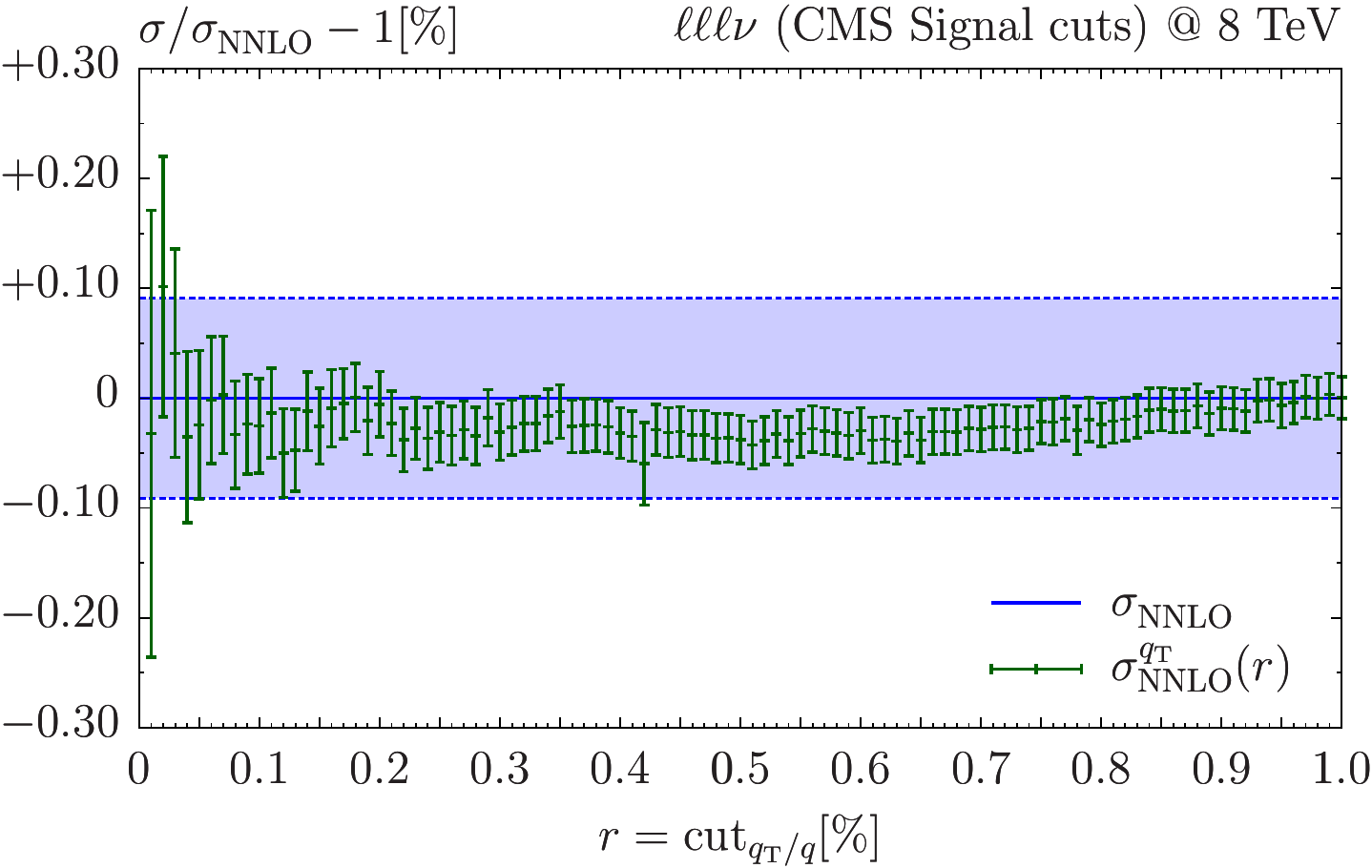}\\[1ex]
\caption[]{\label{fig:stability}{
Dependence of the $pp\to\llln+X$ cross sections on the \qt{}-subtraction cut, $\rcut{}$,
for both NLO (left plots) and NNLO (right plots) results in the ATLAS signal region at 13\,TeV (upper plots) and in the CMS signal region at 8\,TeV cuts (lower plots). 
 NLO results are normalized to the \rcut{}-independent NLO cross section computed with Catani--Seymour subtraction, 
and the NNLO results are normalized to their values at $\rcut\to0$, with a conservative extrapolation error indicated by the blue bands.
}
}
\end{center}
\end{figure}

In the following we investigate the stability of the \qt{}-subtraction
approach for $pp\to\llln+X$.  To this end, in~\fig{fig:stability}
we plot the NLO and NNLO cross sections as functions of the
\qt{}-subtraction cut, $\rcut$, which acts on the dimensionless variable
$r=\ptllln/\mllln$.  
Sample validation plots are presented for two
scenarios investigated in this paper, namely the ATLAS analysis at 13\,TeV and the CMS analysis at 8\,TeV (see \refse{sec:fiducial}), summed over all leptonic channels contributing to the \genllln{} final state.
All other scenarios considered in the paper lead essentially to the same conclusions.

At NLO the \rcut-independent cross section obtained with Catani--Seymour subtraction 
is used as a reference for the validation of the \qt{}-subtraction result.  
The comparison of the NLO cross sections in the left panels of \fig{fig:stability} 
demonstrates
that \qt{} subtraction agrees on the sub-permille level with the \rcut{}-independent result. This is true
already at the moderate value of $\rcut=1\%$. 

At NNLO, where an \rcut{}-independent control result is not available, 
we observe no significant, i.e.\ beyond the numerical uncertainties, \rcut{} dependence below about $\rcut=1\%$;
we thus use the finite-\rcut{} results to extrapolate to $\rcut=0$, 
taking into account the breakdown of predictivity for very low $\rcut$ values,
and conservatively estimate a numerical error due to the \rcut{} dependence of our results.\footnote{In the NNLO calculation the ${\cal O}(\as)$ contributions are evaluated by using Catani--Seymour subtraction.} This procedure allows us
to control all NNLO predictions for fiducial cross sections presented 
in \refse{sec:results} to better than one per mille
in terms of numerical uncertainties.
An analogous bin-wise extrapolation procedure was also performed for all distributions 
under consideration in \refse{sec:results}, and no significant dependence on \rcut{} 
was found, thus confirming the robustness of our results also at the differential level.

\section{Results}
\label{sec:results}

In this section we present our results on fiducial cross sections and distributions for \wz{} production in
proton--proton collisions defined in \eqn{eq:process}. We thus consider the inclusive production of three leptons 
and one neutrino including all possible flavour combinations, apart from channels involving $\tau$ leptons. 
In particular, this involves the SF channels $e^\pm e^+ e^-$ and $\mu^\pm \mu^+ \mu^-$ as well as the DF channels
$\mu^\pm e^+ e^-$ and $e^\pm \mu^+ \mu^-$.
Because of the availability of experimental results we consider LHC energies 
of $8$ and $13$ TeV and compare our predictions to the respective measurements by ATLAS and CMS. 
We finally study the impact of QCD radiative corrections when selection cuts designed for new physics searches
are applied.

For the input of the weak parameters we apply the $G_\mu$ scheme
with complex \w{}- and \z{}-boson masses to define the EW mixing angle as 
$\cos\theta_W^2=(m_W^2-i\Gamma_W\,m_W)/(m_Z^2-i\Gamma_Z\,m_Z)$. 
We use the PDG~\cite{Agashe:2014kda} values $G_F = 1.16639\times 10^{-5}$\,GeV$^{-2}$, 
$m_W=80.385$\,GeV, $\Gamma_W=2.0854$\,GeV, 
$m_Z = 91.1876$\,GeV, $\Gamma_Z=2.4952$\,GeV, and
$m_t = 173.2$\,GeV. 
The CKM matrix is set to unity.\footnote{The numerical effect of the CKM matrix up to NLO is to reduce the cross section by less than $1\%$. 
K-factors are generally affected below the numerical uncertainties.}

We consider $N_f=5$ massless quark flavours, and we use the corresponding NNPDF3.0~\cite{Ball:2014uwa} sets of parton distributions (PDFs) with $\as(m_Z)=0.118$. In particular N$^n$LO ($n=0,1,2$) predictions are obtained by using
PDFs at the respective perturbative order and the evolution of $\as$ at $(n+1)$-loop order, as provided by the PDF set.
Our reference choice for renormalization ($\mu_R$) and factorization ($\mu_F$) scales
is $\mu_R=\mu_F=\mu_0\equiv\frac{1}{2}(m_Z+m_W)=85.7863$\,GeV. Uncertainties from missing higher-order contributions
are estimated as usual
by independently varying $\mu_R$ and $\mu_F$ in the range
$0.5\mu_0\le \mu_R,\mu_F\le 2 \mu_0$, with the constraint $0.5\le \mu_R/\mu_F\le 2$.
We note that a fixed scale choice is only adequate as long as 
the scales in the kinematic distributions do not become too large, 
which is indeed the case in the fiducial phase-space regions of 
\wz{} measurements (see \refsetwo{sec:fiducial}{sec:distributions}).
As background in new-physics searches, on the other hand, that typically focus on the high-$\pT$ tails of distributions, a dynamic
scale is more appropriate, as discussed and applied in \refse{sec:results-np}.\footnote{In 
\wz{} measurements the tails of 
the $p_{T,Z}$ and $m_{T,WZ}$ (see \eqn{eq:mTWZ}) distributions are particularly sensitive to
triple-gauge couplings. In such high-$\pT$ regions, where also EW corrections play a non-negligible role, 
the choice of a dynamical scale turns out to be more appropriate. 
The extraction of the triple-gauge couplings, however, is not considered in the 
present paper.}

\subsection{Fiducial cross sections}
\label{sec:fiducial}

We start the presentation of our results by considering fiducial cross sections. We compute the \wz{} cross section up to NNLO in the same phase space
defined by the LHC experiments and compare our results with ATLAS data at 8 \cite{Aad:2016ett} and 13\,TeV \cite{Aaboud:2016yus},
and with CMS data at 13\,TeV \cite{Khachatryan:2016tgp}.
The selection cuts defining the ATLAS and CMS fiducial volumes are summarized in \tab{tab:cuts}.

\renewcommand{\baselinestretch}{1.5}
\newcommand{\lw}{\ensuremath{\ell_{\textrm{w}}}}
\newcommand{\lpw}{\ensuremath{\ell^+_{\textrm{w}}}}
\newcommand{\lmw}{\ensuremath{\ell^-_{\textrm{w}}}}
\newcommand{\lpmw}{\ensuremath{\ell^{\pm}_{\textrm{w}}}}

\newcommand{\lz}{\ensuremath{\ell_{\textrm{z}}}}
\newcommand{\lpz}{\ensuremath{\ell^+_{\textrm{z}}}}
\newcommand{\lmz}{\ensuremath{\ell^-_{\textrm{z}}}}
\newcommand{\lpmz}{\ensuremath{\ell^{\pm}_{\textrm{z}}}}
\newcommand{\lzlead}{\ensuremath{\ell_{\textrm{z},1}}}
\newcommand{\lzsubl}{\ensuremath{\ell_{\textrm{z},2}}}

\newcommand{\ptlz}{\ensuremath{p_{T,\lz}}}
\newcommand{\ptlw}{\ensuremath{p_{T,\lw}}}
\newcommand{\ptlzlead}{\ensuremath{p_{T,\lzlead}}}
\newcommand{\ptlzsubl}{\ensuremath{p_{T,\lzsubl}}}

\newcommand{\mll}{\ensuremath{m_{\ell\ell}}}
\newcommand{\mlll}{\ensuremath{m_{\ell\ell\ell}}}
\newcommand{\mwz}{\ensuremath{m_{WZ}}}
\newcommand{\mtw}{\ensuremath{m_{T,W}}}
\newcommand{\ptwz}{\ensuremath{p_{T,WZ}}}
\newcommand{\ptw}{\ensuremath{p_{T,W}}}
\newcommand{\ptz}{\ensuremath{p_{T,Z}}}
\newcommand{\ptl}{\ensuremath{p_{T,\ell}}}
\newcommand{\ptlp}{\ensuremath{p_{T,\ell'}}}
\newcommand{\ptlone}{\ensuremath{p_{T,\ell_1}}}
\newcommand{\ptltwo}{\ensuremath{p_{T,\ell_2}}}
\newcommand{\ptlsub}{\ensuremath{p_{T,\ell_{\ge2}}}}
\newcommand{\ptmiss}{\ensuremath{p_{T}^{\text{miss}}}}
\newcommand{\dphill}{\ensuremath{\Delta\phi_{\ell\ell}}}
\newcommand{\dRll}{\ensuremath{\Delta R_{\ell\ell}}}
\newcommand{\dRllp}{\ensuremath{\Delta R_{\ell\ell'}}}
\newcommand{\etal}{\ensuremath{\eta_{\ell}}}
\newcommand{\etalp}{\ensuremath{\eta_{\ell'}}}

\begin{table}
\begin{center}
\begin{tabular}{c|c}
\toprule
& definition of the fiducial volume for $pp\to \ell'^\pm{\nu}_{\ell^\prime} \ell^+\ell^-+X,\quad \ell,\ell'\in\{e,\mu\}$\\
\midrule
ATLAS 8/13 TeV & $\ptlz>15$\,GeV, \quad$\ptlw>20$\,GeV, \quad$|\etal|<2.5$, \\
(cf. \citere{Aaboud:2016yus,Aad:2016ett})& $|m_{\lz\lz}-m_Z|<10$\,GeV,\quad $\mtw> 30$\,GeV, \quad $\Delta R_{\lz\lz} >0.2$, \quad $\Delta R_{\lz\lw}>0.3$\\
\midrule
CMS 13 TeV & $\ptlzlead>20$\,GeV, \quad$\ptlzsubl>10$\,GeV, \quad$\ptlw>20$\,GeV, \quad$|\etal|<2.5$,\\
(cf. \citere{Khachatryan:2016tgp}) & $60$\,GeV$<m_{\lz\lz}<120$\,GeV, \quad $m_{\ell^+\ell^-}>4$\,GeV\\
\bottomrule
\end{tabular}
\end{center}
\renewcommand{\baselinestretch}{1.0}
\caption{\label{tab:cuts} Definition of the fiducial volume of the \wz{} measurements by ATLAS and CMS. While $\ell$ refers to all charged leptons, $\lz$ and $\lw$ denote the leptons assigned to the $Z$ and $W$ boson decay, according to the procedure described in the text. Numbers in indices refer to $\pT$-ordered particles of the respective group.}
\end{table}

\renewcommand{\baselinestretch}{1.0}

The fiducial cuts used by ATLAS are identical at both collider energies, and they are close to the applied 
event-selection cuts \cite{Aad:2016ett,Aaboud:2016yus}.
The cuts require an identification of the leptons stemming from the $Z$ and $W$ bosons. 
This is trivial in the DF channel, where they are unambiguously assigned to the parent boson. In the 
SF channel, there are, in a theoretical computation of \genllln{} production, 
two possible combinations of opposite-sign leptons that can be matched to the $Z$ boson. 
ATLAS applies the so-called resonant-shape procedure \cite{Aad:2016ett}, where, among the two possible assignments,
the one that maximizes the estimator
\begin{align}
\label{eq:pestimator}
P = \Bigg|\frac1{m^2_{\ell\ell}-m^2_Z+i\,\Gamma_Z\,m_Z}\Bigg|^2 \,\cdot\Bigg|\frac1{m^2_{\ell'\nu_{\ell'}}-m^2_W+i\,\Gamma_W\,m_W}\Bigg|^2
\end{align}
is chosen. After this identification, the cuts involve standard requirements on the 
transverse momenta and pseudo-rapidities of the leptons as well as lepton 
separations in the $R=\sqrt{\eta^2+\phi^2}$ plane. 
The latter already regularize all possible divergences from collinear $\gamma^\ast\to\elle^-\elle^+$ splittings by implying 
an effective invariant-mass cut on each OSSF lepton pair.
The invariant mass of the lepton pair assigned to the $Z$-boson decay is further 
required not to deviate by more than $10$\,GeV from the $Z$-boson mass,
and the transverse mass of the $W$ boson, defined as
\begin{align}
\label{eq:mTW}
m_{T,W} =\sqrt{ \left(E_{T,{\lw}}+E_{T,\nu_{\lw}}\right)^2 - p_{T,(\lw\nu_{\lw})= W}^2}\quad \mathrm{with} \quad E_{T,x}^2=m_x^2+p_{T,x}^2,
\end{align}
is bounded from below.

A CMS measurement of the fiducial cross section is available only at 13\,TeV \cite{Khachatryan:2016tgp}. The analysis applies a simple 
identification of the leptons in the SF channel by associating the lepton pair 
whose invariant mass is closest to the $Z$-boson mass with the $Z$-boson decay. 
The leptons then must meet standard requirements on 
their transverse momenta and pseudo-rapidities, which are chosen differently 
for the hardest and second-hardest lepton assigned to the $Z$-boson decay and 
for the lepton from the $W$ boson. 
Additionally, the invariant mass of the lepton pair associated with 
the $Z$ boson is required to be in a fixed range around the $Z$-boson mass. To guarantee infrared safety in the SF channel
in spite of possible divergences from collinear $\gamma^\ast\to\elle^-\elle^+$ splittings, this requirement is supplemented 
by a lower 4\,GeV cut on the invariant mass of any OSSF lepton pair. 

We note that the CMS selection cuts at the detector level
are somewhat different from those defining the 
fiducial volume \cite{Khachatryan:2016tgp}. In particular, the invariant-mass cut on the identified lepton pair from the $Z$ 
boson is much tighter than in the fiducial volume, and 
a $b$-jet veto is applied at detector level, which is absent in the definition of the fiducial phase space. 
As a meaningful comparison to theoretical predictions can only be pursued at the fiducial level, these 
differences require an extrapolation from the detector to the fiducial level, which could lead
to additional theoretical uncertainties.

\subsubsection{ATLAS 8\,TeV}
\label{sec:atlas8}

\renewcommand{\baselinestretch}{1.5}
\begin{table}[t]
\begin{center}
\resizebox{\columnwidth}{!}{%
\begin{tabular}{c c c c c}
\toprule
channel
& $\sigma_{\textrm{LO}}$ [fb]
& $\sigma_{\textrm{NLO}}$ [fb]
& $\sigma_{\textrm{NNLO}}$ [fb]
& $\sigma_{\textrm{ATLAS}}$ [fb]\\
\bottomrule
$\mu^+ e^+e^-$ & \multirow{ 2}{*}{$11.59(0)_{-3.0\%}^{+2.2\%}$} & \multirow{ 2}{*}{$20.42(0)_{-4.0\%}^{+5.3\%}$} & \multirow{ 2}{*}{$22.11(1)_{-1.9\%}^{+1.8\%}$} & $23.9\,\pm 6.5\%{\rm (stat)}\pm 2.5\%{\rm (syst)}\pm 2.2\%{\rm (lumi)}$\Bstrut\\
$e^+ \mu^+\mu^-$ &  &  &  &   $19.9\,\pm 7.2\%{\rm (stat)}\pm 3.5\%{\rm (syst)}\pm 2.2\%{\rm (lumi)}$\Bstrut\\
$e^+ e^+e^-$ & \multirow{ 2}{*}{$11.62(0)_{-3.0\%}^{+2.2\%}$} & \multirow{ 2}{*}{$20.48(0)_{-4.0\%}^{+5.3\%}$} & \multirow{ 2}{*}{$22.17(1)_{-1.9\%}^{+1.8\%}$} & $22.6\,\pm 8.0\%{\rm (stat)}\pm 4.4\%{\rm (syst)}\pm 2.2\%{\rm (lumi)}$\Bstrut\\
$\mu^+ \mu^+\mu^-$ &  &  &  &   $19.8\,\pm 6.0\%{\rm (stat)}\pm 2.5\%{\rm (syst)}\pm 2.2\%{\rm (lumi)}$\Bstrut\\
\midrule
combined & $11.60(0)_{-3.0\%}^{+2.2\%}$ & $20.45(0)_{-4.0\%}^{+5.3\%}$ & $22.14(1)_{-1.9\%}^{+1.8\%}$ & $21.2\,\pm 3.4\%{\rm (stat)}\pm 2.3\%{\rm (syst)}\pm 2.2\%{\rm (lumi)}$\Bstrut\\
\bottomrule
$\mu^- e^+e^-$ & \multirow{ 2}{*}{$6.732(1)_{-3.4\%}^{+2.4\%}$} & \multirow{ 2}{*}{$12.35(0)_{-4.3\%}^{+5.7\%}$} & \multirow{ 2}{*}{$13.42(1)_{-1.9\%}^{+1.9\%}$} & $12.4\,\pm 9.5\%{\rm (stat)}\pm 3.1\%{\rm (syst)}\pm 2.3\%{\rm (lumi)}$\Bstrut\\
$e^- \mu^+\mu^-$ &  &  &  &   $15.7\,\pm 7.5\%{\rm (stat)}\pm 2.8\%{\rm (syst)}\pm 2.3\%{\rm (lumi)}$\Bstrut\\
$e^- e^+e^-$ & \multirow{ 2}{*}{$6.750(1)_{-3.4\%}^{+2.4\%}$} & \multirow{ 2}{*}{$12.38(0)_{-4.3\%}^{+5.7\%}$} & \multirow{ 2}{*}{$13.47(1)_{-2.0\%}^{+1.9\%}$} & $15.4\,\pm 9.8\%{\rm (stat)}\pm 5.0\%{\rm (syst)}\pm 2.3\%{\rm (lumi)}$\Bstrut\\
$\mu^- \mu^+\mu^-$ &  &  &  &   $13.4\,\pm 7.5\%{\rm (stat)}\pm 2.8\%{\rm (syst)}\pm 2.3\%{\rm (lumi)}$\Bstrut\\
\midrule
combined & $6.741(1)_{-3.4\%}^{+2.4\%}$ & $12.36(0)_{-4.3\%}^{+5.7\%}$ & $13.45(1)_{-2.0\%}^{+1.9\%}$ & $14.0\,\pm 4.3\%{\rm (stat)}\pm 2.8\%{\rm (syst)}\pm 2.3\%{\rm (lumi)}$\Bstrut\\
\bottomrule
$\mu^\pm e^+e^-$ & \multirow{ 2}{*}{$18.32(0)_{-3.2\%}^{+2.3\%}$} & \multirow{ 2}{*}{$32.76(1)_{-4.1\%}^{+5.4\%}$} & \multirow{ 2}{*}{$35.53(2)_{-1.9\%}^{+1.8\%}$} & $36.3\,\pm 5.4\%{\rm (stat)}\pm 2.6\%{\rm (syst)}\pm 2.2\%{\rm (lumi)}$\Bstrut\\
$e^\pm \mu^+\mu^-$ &  &  &  &   $35.7\,\pm 5.3\%{\rm (stat)}\pm 3.7\%{\rm (syst)}\pm 2.2\%{\rm (lumi)}$\Bstrut\\
$e^\pm e^+e^-$ & \multirow{ 2}{*}{$18.37(0)_{-3.2\%}^{+2.3\%}$} & \multirow{ 2}{*}{$32.85(1)_{-4.1\%}^{+5.4\%}$} & \multirow{ 2}{*}{$35.64(2)_{-1.9\%}^{+1.8\%}$} &  $38.1\,\pm 6.2\%{\rm (stat)}\pm 4.5\%{\rm (syst)}\pm 2.2\%{\rm (lumi)}$\Bstrut\\
$\mu^\pm \mu^+\mu^-$ &  &  &  &  $33.3\,\pm 4.7\%{\rm (stat)}\pm 2.5\%{\rm (syst)}\pm 2.2\%{\rm (lumi)}$\Bstrut\\
\midrule
combined & $18.35(0)_{-3.2\%}^{+2.3\%}$ & $32.81(1)_{-4.1\%}^{+5.4\%}$ & $35.59(2)_{-1.9\%}^{+1.8\%}$ & $35.1\,\pm 2.7\%{\rm (stat)}\pm 2.4\%{\rm (syst)}\pm 2.2\%{\rm (lumi)}$\Bstrut\\
\bottomrule

\end{tabular}}
\end{center}
\renewcommand{\baselinestretch}{1.0}
\caption{\label{tab:ATLAS8} Fiducial cross sections for ATLAS 8 TeV. Note that due to the flavour-unspecific lepton cuts the theoretical predictions are flavour-blind, which is why the results are symmetric under $e \leftrightarrow \mu$ exchange. The available ATLAS data from \citere{Aad:2016ett} are also shown. ``Combined'' refers to the \textit{average} of different lepton channels.}
\end{table}

\renewcommand{\baselinestretch}{1.0}

ATLAS presents their fiducial results split into both 
SF/DF channels and $W^-Z$/$W^+Z$ production \cite{Aad:2016ett}. In \tab{tab:ATLAS8}
we compare our theoretical predictions for the fiducial rates at LO, NLO and NNLO at 8\,TeV 
to the measured cross sections. Since the cuts do not depend on the lepton flavour, the theoretical 
predictions are identical when exchanging electrons and muons, e.g. 
$\sigma(\mu^+\nu_\mu e^+e^-)\equiv\sigma(e^+\nu_e \mu^+\mu^-)$.
The statistical uncertainties of the experimental results are strongly reduced 
upon combination, from $\sim 5\%-10\%$ for the individual channels to $3\%-4\%$ when combined.

For proton--proton collisions the cross sections in the $W^+Z$ and $W^-Z$ channels 
are different due to their charge-conjugate partonic initial states:
The $W^+Z$ final state is mainly produced through $u{\bar d}$ scattering (see \fig{fig:Borndiagrams}), while $W^-Z$ originates from ${\bar u}d$ scattering. 
Roughly speaking, the $u$ valence density is larger than the $d$ valence density and ${\bar u}\sim {\bar d}$, so we have $\sigma_{W^+Z}>\sigma_{W^-Z}$.

It is clear from \tab{tab:ATLAS8} that the inclusion of higher-order corrections is crucial for a proper 
prediction of the fiducial cross sections. NLO corrections have the effect of increasing
the corresponding LO results by 
up to $85\%$, and the NNLO effects further increase the NLO result by about $10\%$. 
The LO cross section is thus increased by almost a factor of two upon inclusion of higher-order 
corrections. The scale uncertainties are reduced from about $4\%-6\%$ at NLO to only about $2\%$ at NNLO.
The inclusion of NNLO corrections nicely
improves the agreement between the theoretical predictions and the data, which are largely consistent within the uncertainties.

These observations are irrespective of whether $W^+Z$, $W^-Z$ or their combination is 
considered, and very similar to what has been found for the total inclusive cross sections in \citere{Grazzini:2016swo}.
As pointed out there, the origin of the large radiative corrections is an approximate radiation zero \cite{Baur:1994ia}: 
The LO cross section in the leading helicity amplitude vanishes at a specific scattering angle of the $W$ boson in the centre-of-mass frame.
This phase-space region is filled only upon inclusion of higher-order contributions, 
thereby effectively decreasing the perturbative accuracy in that region by one order.
Therefore, the perturbative uncertainties at LO and NLO,
estimated from scale variations, fail to cover the actual size of missing higher-order corrections.
Nonetheless, the convergence of the perturbative series is noticeably improved beyond LO, and
we expect NNLO scale uncertainties to provide the correct size of yet uncalculated perturbative contributions.

\subsubsection{ATLAS 13\,TeV}

\renewcommand{\baselinestretch}{1.5}
\begin{table}
\begin{center}
\resizebox{\columnwidth}{!}{%
\begin{tabular}{c c c c c}
\toprule
channel
& $\sigma_{\textrm{LO}}$ [fb]
& $\sigma_{\textrm{NLO}}$ [fb]
& $\sigma_{\textrm{NNLO}}$ [fb]
& $\sigma_{\textrm{ATLAS}}$ [fb]\\
\bottomrule

$\mu^+ e^+e^-$ & \multirow{ 2}{*}{$17.33(0)_{-6.3\%}^{+5.3\%}$} & \multirow{ 2}{*}{$34.12(1)_{-4.3\%}^{+5.3\%}$} & \multirow{ 2}{*}{$37.75(2)_{-2.0\%}^{+2.3\%}$}& $32.2\,\pm 14.4\%{\rm (stat)}\pm 5.0\%{\rm (syst)}\pm 2.4\%{\rm (lumi)}$\Bstrut\\
$e^+ \mu^+\mu^-$ &  &  &  &   $45.0\,\pm 12.1\%{\rm (stat)}\pm 4.6\%{\rm (syst)}\pm 2.3\%{\rm (lumi)}$\Bstrut\\
$e^+ e^+e^-$ & \multirow{ 2}{*}{$17.37(0)_{-6.3\%}^{+5.3\%}$} & \multirow{ 2}{*}{$34.21(1)_{-4.3\%}^{+5.3\%}$} & \multirow{ 2}{*}{$37.84(2)_{-2.0\%}^{+2.2\%}$} & $28.0\,\pm 19.2\%{\rm (stat)}\pm 11.2\%{\rm (syst)}\pm 2.4\%{\rm (lumi)}$\Bstrut\\
$\mu^+ \mu^+\mu^-$ &  &  &  &   $36.5\,\pm 11.6\%{\rm (stat)}\pm 4.1\%{\rm (syst)}\pm 2.3\%{\rm (lumi)}$\Bstrut\\
\midrule
combined & $17.35(0)_{-6.3\%}^{+5.3\%}$ & $34.16(1)_{-4.3\%}^{+5.3\%}$ & $37.80(2)_{-2.0\%}^{+2.2\%}$ & $36.7\,\pm 6.7\%{\rm (stat)}\pm 3.9\%{\rm (syst)}\pm 2.3\%{\rm (lumi)}$\Bstrut\\
\bottomrule
$\mu^- e^+e^-$ & \multirow{ 2}{*}{$11.50(0)_{-6.8\%}^{+5.7\%}$} & \multirow{ 2}{*}{$23.57(1)_{-4.5\%}^{+5.5\%}$} & \multirow{ 2}{*}{$26.18(1)_{-2.1\%}^{+2.3\%}$}  & $22.9\,\pm 17.5\%{\rm (stat)}\pm 5.8\%{\rm (syst)}\pm 2.4\%{\rm (lumi)}$\Bstrut\\
$e^- \mu^+\mu^-$ &  &  &  &    $30.2\,\pm 15.2\%{\rm (stat)}\pm 6.9\%{\rm (syst)}\pm 2.3\%{\rm (lumi)}$\Bstrut\\
$e^- e^+e^-$ & \multirow{ 2}{*}{$11.53(0)_{-6.8\%}^{+5.7\%}$} & \multirow{ 2}{*}{$23.63(0)_{-4.5\%}^{+5.5\%}$} & \multirow{ 2}{*}{$26.25(1)_{-2.1\%}^{+2.2\%}$} & $22.5\,\pm 21.0\%{\rm (stat)}\pm 10.5\%{\rm (syst)}\pm 2.4\%{\rm (lumi)}$\Bstrut\\
$\mu^- \mu^+\mu^-$ &  &  &  &   $27.1\,\pm 13.7\%{\rm (stat)}\pm 5.0\%{\rm (syst)}\pm 2.4\%{\rm (lumi)}$\Bstrut\\
\midrule
combined & $11.51(0)_{-6.8\%}^{+5.7\%}$ & $23.60(1)_{-4.5\%}^{+5.5\%}$ & $26.22(1)_{-2.1\%}^{+2.3\%}$ & $26.1\,\pm 8.1\%{\rm (stat)}\pm 4.7\%{\rm (syst)}\pm 2.4\%{\rm (lumi)}$\Bstrut\\
\bottomrule
$\mu^\pm e^+e^-$ & \multirow{ 2}{*}{$28.83(0)_{-6.5\%}^{+5.4\%}$} & \multirow{ 2}{*}{$57.69(1)_{-4.3\%}^{+5.4\%}$} & \multirow{ 2}{*}{$63.93(3)_{-2.1\%}^{+2.3\%}$}  & $55.1\,\pm 11.1\%{\rm (stat)}\pm 5.1\%{\rm (syst)}\pm 2.4\%{\rm (lumi)}$\Bstrut\\

$e^\pm \mu^+\mu^-$ &  &  &  &   $75.2\,\pm 9.5\%{\rm (stat)}\pm 5.3\%{\rm (syst)}\pm 2.3\%{\rm (lumi)}$\Bstrut\\
$e^\pm e^+e^-$ & \multirow{ 2}{*}{$28.90(0)_{-6.5\%}^{+5.4\%}$} & \multirow{ 2}{*}{$57.84(1)_{-4.3\%}^{+5.4\%}$} & \multirow{ 2}{*}{$64.09(3)_{-2.1\%}^{+2.2\%}$}  & $50.5\,\pm 14.2\%{\rm (stat)}\pm 10.6\%{\rm (syst)}\pm 2.4\%{\rm (lumi)}$\Bstrut\\
$\mu^\pm \mu^+\mu^-$ &  &  &  &   $63.6\,\pm 8.9\%{\rm (stat)}\pm 4.1\%{\rm (syst)}\pm 2.3\%{\rm (lumi)}$\Bstrut\\
\midrule
combined & $28.86(0)_{-6.5\%}^{+5.4\%}$ & $57.76(1)_{-4.3\%}^{+5.4\%}$ & $64.01(3)_{-2.1\%}^{+2.3\%}$ & $63.2\,\pm 5.2\%{\rm (stat)}\pm 4.1\%{\rm (syst)}\pm 2.4\%{\rm (lumi)}$\Bstrut\\
\bottomrule

\end{tabular}}
\end{center}
\renewcommand{\baselinestretch}{1.0}
\caption{\label{tab:ATLAS13} Fiducial cross sections for ATLAS 13 TeV. Note that due to the flavour-unspecific lepton cuts the theoretical predictions are flavour-blind, which is why the results are symmetric under $e \leftrightarrow \mu$ exchange. The available ATLAS data from \citere{Aaboud:2016yus} are also shown. ``Combined'' refers to the \textit{average} of different lepton channels.}
\vspace{0.75cm}
\renewcommand{\baselinestretch}{1.0}
\renewcommand{\baselinestretch}{1.5}
\begin{center}
\resizebox{\columnwidth}{!}{%
\begin{tabular}{c c c c c}
\toprule
channel
& $\sigma_{\textrm{LO}}$ [fb]
& $\sigma_{\textrm{NLO}}$ [fb]
& $\sigma_{\textrm{NNLO}}$ [fb]
& $\sigma_{\textrm{CMS}}$ [fb]\\
\midrule
combined & $148.4(0)^{+5.4\%}_{-6.4\%}$ & $301.4(1)^{+5.5\%}_{-4.5\%}$ & $334.3(2)^{+2.3\%}_{-2.1\%}$ & \quad\,$258\,\pm 8.1\%{\rm (stat)}^{+7.4\%}_{-7.7\%}{\rm (syst)}\pm 3.1{\rm (lumi)}$\quad\,\Bstrut\\\bottomrule
\end{tabular}}
\end{center}
\renewcommand{\baselinestretch}{1.0}
\caption{\label{tab:CMS13} Fiducial cross sections for CMS 13 TeV. 
The available CMS data from \citere{Khachatryan:2016tgp} are also shown. ``Combined'' refers to the \textit{sum} of all separate contributions. Our theoretical predictions for all individual channels for CMS at 8 TeV and 13 TeV can be found in \app{app:rates_full}.}
\end{table}

\renewcommand{\baselinestretch}{1.0}

ATLAS has reported experimental results of the fiducial \wz{} cross section also for the early 13\,TeV data set collected in 2015 \cite{Aaboud:2016yus}.
At the level of the inclusive cross section very good agreement with our NNLO computation of \citere{Grazzini:2016swo} is quoted.
\tab{tab:ATLAS13} confirms that agreement also for the fiducial cross sections.
There is also a marked improvement of the accuracy 
of the NNLO cross section regarding its scale uncertainties, which have been reduced to $\sim 2\%$ from $\sim 4\%-6\%$ 
at NLO. Overall, the findings at 13\,TeV draw essentially the same picture as those at 8\,TeV discussed in the previous section.

\subsubsection{CMS 13\,TeV}
CMS provides a cross-section measurement in the fiducial phase space for \wz{} production 
only for their 13\,TeV analysis, and summed over all individual lepton 
channels \cite{Khachatryan:2016tgp}.\footnote{The 8\,TeV \wz{} measurement by CMS \cite{Khachatryan:2016poo} 
does not provide fiducial cross sections, and the differential results are extrapolated 
to the full phase space. Since such results depend 
on the underlying Monte Carlo used for the extrapolation, we refrain from 
including them in our comparison. The full set of predictions for all individual channels 
for CMS at 8 TeV and 13 TeV are reported in \app{app:rates_full}.}
\tab{tab:CMS13} contains our theoretical predictions at LO, NLO and NNLO for the 
combination of all leptonic channels. The cuts are 
looser as compared to the ones applied by ATLAS, but the relative size of radiative corrections is rather 
similar. The comparison to the fiducial cross section measured by CMS 
shows quite a large discrepancy:
The theoretical prediction is $2.6\sigma$ above the experimental result. 
We point out that
CMS uses fiducial cuts that are quite different from those used in their event-selection. 
This comes at the price that the extrapolation from the CMS selection cuts 
to the fiducial phase space is affected by an uncertainty from the
employed Monte Carlo generator.
The observed discrepancy, however, might well be due to 
a statistical fluctuation of the limited dataset used in this early measurement.
Further data collection at 13\,TeV will hopefully clarify this issue.

\subsection{Distributions in the fiducial phase space}
\label{sec:distributions}
\begin{figure}
\begin{center}
\begin{tabular}{cc}
\hspace*{-0.17cm}
\includegraphics[trim = 7mm -7mm 0mm 0mm, width=.33\textheight]{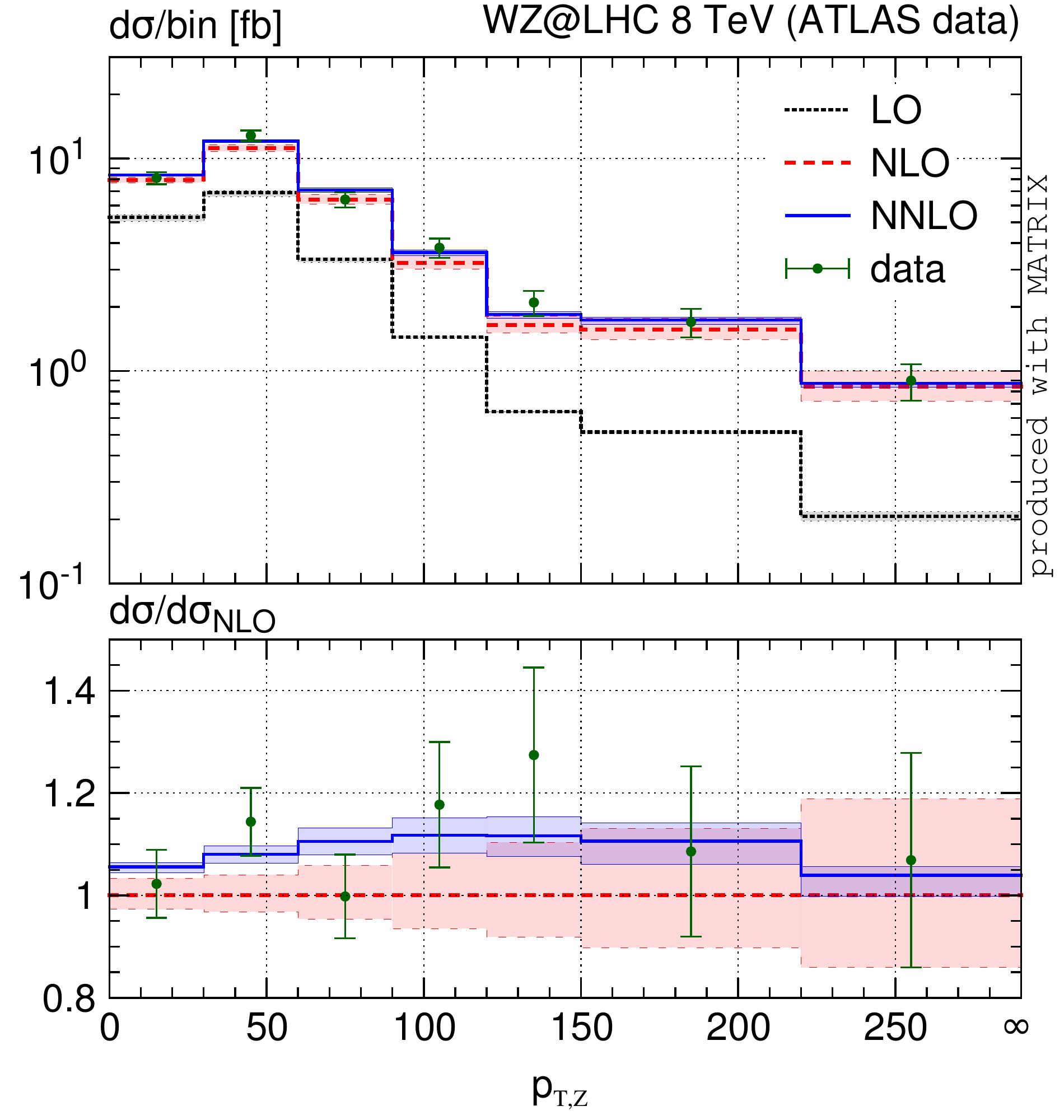} &
\includegraphics[trim = 7mm -7mm 0mm 0mm, width=.33\textheight]{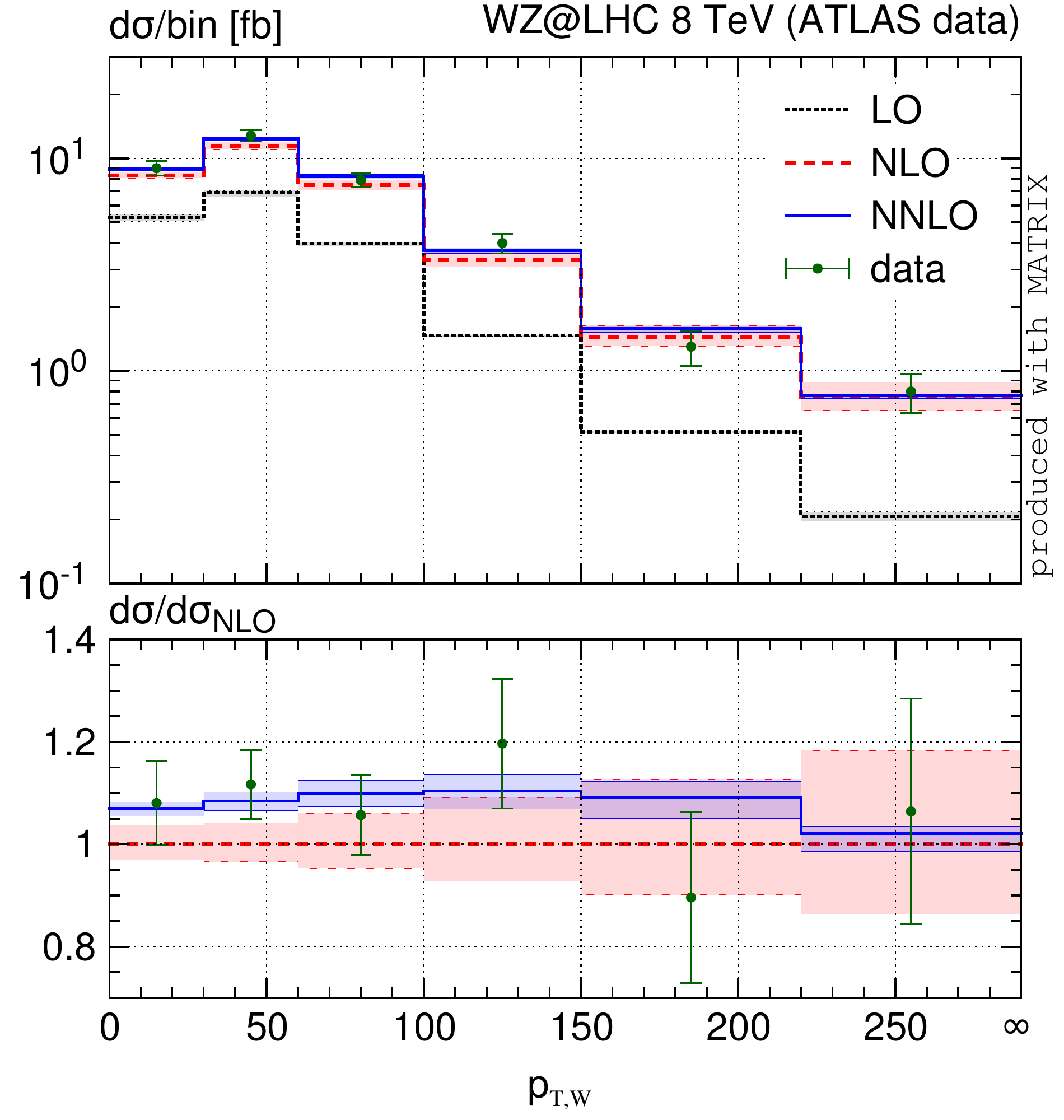} \\[-1em]
\hspace{0.6em} (a) & \hspace{1em}(b)
\end{tabular}
\caption[]{\label{fig:pTZW}{
Distribution in the transverse momentum of the reconstructed (a) $Z$ and (b) $W$ bosons at LO (black, dotted), NLO 
(red, dashed) and NNLO (blue, solid) compared to the corresponding ATLAS data at 8\,TeV (green points with error bars). 
The lower panel shows the ratio over the NLO prediction.}}
\end{center}
\end{figure}

We now turn to the discussion of differential observables in the fiducial phase space. 
In \reffis{fig:pTZW}{fig:dyZlWnjet} we consider 
predictions up to NNLO accuracy for various distributions that have been measured by ATLAS at 8\,TeV \cite{Aad:2016ett}. The fiducial phase-space definition 
is discussed in \sct{sec:fiducial}, see also \tab{tab:cuts}. All figures have the identical layout:
The main frame shows the predictions at LO (black dotted histogram), 
NLO (red dashed histogram) and NNLO (blue solid histogram) with their absolute normalization 
as cross section per bin (i.e.\ the sum of the bins is equal to the fiducial cross section),
compared to the cross sections measured by ATLAS (green data points with error bars). The lower panel 
displays the respective bin-by-bin ratios normalized to the NLO prediction (LO is not shown here).
The shaded uncertainty bands of the theoretical predictions correspond to scale variations as discussed 
above, and the error bars are the combined experimental uncertainties quoted by ATLAS.
Unless stated otherwise, all distributions include the combination of all relevant 
leptonic channels (SF/DF channels and $W^+Z/W^-Z$ production).
Note that, in order to compare to ATLAS results, we combine different lepton channels by averaging them
for both the fiducial cross sections and distributions, while summing the cross sections for $W^+Z$ and $W^-Z$ production.

Some general statements regarding the scale uncertainties which are common to all 
subsequent plots are in order: NNLO corrections further reduce the scale dependence of the NLO cross sections in 
all distributions. In absolute terms, the NLO uncertainties generally vary within 5\%$-$10\%, and reach 
up to $20\%$ only in the tails of some transverse-momentum distributions. The NNLO uncertainties, 
on the other hand, hardly ever exceed $5\%$ in all differential observables.
Correspondingly, given that the NNLO corrections on the fiducial rate are about $+8.5\%$, NLO and NNLO scale-uncertainty bands mostly do not overlap, in particular in the bins that provide the bulk of the cross section. Nonetheless, we expect NNLO uncertainties to generally provide the correct size of missing higher-order contributions (see our corresponding comments at the end of \refse{sec:atlas8}).

\fig{fig:pTZW} shows the transverse-momentum spectra of the reconstructed $Z$ and $W$ bosons, which both 
peak around $p_{\mathrm{T,}V}\sim 30$\,GeV. 
As can be seen from the ratio plots,
the inclusion of NNLO corrections affects the shapes of both distributions at the 10\% level,
the effect being largest in the region $p_{\mathrm{T,}V}\ltap 150$\,GeV. The
comparison with the data is good already at NLO, but it is further improved,
in particular in terms of shape, at NNLO. 
All data points agree within roughly $1\sigma$ with the NNLO predictions.

\begin{figure}
\begin{center}
\begin{tabular}{cc}
\hspace*{-0.17cm}
\includegraphics[trim = 7mm -7mm 0mm 0mm, width=.33\textheight]{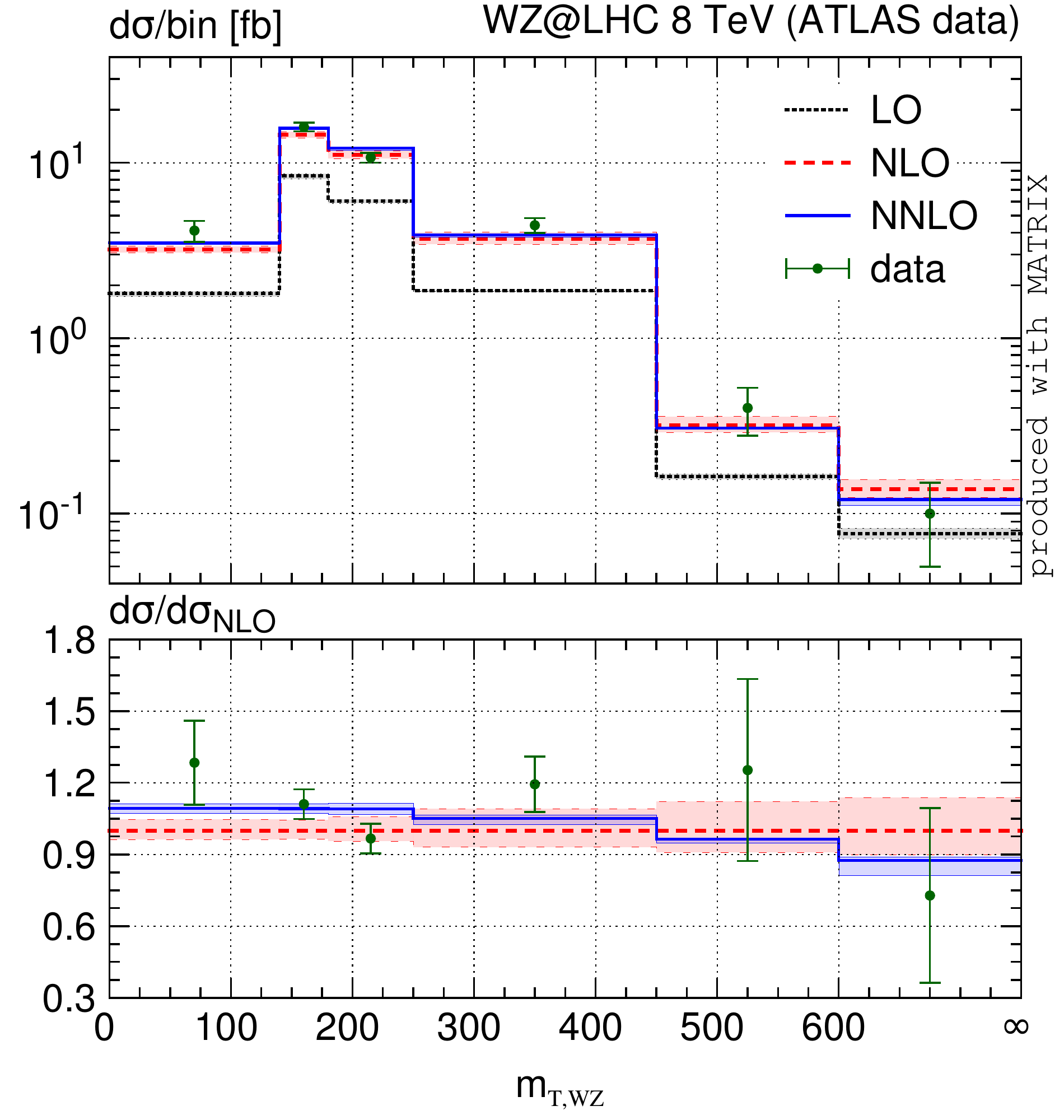} &
\includegraphics[trim = 7mm -7mm 0mm 0mm, width=.33\textheight]{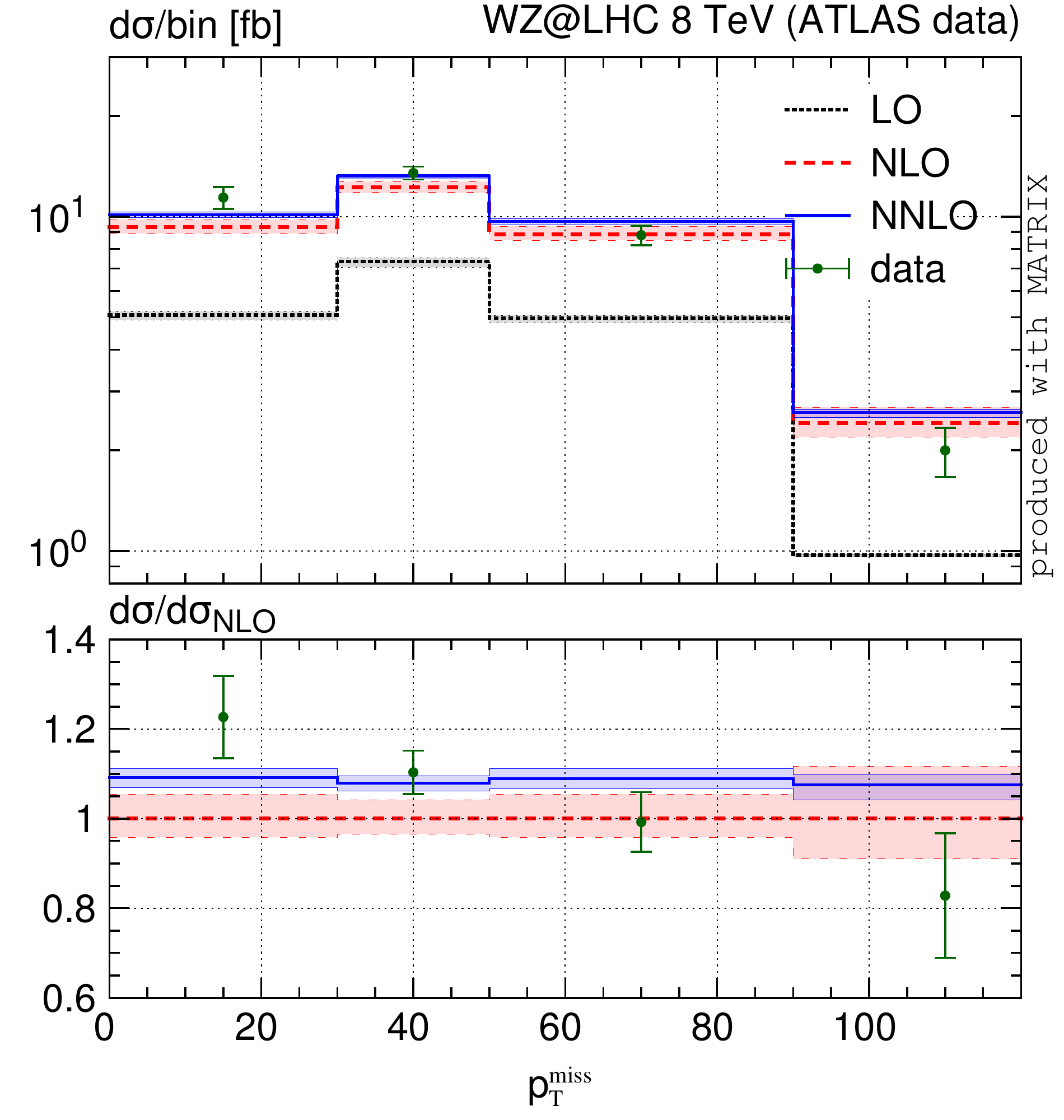} \\[-1em]
\hspace{0.6em} (a) & \hspace{1em}(b)
\end{tabular}
\caption[]{\label{fig:pTmissmTWZ}{Same as \fig{fig:pTZW}, but for (a) the transverse
mass of the $WZ$ system as defined in \eqn{eq:mTWZ} and (b) the missing transverse energy.}}
\end{center}
\end{figure}

\begin{figure}
\begin{center}
\begin{tabular}{cc}
\hspace*{-0.17cm}
\includegraphics[trim = 7mm -7mm 0mm 0mm, width=.33\textheight]{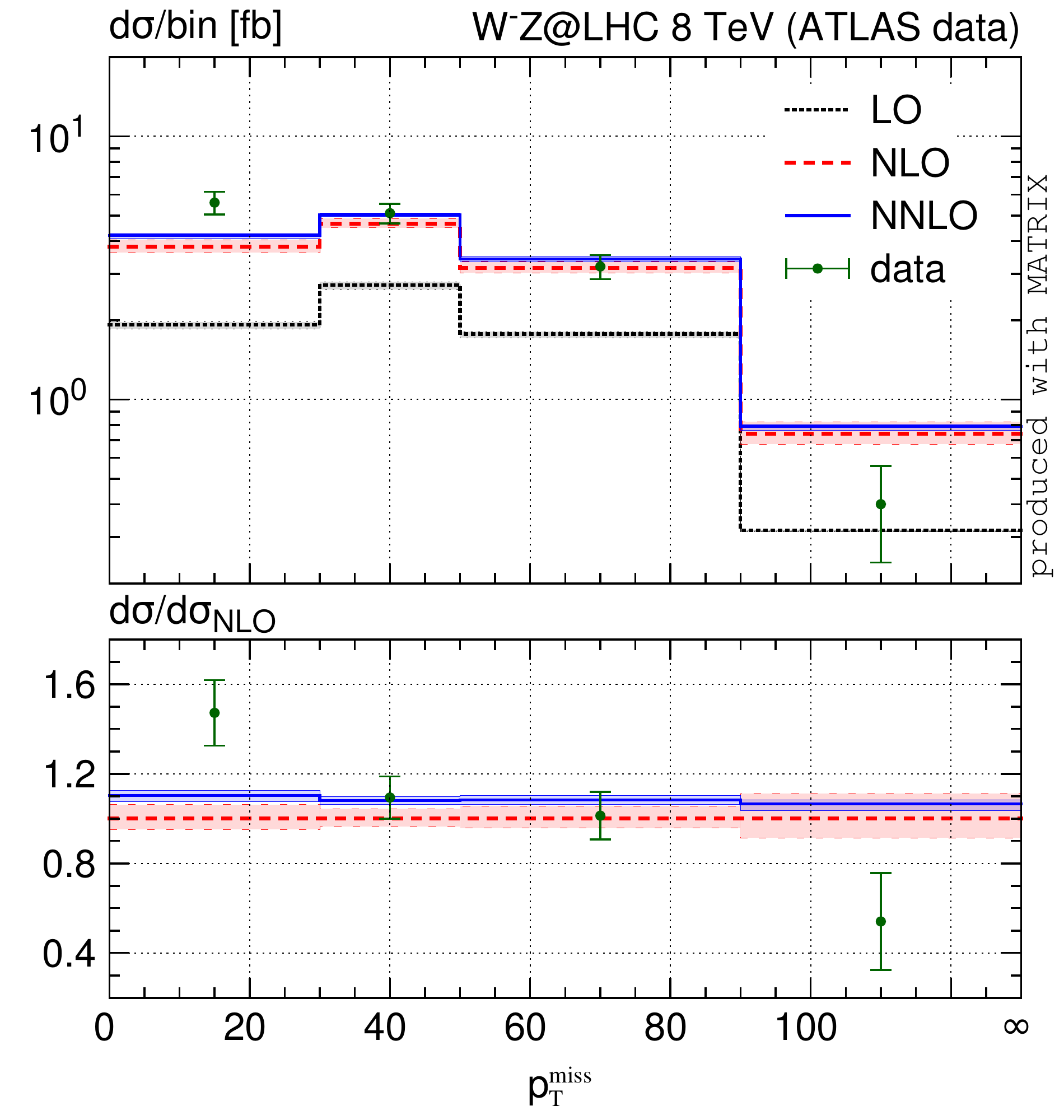} &
\includegraphics[trim = 7mm -7mm 0mm 0mm, width=.33\textheight]{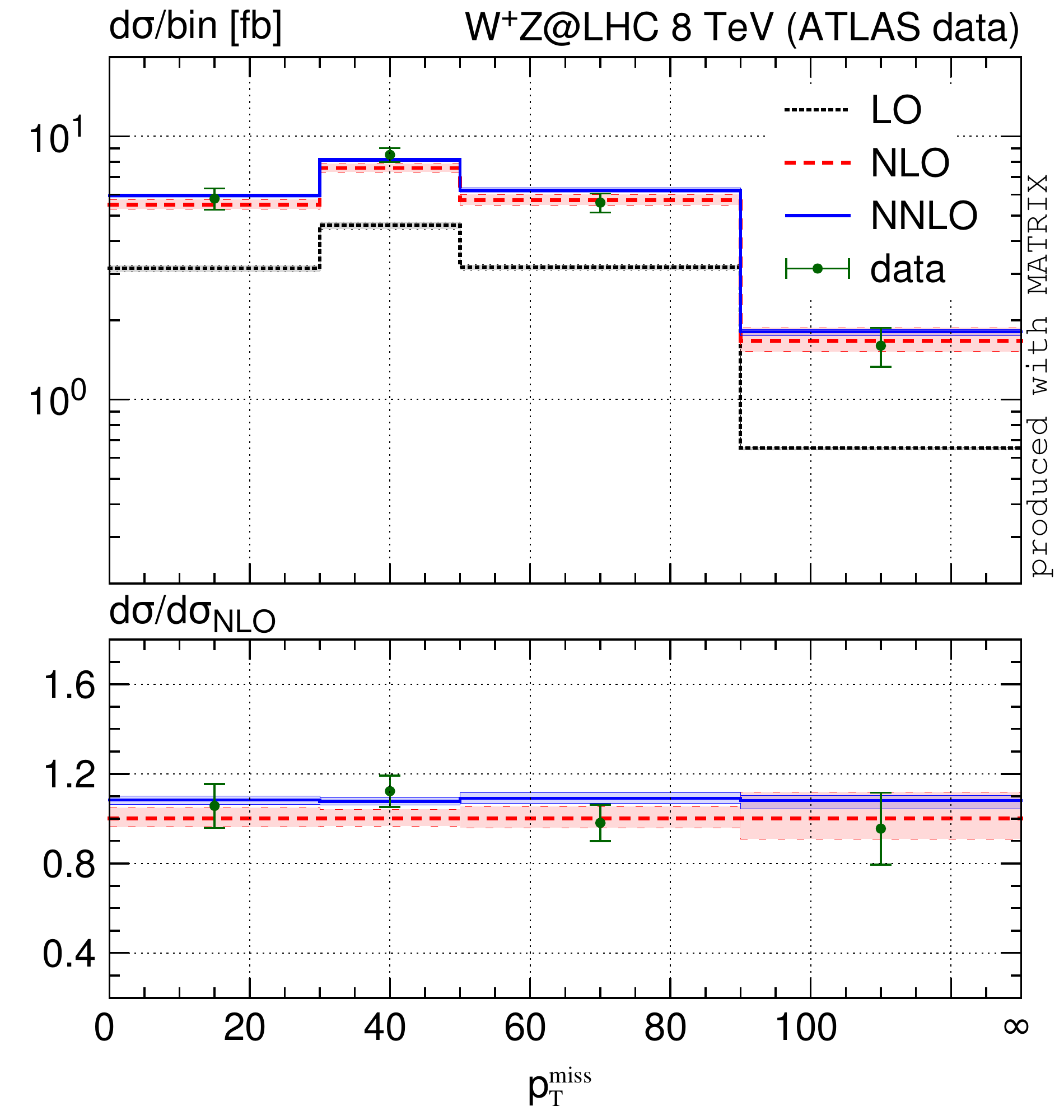} \\[-1em]
\hspace{0.6em} (a) & \hspace{1em}(b)
\end{tabular}
\caption[]{\label{fig:pTmissWmWp}{Same as \fig{fig:pTmissmTWZ}\,(b), but separated by (a) $W^-Z$ and (b) $W^+Z$ production.}}
\end{center}
\end{figure}

In \fig{fig:pTmissmTWZ}\,(a), we consider the distribution in the transverse mass of the $WZ$ system, defined by
\begin{align}
\label{eq:mTWZ}
m_{T,WZ} = 
\sqrt{\left(E_{T,\lw}+E_{T,\nu_{\lw}}+E_{T,\lpz}+E_{T,\lmz}\right)^2 
- p_{T,\lw\nu_{\lw}\lpz\lmz}^2}\quad \mathrm{with} \quad E_{T,x}^2=m_x^2+p_{T,x}^2\,.
\end{align}
With shape effects of about $15\%$,
the NNLO corrections significantly soften the spectrum.
Already the NLO prediction is in good agreement with data, and the NNLO corrections tend to slightly improve that agreement mainly due to the shape correction,
so that the measured results are well described by the theoretical predictions within roughly $1\sigma$ of the experimental errors.

The ATLAS result for the missing transverse energy distribution in \fig{fig:pTmissmTWZ}\,(b) shows some discrepancy 
in shape compared to the NLO prediction. The NNLO corrections are essentially flat, so they cannot 
account for that difference. Overall, the uncertainties of the measured results are still rather large, such that 
the deviation of the predicted cross section in each bin stays within $1\sigma-2\sigma$. Looking at \reffi{fig:pTmissWmWp} 
where we plot the missing transverse energy distribution separately for $W^-Z$ and $W^+Z$ production, we see that 
the observed discrepancy between theory and data appears only for $W^-Z$ production, where it extends 
up to roughly $2\sigma-3\sigma$ for the lowest and highest \ptmiss{} bins. 
To clarify the origin of this discrepancy more precise data are needed,
given that only four separate bins are measured at the moment.

\begin{figure}[tp]
\begin{center}
\begin{tabular}{cc}
\hspace*{-0.17cm}
\includegraphics[trim = 7mm -7mm 0mm 0mm, width=.33\textheight]{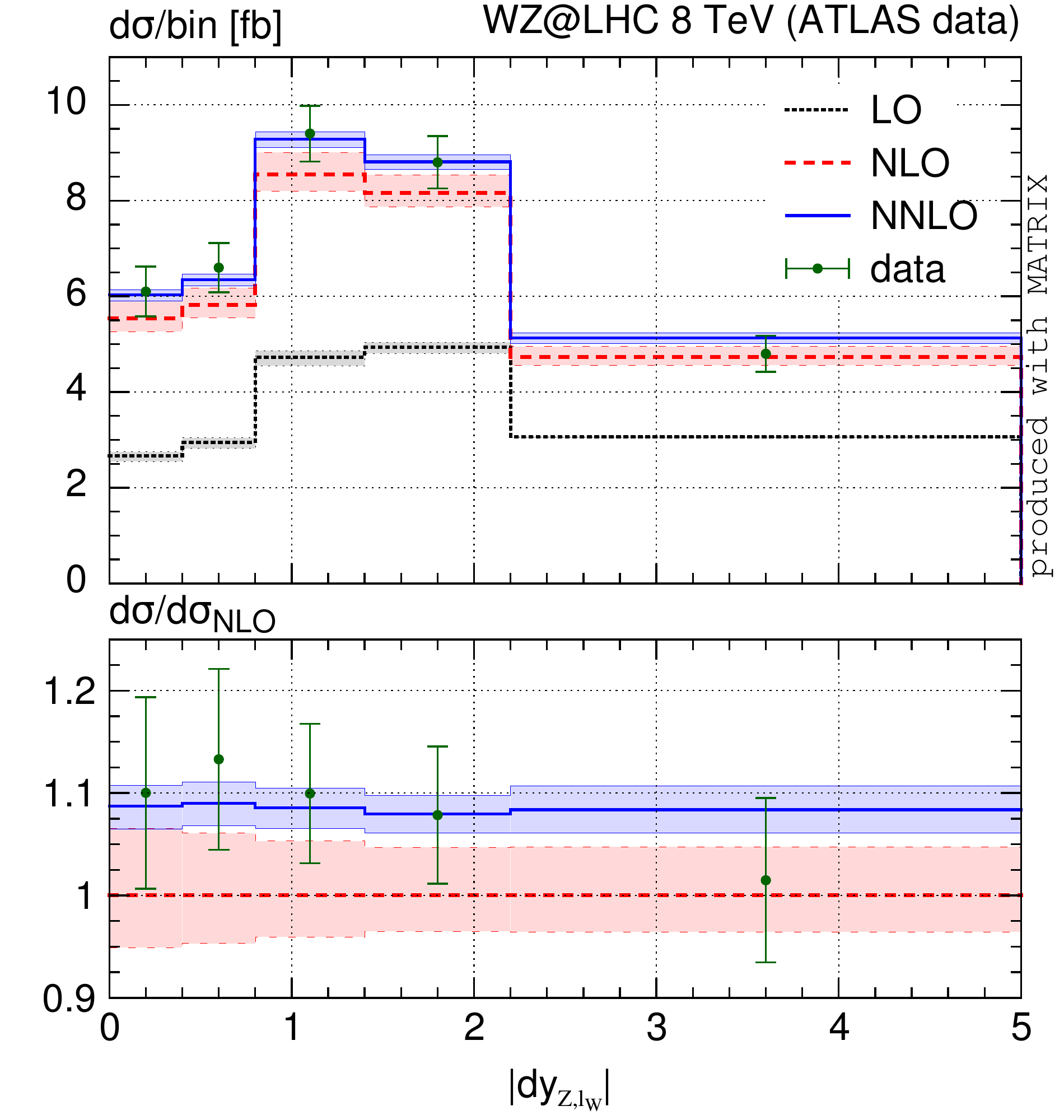} &
\includegraphics[trim = 7mm -7mm 0mm 0mm, width=.33\textheight]{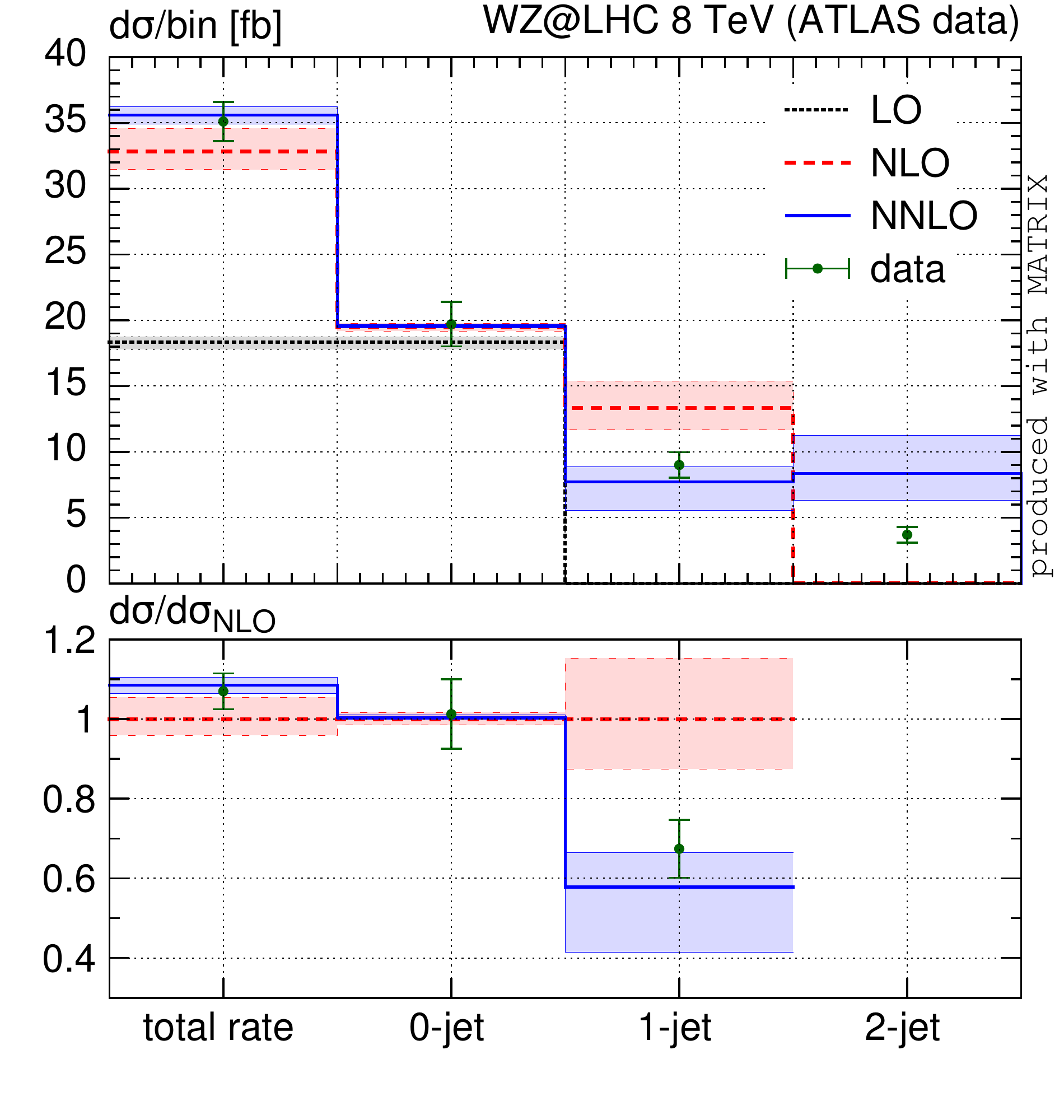} \\[-1em]
\hspace{0.6em} (a) & \hspace{1em}(b)
\end{tabular}
\caption[]{\label{fig:dyZlWnjet}{Same as \fig{fig:pTZW}, but for (a)  the
absolute rapidity separation between the reconstructed $Z$ boson and the lepton from the $W$-boson decay, and (b) the number of jets.}}
\end{center}
\end{figure}

Next, we discuss the absolute rapidity difference between the reconstructed $Z$ boson and the lepton 
associated with the $W$-boson decay, shown in \reffi{fig:dyZlWnjet}\,(a). 
This $|\mathrm{d}y_{Z,\ell_W}|$ distribution has a distinctive
shape, with a dip at vanishing rapidity difference and a maximum around $|\mathrm{d}y_{Z,\ell_W}|=0.8$,
and it is sensitive to the approximate radiation zero \cite{Baur:1994ia} mentioned before. As expected, the 
LO prediction does not describe the data in any sensible way.
The NLO prediction already captures the dominant  
shape effects. The NNLO corrections are rather flat and are
consistent within uncertainties with (and in most cases right on top of) the data,
thanks to the improved normalization.

Finally, \reffi{fig:dyZlWnjet}\,(b) shows the distribution in the jet multiplicity.
Jets are defined with the anti-$k_T$ algorithm \cite{Cacciari:2008gp} with radius parameter $R=0.4$.
A jet must have a minimum transverse momentum of $25$\,GeV and a maximum pseudo-rapidity 
of $4.5$. 
We already know that the measured fiducial cross section is in excellent agreement with the 
NNLO prediction. As expected, radiative corrections are strongly reduced when considering a jet veto ($0$-jet bin).
NLO and NNLO predictions are essentially indistinguishable, apart from the reduction of the theoretical uncertainties 
when going from NLO to NNLO. The experimental result is right on top of them. In the exclusive $1$-jet bin NLO (NNLO) predictions 
are formally only LO (NLO) accurate. It is well-known that LO-accurate predictions tend to underestimate the 
uncertainties. The blue solid NNLO result has the effect of decreasing the cross section in that bin by almost a factor of 
two with respect to NLO, well beyond the given uncertainties. 
The data point is significantly closer to the NNLO prediction and fully 
consistent with it within uncertainties. Finally, in the $2$-jet bin even the NNLO contribution is effectively 
only LO, and our computation cannot provide a reliable prediction. 
Indeed, it significantly overestimates the measured cross section. A more accurate 
description of the $2$-jet bin requires at least NLO QCD corrections 
to the $\wz{}+2$ jets process \cite{Campanario:2013qba}.

\begin{figure}
\begin{center}
\begin{tabular}{cc}
\hspace*{-0.17cm}
\includegraphics[trim = 7mm -7mm 0mm 0mm, width=.33\textheight]{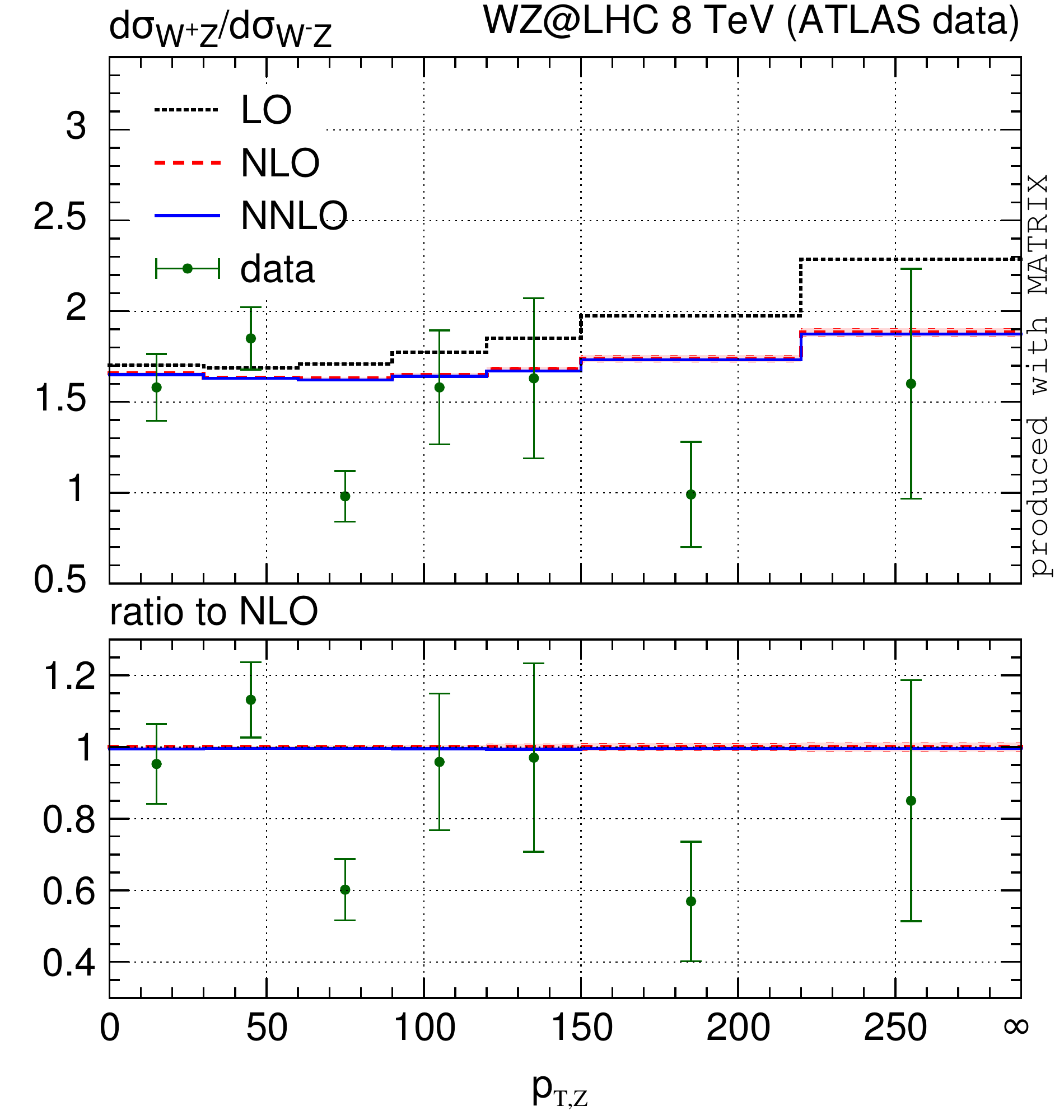} &
\includegraphics[trim = 7mm -7mm 0mm 0mm, width=.33\textheight]{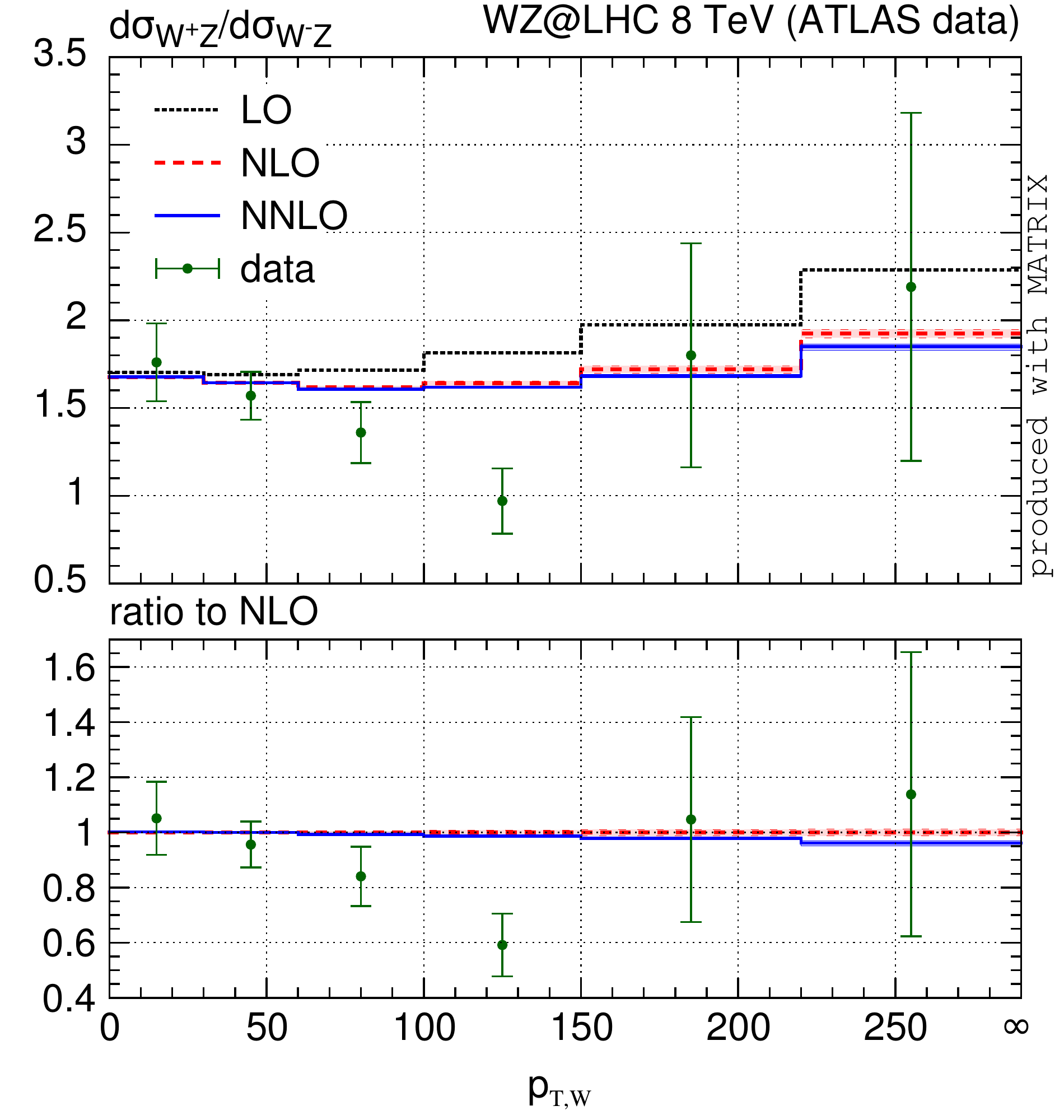} \\[-1em]
\hspace{0.6em} (a) & \hspace{1em}(b)
\end{tabular}
\caption[]{\label{fig:ratiopTZW}{Same as \fig{fig:pTZW}, but shows the 
ratio of cross sections for $W^+Z$ and $W^-Z$ production.}}
\end{center}
\end{figure}

\begin{figure}
\begin{center}
\begin{tabular}{cc}
\hspace*{-0.17cm}
\includegraphics[trim = 7mm -7mm 0mm 0mm, width=.33\textheight]{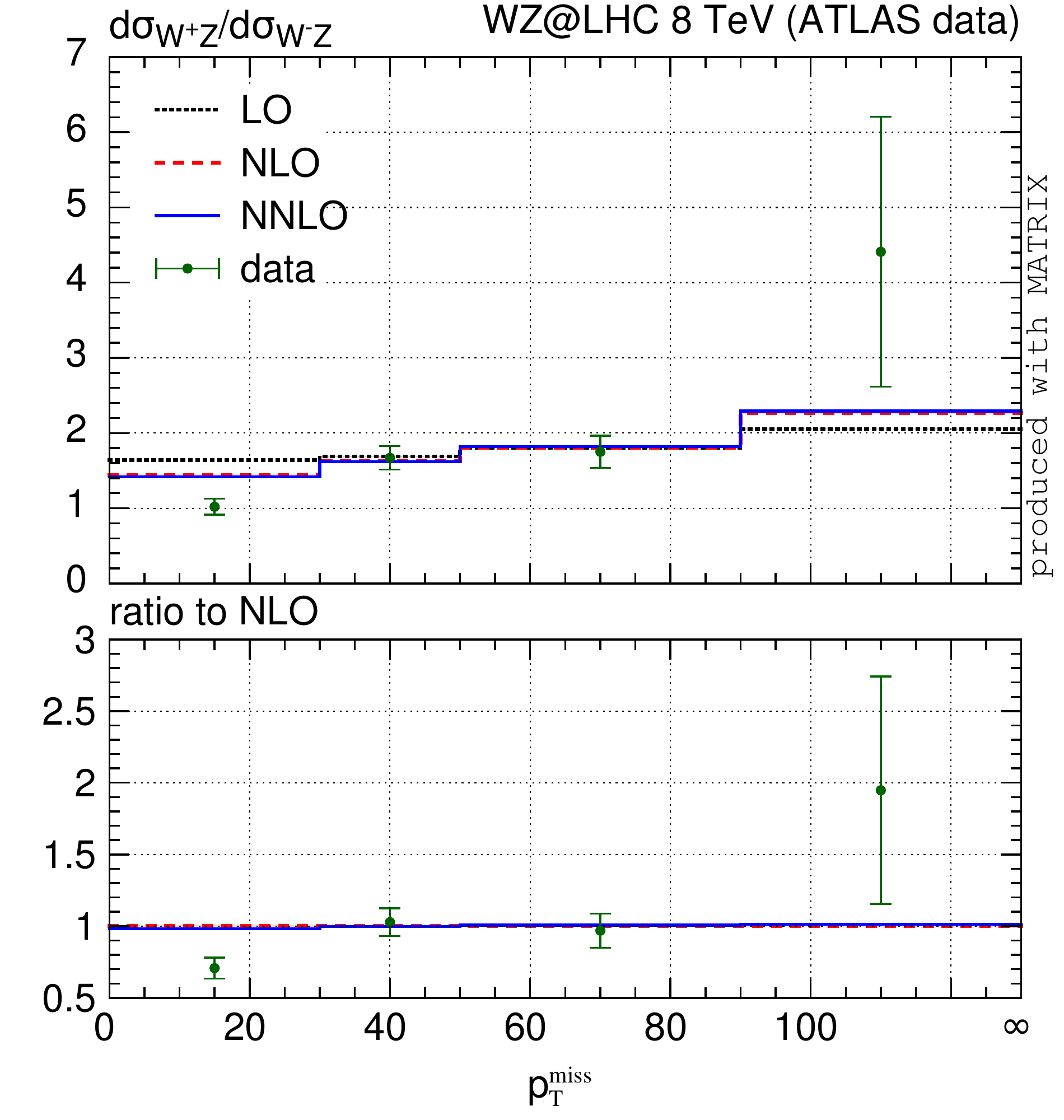} &
\includegraphics[trim = 7mm -7mm 0mm 0mm, width=.33\textheight]{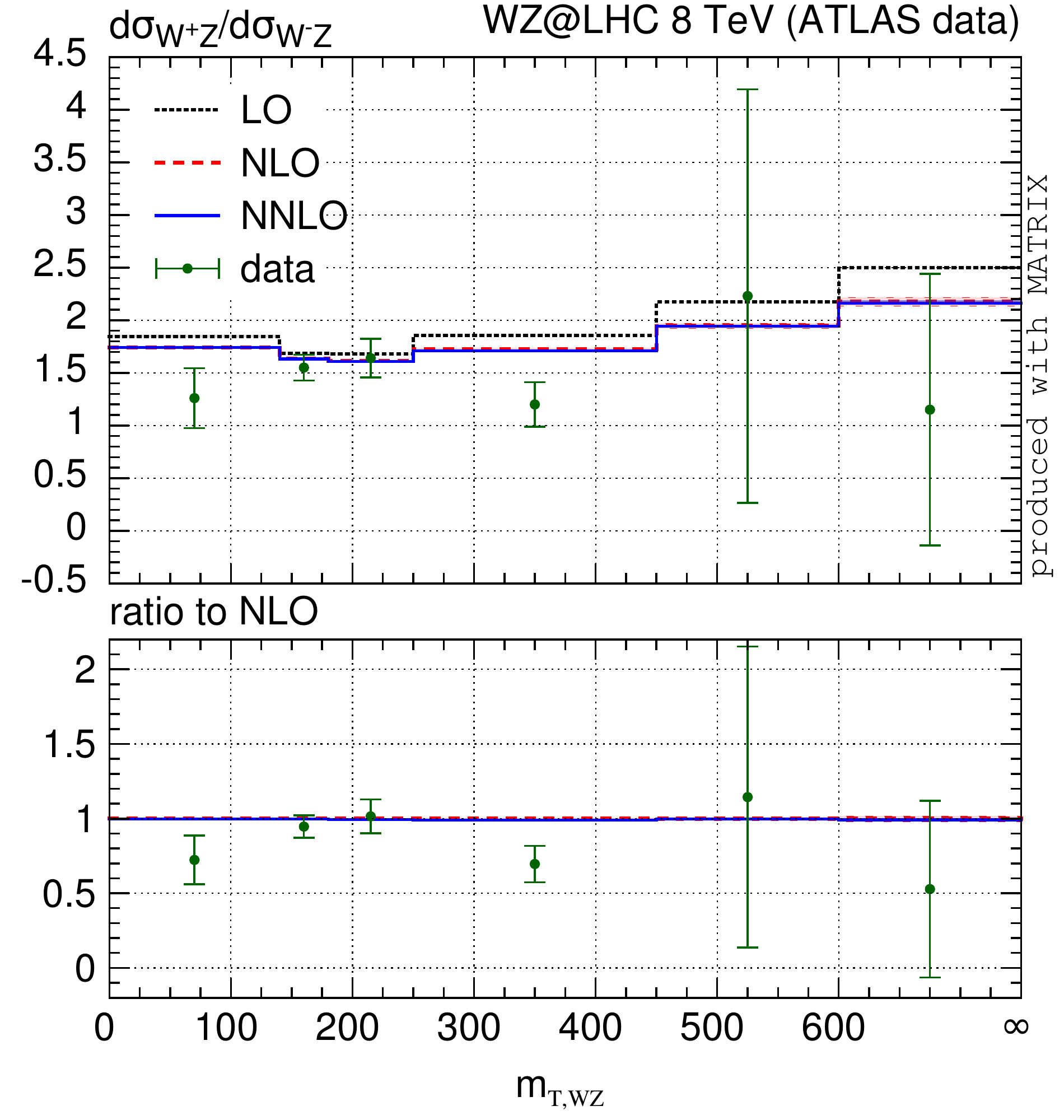} \\[-1em]
\hspace{0.6em} (a) & \hspace{1em}(b)
\end{tabular}
\caption[]{\label{fig:ratiopTmissmTWZ}{Same as \fig{fig:pTmissmTWZ}, but shows the 
ratio of cross sections for $W^+Z$ and $W^-Z$ production.}}
\end{center}
\end{figure}

\begin{figure}[tp]
\begin{center}
\begin{tabular}{cc}
\hspace*{-0.17cm}
\includegraphics[trim = 7mm -7mm 0mm 0mm, width=.33\textheight]{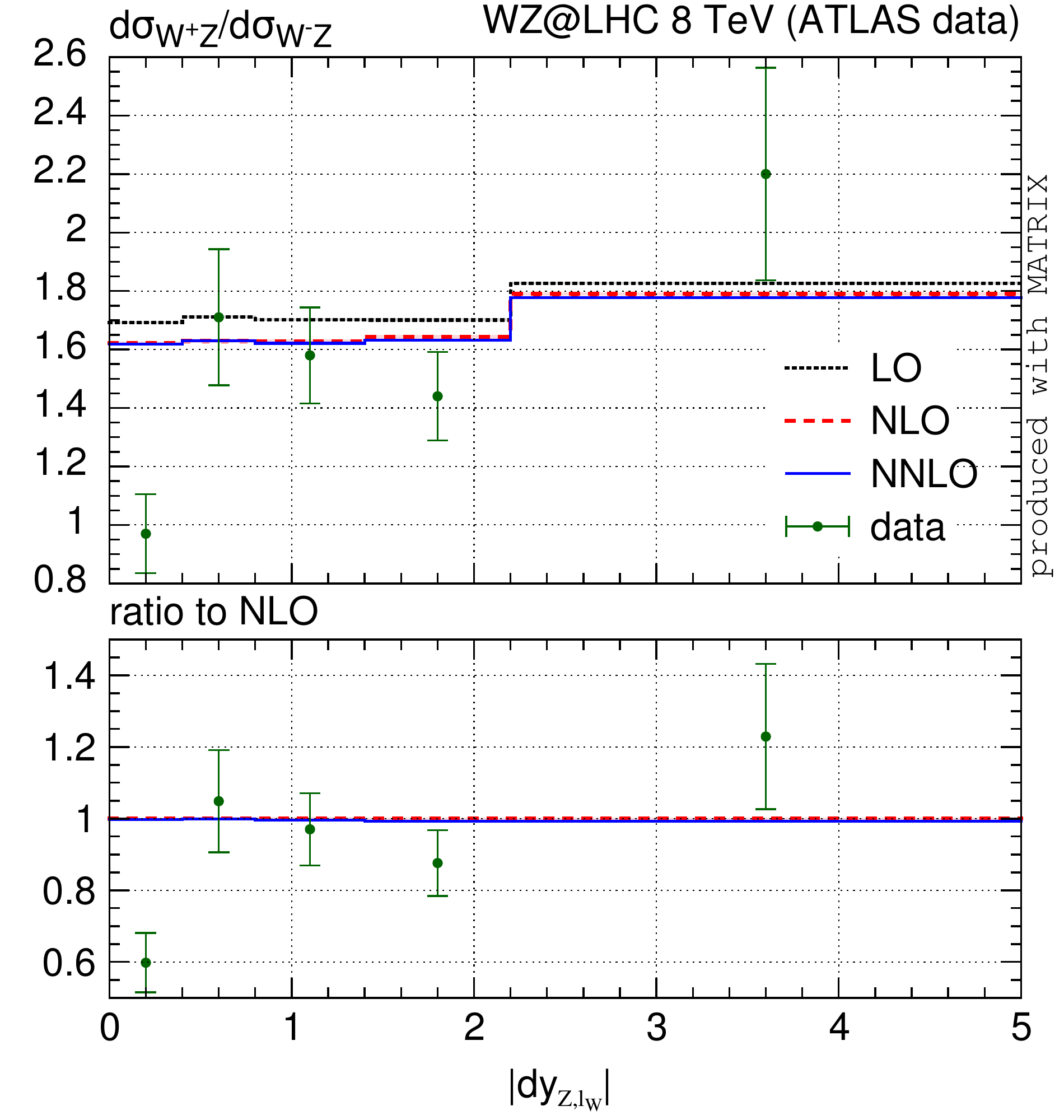} &
\includegraphics[trim = 7mm -7mm 0mm 0mm, width=.33\textheight]{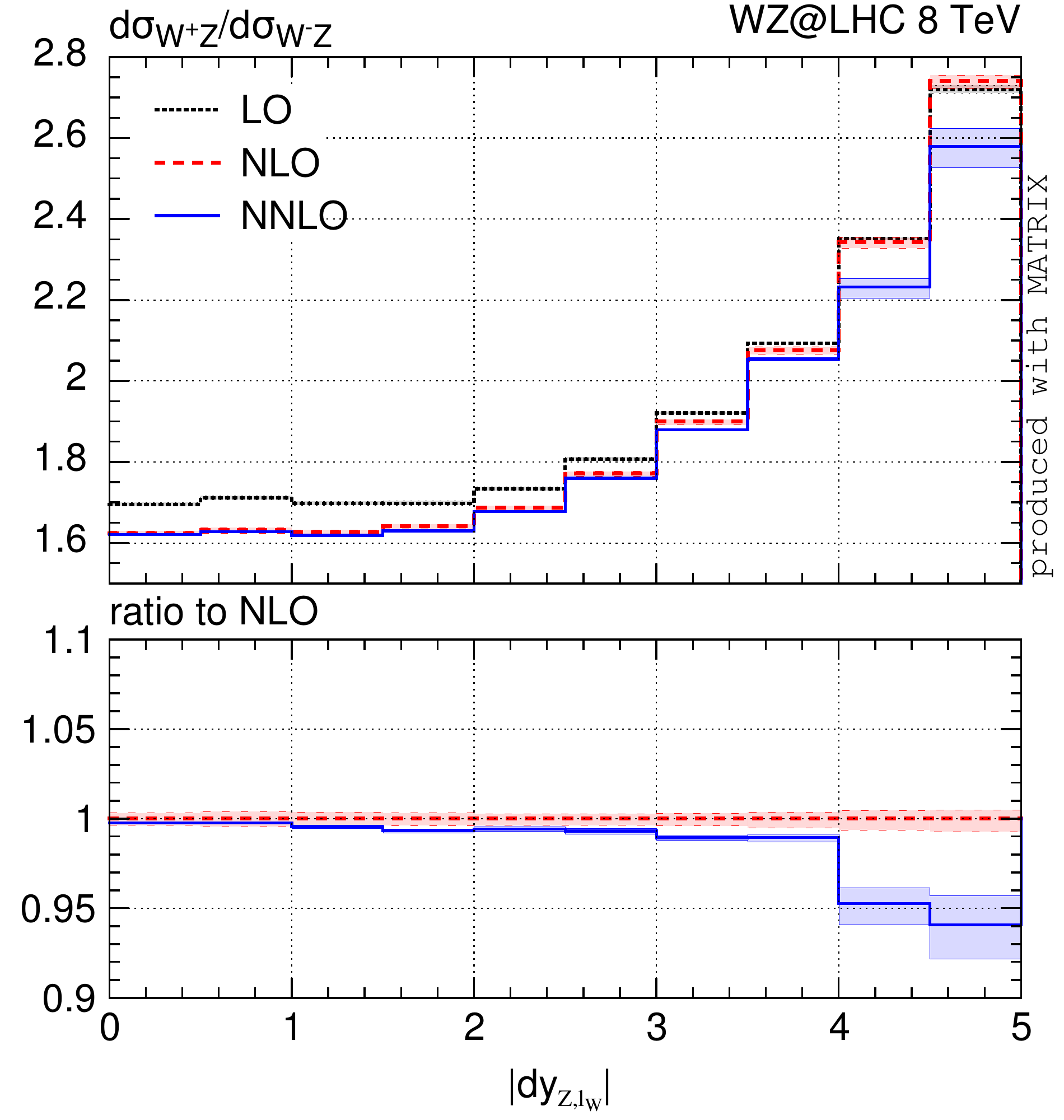} \\[-1em]
\hspace{0.6em} (a) & \hspace{1em}(b)
\end{tabular}
\caption[]{\label{fig:ratiodyZlW}{(a) Same as \fig{fig:dyZlWnjet}\,(a), but 
shows the ratio of cross sections for $W^+Z$ and $W^-Z$ production, and (b) 
same plot with a different binning and without data.}}
\end{center}
\end{figure}

We conclude our discussion of differential distributions by considering ratios of 
$W^+Z$ over $W^-Z$ cross sections. In \figs{fig:ratiopTZW}$-$\ref{fig:ratiodyZlW}\,(a) such ratios are 
compared to the ATLAS 8 TeV data. Otherwise, these plots 
have exactly the same structure as the previous ones. 
The uncertainty bands are computed by taking fully correlated scale variations, i.e., using the same scale in 
numerator and denominator. The ensuing bands are extremely small, with 
relative uncertainties never exceeding
$\sim1\%-2\%$ both at NLO and NNLO. In most cases the perturbative 
computation of the ratios is very stable and in particular NNLO corrections 
are very small, which justifies fully correlated scale variations to estimate 
the perturbative uncertainties. Nevertheless, some observables are affected 
by $\mathcal{O}(\as^2)$ corrections beyond the residual uncertainty bands:
Such cases are discussed at the end of this section.

By and large, we find  
reasonable agreement between the predicted and the measured ratios in all 
distributions under consideration, which is, in part, due to the relatively large 
experimental uncertainties. The latter prevent to clearly discriminate whether 
NNLO corrections improve the agreement with data. Nevertheless, for each 
distribution at least one data point deviates from the prediction by more than 
$2\sigma$, some of which appear even quite significant. For example, 
in \fig{fig:ratiopTZW}\,(a) there is one bin in the transverse-momentum 
spectrum of the reconstructed $Z$ boson with a discrepancy of roughly 
$4\sigma$ and another one with more than $2\sigma$. However, the experimental 
results fluctuate too much to claim that these are genuine effects beyond statistics. 
In fact, similar differences as we observe here 
are evident also in the ATLAS study \cite{Aad:2016ett} when 
data are compared to NLO+PS predictions.
Only higher experimental accuracy, to become available at 13 TeV soon, will 
allow for a more conclusive comparison in these cases. Indeed, even the distribution in the 
missing transverse energy in \fig{fig:ratiopTmissmTWZ}, where we found some 
apparent difference in the shape for $W^-Z$, but not for $W^+Z$ production (see \fig{fig:pTmissWmWp}), 
does not seem to be particularly (more) significant when considering the 
$W^+Z$/$W^-Z$ ratio due to the large experimental errors. 

\begin{figure}[tp]
\begin{center}
\begin{tabular}{cc}
\hspace*{-0.17cm}
\includegraphics[trim = 7mm -7mm 0mm 0mm, width=.33\textheight]{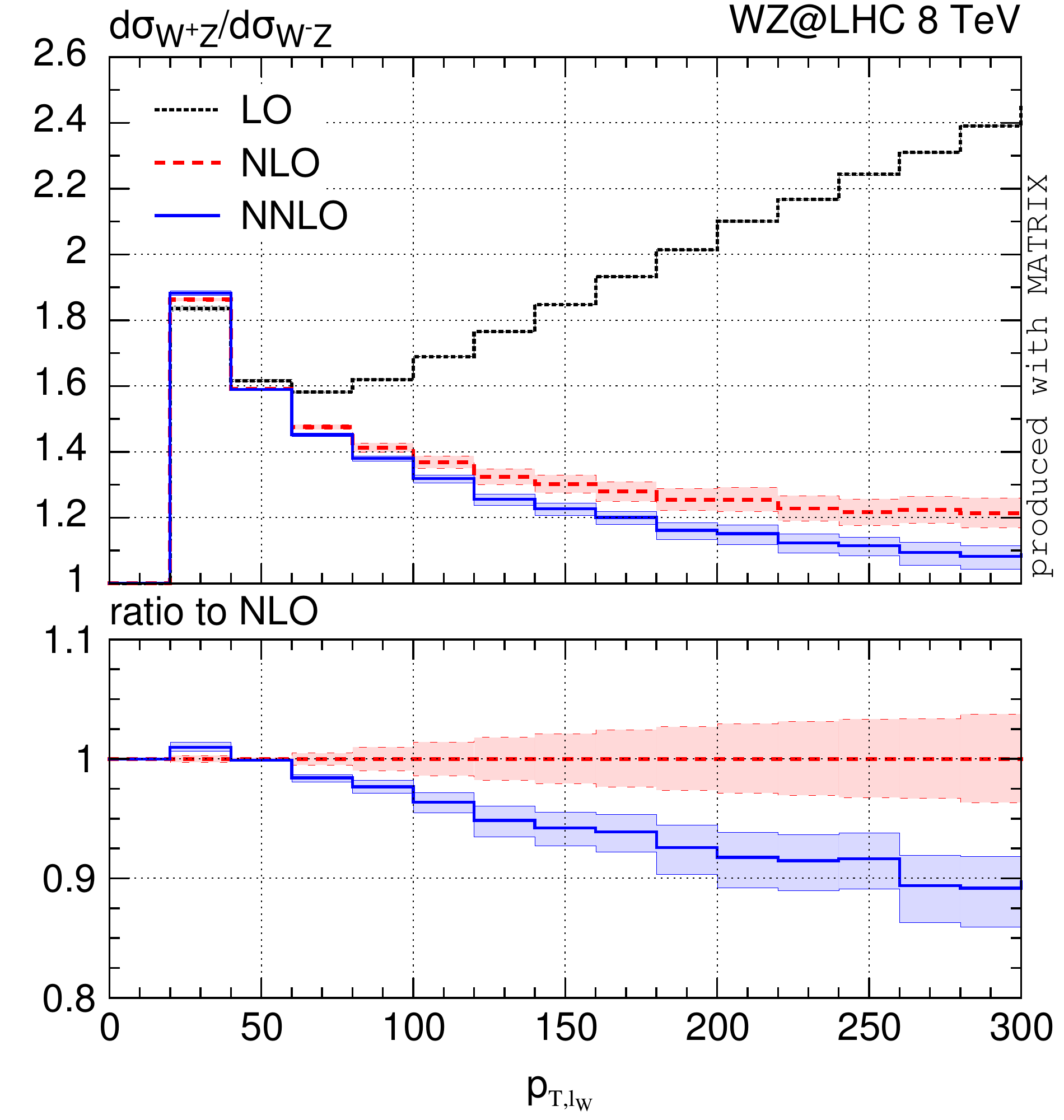} &
\includegraphics[trim = 7mm -7mm 0mm 0mm, width=.33\textheight]{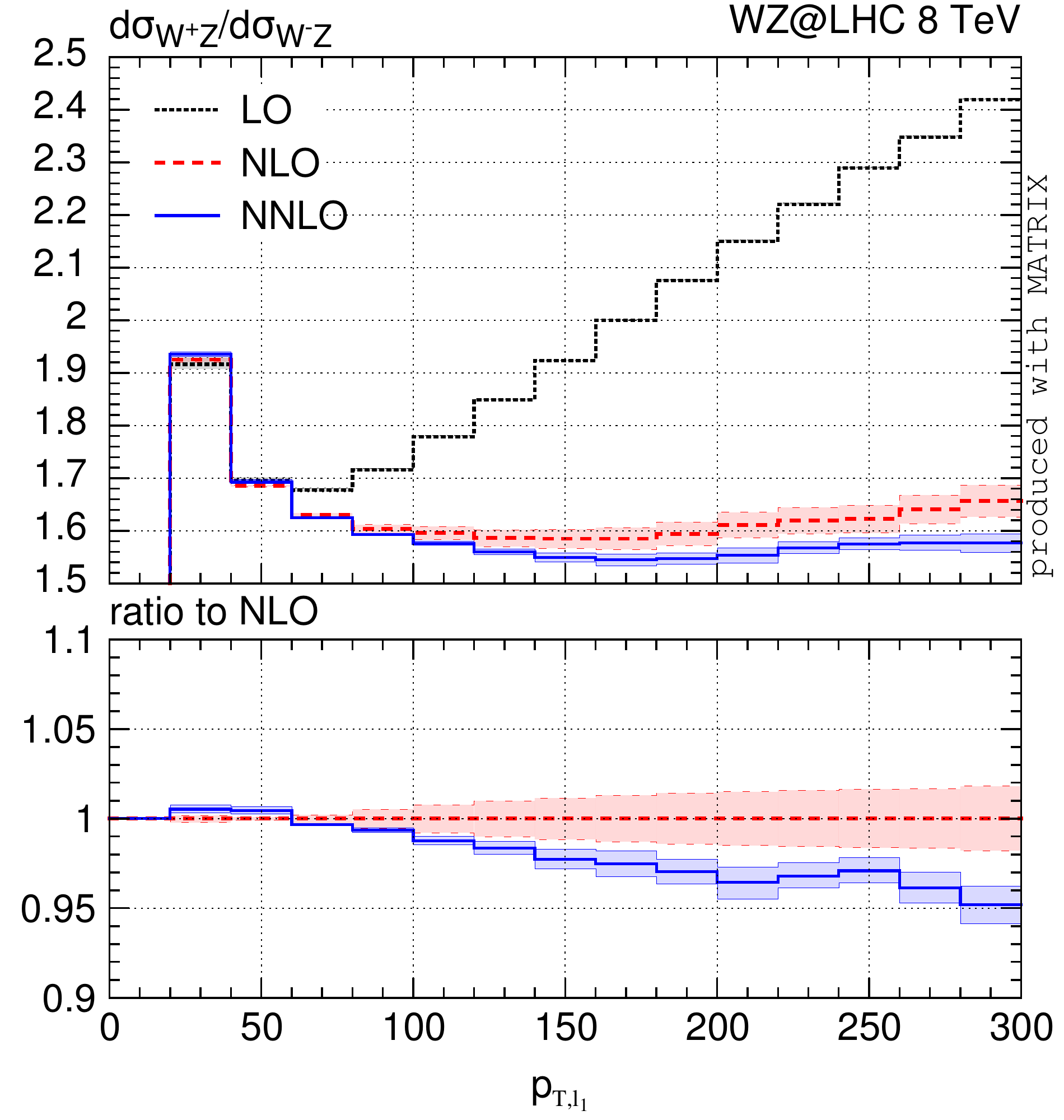} \\[-1em]
\hspace{0.6em} (a) & \hspace{1em}(b)
\end{tabular}
\caption[]{\label{fig:ratiopTlepWpTlepone}{Ratio of $W^+Z$ and $W^-Z$ distributions in (a) the
transverse momentum of the lepton associated with the $W$ decay and (b) the transverse momentum of the hardest lepton.}}
\end{center}
\end{figure}

Finally, we point out certain distributions which show prominent shape 
differences between $W^+Z$ and $W^-Z$ production, while featuring 
visible effects from the NNLO corrections. Several distributions 
exist, see, e.g., \figs{fig:ratiodyZlW}\,(b)$-$\ref{fig:ratiopTlepWpTlepone}, which depend rather strongly on the charge of 
the $W$ boson. Unfortunately, large NNLO effects often appear only 
in corners of phase space that are strongly suppressed and thus
have low experimental sensitivity. One example is the absolute rapidity difference  
between the reconstructed $Z$ boson and the lepton associated with the 
$W$-boson decay, which is 
compared to data in \fig{fig:ratiodyZlW}\,(a), but shown with a finer binning 
in \fig{fig:ratiodyZlW}\,(b): The effect of 
NNLO corrections in the forward region is manifest, but it is entirely due to differences between NLO and NNLO PDFs
\footnote{We have checked that 
by using the NNLO set also for the NLO predictions the difference disappears.}.

\begin{figure}[tp]
\begin{center}
\begin{tabular}{cc}
\hspace*{-0.17cm}
\includegraphics[trim = 7mm -7mm 0mm 0mm, width=.33\textheight]{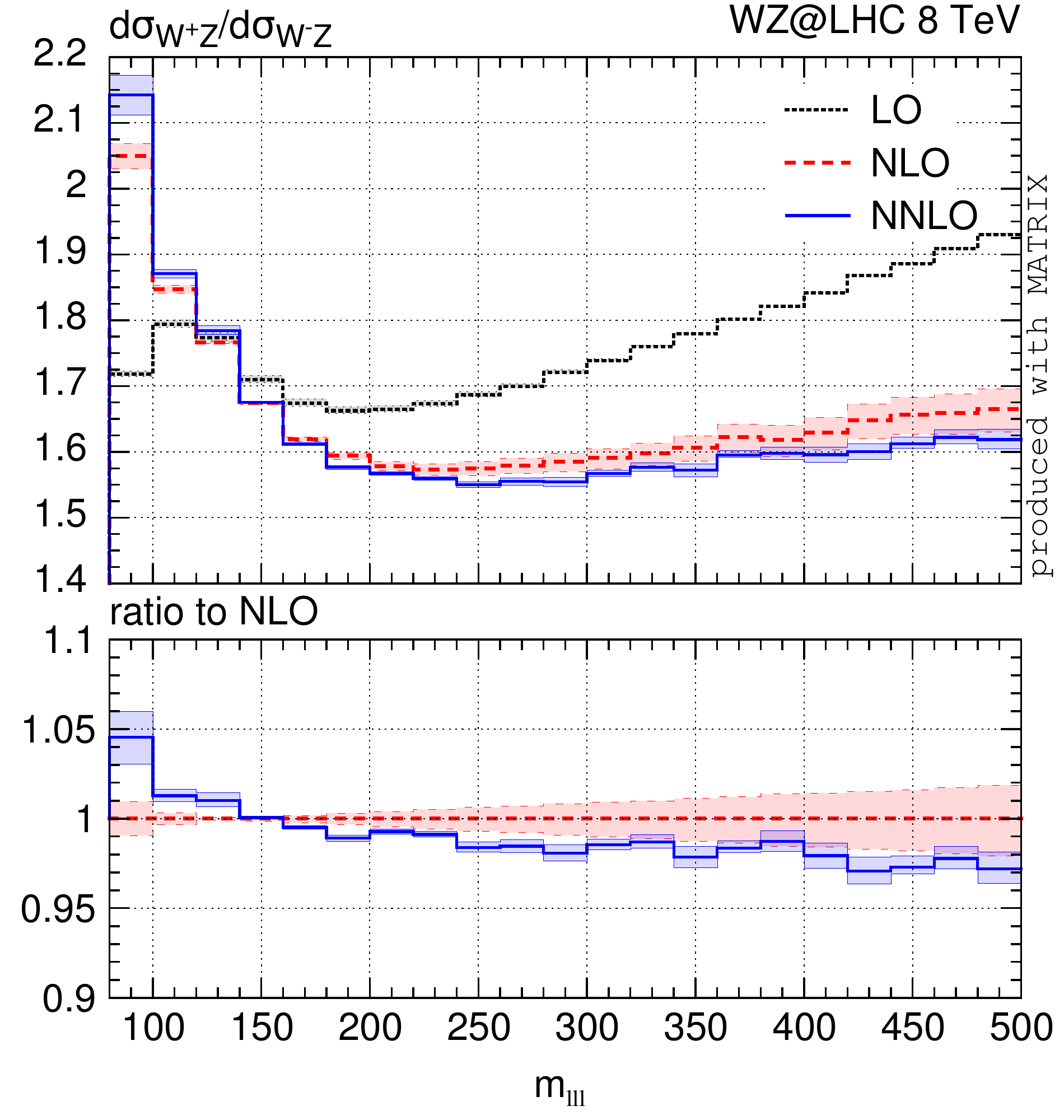} &
\includegraphics[trim = 7mm -7mm 0mm 0mm, width=.33\textheight]{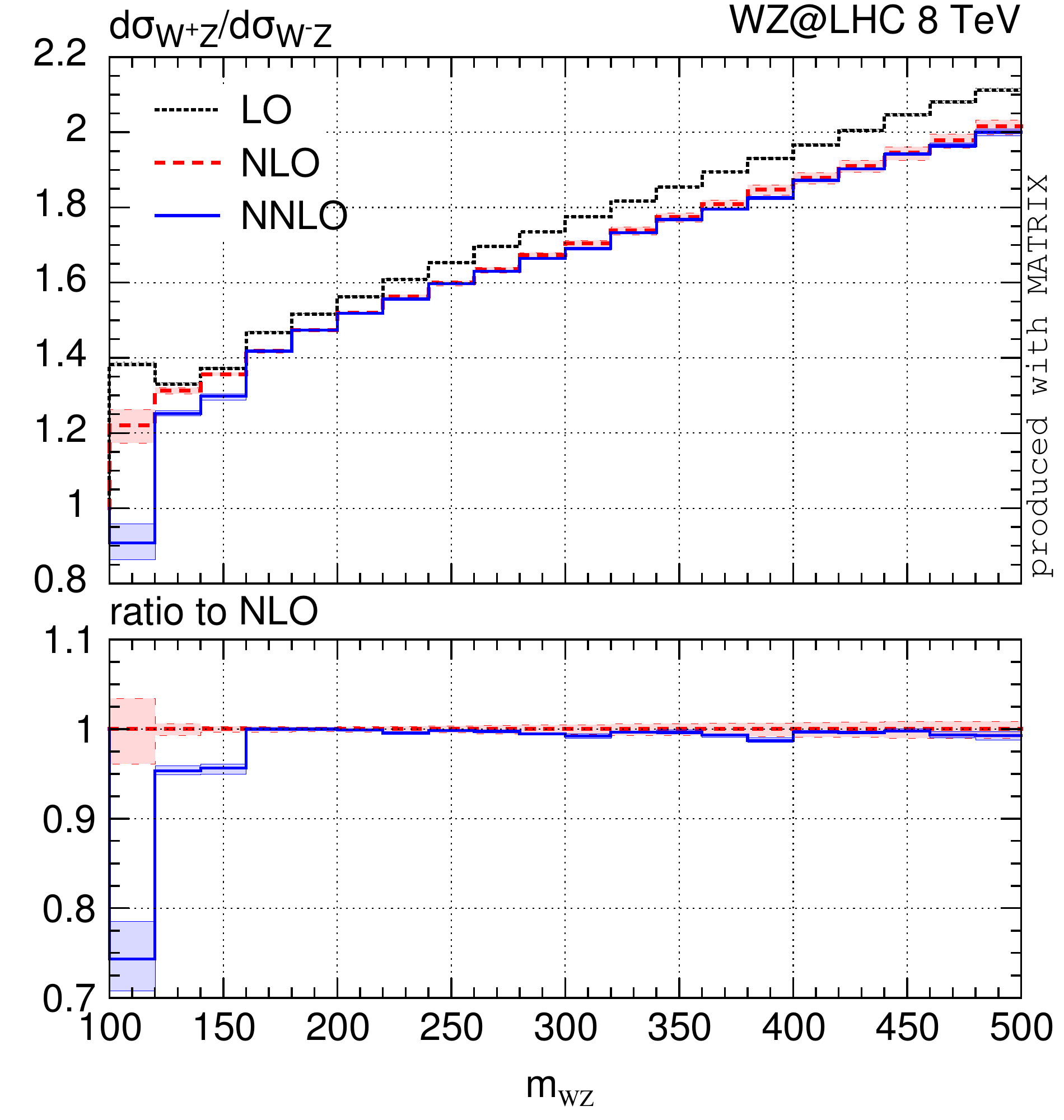} \\[-1em]
\hspace{0.6em} (a) & \hspace{1em}(b)
\end{tabular}
\caption[]{\label{fig:ratioinv3linvWZ}{Ratio of $W^+Z$ and $W^-Z$ distributions in the (a)
invariant mass of the three leptons and (b) invariant mass of the $WZ$ system.}}
\end{center}
\end{figure}

There are, however, examples where the effects of NNLO corrections 
on the $W^+Z$/$W^-Z$ ratio are evident 
already in the bulk region of the distribution.
Such examples are given in \figs{fig:ratioinv3linvWZ}$-$\ref{fig:ratiopTlepWpTlepone}. The $W^+Z$/$W^-Z$ ratio for the invariant 
mass of the three leptons in \fig{fig:ratioinv3linvWZ}\,(a) evidently 
increases for small $m_{\ell\ell\ell}$ values and decreases in the tail 
of the distribution upon inclusion of higher-order corrections, the effect being at the $5\%$ level.
Also the $W^+Z$/$W^-Z$ ratio as a function of the invariant mass of the \wz{} pair 
in \fig{fig:ratioinv3linvWZ}\,(b) shows a large impact of NNLO corrections, although this is close to the kinematical boundary where the
cross section is strongly suppressed.

\begin{figure}[tp]
\begin{center}
\begin{tabular}{cc}
\hspace*{-0.17cm}
\includegraphics[trim = 7mm -7mm 0mm 0mm, width=.33\textheight]{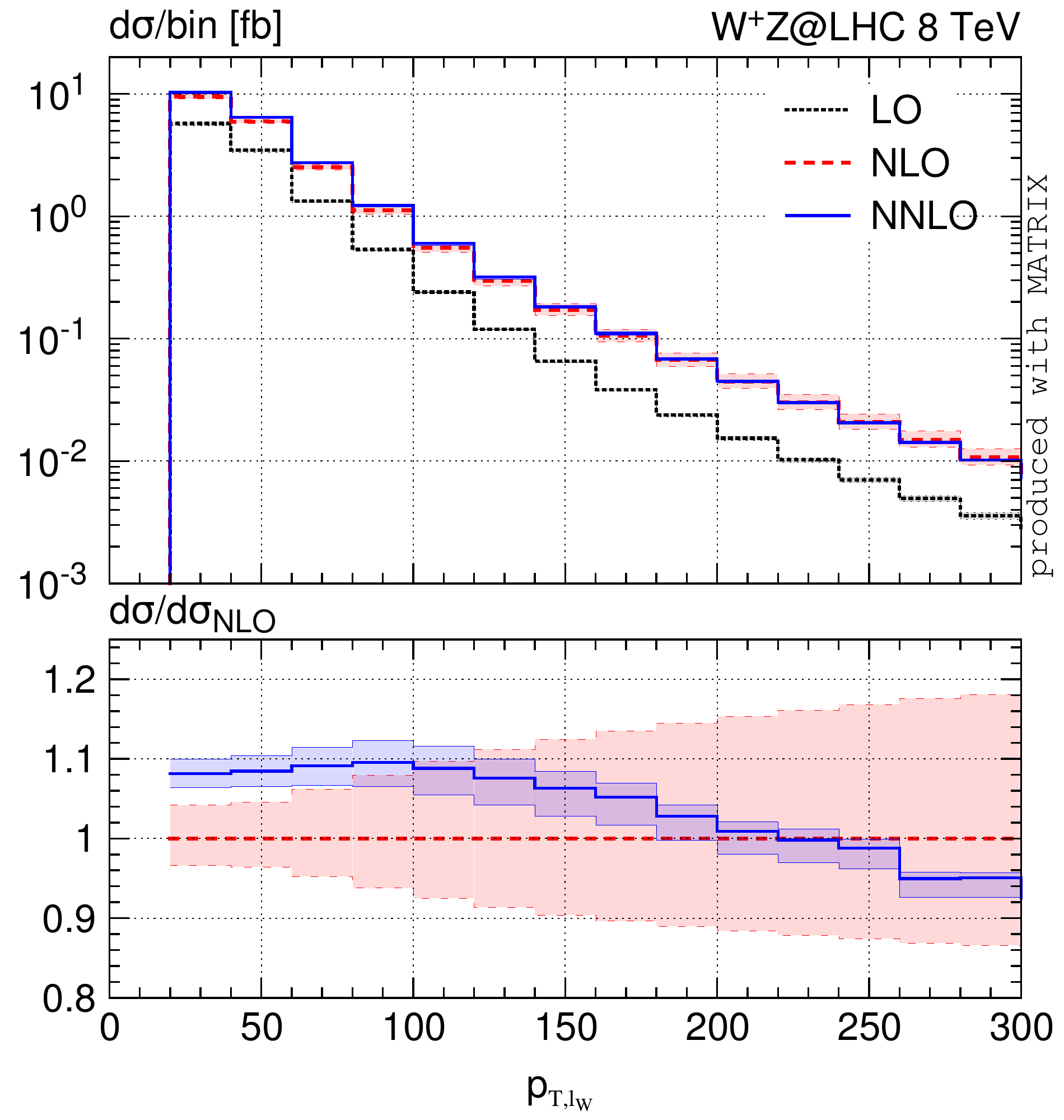} &
\includegraphics[trim = 7mm -7mm 0mm 0mm, width=.33\textheight]{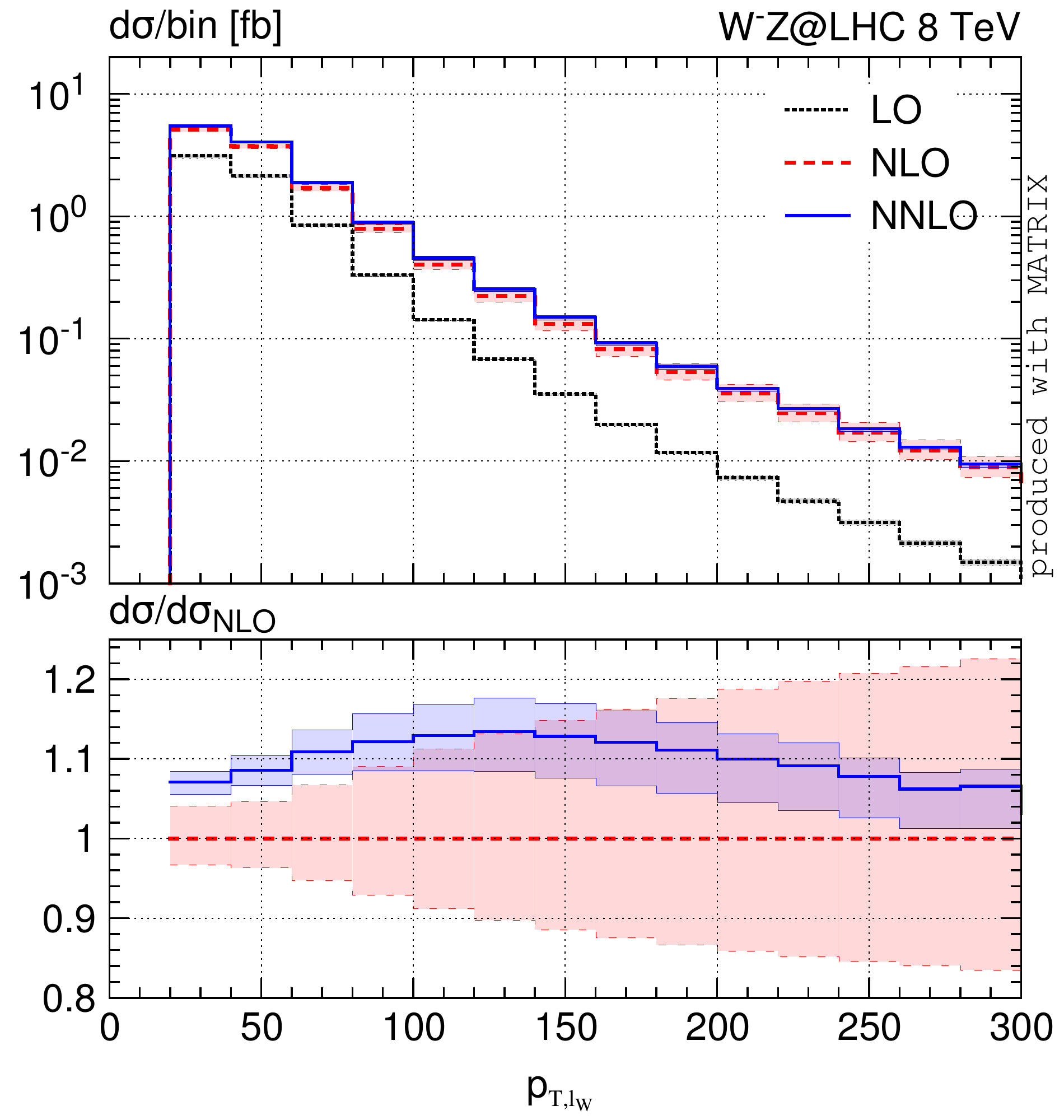} \\[-1em]
\hspace{0.6em} (a) & \hspace{1em}(b)
\end{tabular}
\caption[]{\label{fig:ratioW+W-pTlw}{Distributions in the
transverse momentum of the lepton associated with the $W$ decay,  
but separately for (a) $W^+Z$ and (b) $W^-Z$ production (corresponding
to the ratio in \fig{fig:ratiopTlepWpTlepone}\,(a)).}}
\end{center}
\end{figure}

The largest impact of NNLO corrections on the ratio of $W^+Z$ and $W^-Z$ 
cross sections is found for the distribution in the transverse momentum of the 
lepton associated with the $W$-boson decay (\ptlw{}) in \fig{fig:ratiopTlepWpTlepone}\,(a). 
The shape of the ratio significantly changes when going from NLO to NNLO, the effects being
more than $10\%$ in the tail of the distribution. 
Qualitatively similar, though smaller, effects can be observed in \fig{fig:ratiopTlepWpTlepone}\,(b)
for the leading-lepton \pt{}.

We conclude our presentation of the differential distributions with a comment on the
perturbative uncertainties affecting the $W^+Z$/$W^-Z$ ratios. The NLO uncertainties reported
in  \reffis{fig:ratioinv3linvWZ}{fig:ratiopTlepWpTlepone} underestimate the 
actual size of the NNLO corrections in certain phase-space regions. We note, however, 
that such uncertainties are computed by performing fully correlated variations. While in the majority of the cases
this procedure is justified by the small size of perturbative corrections, in some phase space regions
independent scale variations in numerator and denominator would 
be more appropriate to obtain realistic perturbative uncertainties.
This is demonstrated in \fig{fig:ratioW+W-pTlw}, which separately shows 
the absolute $\ptlw$ distribution for $W^+Z$ and $W^-Z$ production. 
Indeed, the NLO and NNLO predictions are actually  quite 
consistent within uncertainties. Similar conclusions can be drawn also for the 
other observables in \reffis{fig:ratioinv3linvWZ}{fig:ratiopTlepWpTlepone} when 
separately looking at their absolute distributions for $W^+Z$ and $W^-Z$ production.

\subsection{New-physics searches}
\label{sec:results-np}

In \sct{sec:fiducial} and \sct{sec:distributions} we have presented cross sections and distributions
in the fiducial regions defined by ATLAS and CMS to isolate the $W^\pm Z$ signature.
The comparison between theoretical predictions and experimental data in this region is certainly important
to test the SM. The \wz{} signature, however, and, more precisely, the production of three leptons + missing energy,
is important in many BSM searches, for which
the SM prediction provides an irreducible background.
One important example in this respect are searches for 
heavy supersymmetric (SUSY) particles:
The extraction of limits on SUSY 
masses relies on a precise prediction of the SM background.
In the following, we present an illustrative study where we
focus on a definite scenario for SUSY searches, and we study 
the impact of higher-order QCD corrections on both 
cross sections and distributions.

Typical experimental new-physics searches that consider
three leptons plus missing energy apply basic cuts which are rather similar to those 
considered in SM measurements. Here we follow as close as 
possible the selection cuts used in the CMS analysis of \citere{CMS:2016gvu} at 13 TeV. 
The selection cuts are summarized in \tab{tab:SUSYcuts}; they differ 
in some details from those considered in \sct{sec:fiducial}: First of all, 
lepton cuts are chosen differently for electrons and muons. More precisely,
all leptons are first ordered in $\pt$, and then the \pt{} threshold for each 
lepton is set according to its flavour and to whether it is the leading or a subleading 
lepton. Also the pseudo-rapidity cuts are different for 
electrons and muons. These cuts imply that the theoretical prediction of the cross section 
in this case is not symmetric under $e\leftrightarrow\mu$ exchange any more,
and the full set of eight channels must be computed separately for the 
\genllln{} final state.
Furthermore, the invariant mass of the three leptons is required to differ 
by at least $15$\,GeV from the $Z$-boson mass, 
and the invariant mass of every OSSF lepton pair is bounded from below to ensure IR safety.

\renewcommand{\baselinestretch}{1.5}
\begin{table}
\begin{center}
\begin{tabular}{c|c}
\toprule
& definition of the selection cuts for $pp\to \ell'^\pm{\nu}_{\ell^\prime} \ell^+\ell^-+X,\quad \ell,\ell'\in\{e,\mu\}$\\
\midrule
CMS 13 TeV & 
$p_{T,\ell_1}>25(20)$\,GeV if $\ell_1=e(\mu)$, \quad$p_{T,\ell_1}>25$\,GeV if $\ell_1=\mu$ and $\ell_{\ge 2}\neq\mu$\\
(cf. \citere{CMS:2016gvu}) & $p_{T,\ell_{\ge2}}>15(10)$\,GeV if $\ell_{\ge2}=e(\mu)$,\quad$|\eta_{e}|<2.5$, \quad$|\eta_{\mu}|<2.4$,\\
&$|m_{3\ell}-m_Z|>15$\,GeV,\quad 
$m_{\ell^+\ell^-}>12$\,GeV\\
\bottomrule
\end{tabular}
\end{center}
\renewcommand{\baselinestretch}{1.0}
\caption{\label{tab:SUSYcuts} Selection cuts used in our new-physics analysis. 
$\ell$ refers to all charged leptons, and numbers in indices refer to $\pT$-ordered particles of the respective group.}
\end{table}

\renewcommand{\baselinestretch}{1.0}

\setcounter{footnote}{0}

Our goal is to study QCD effects on distributions which are known
to provide a high experimental 
sensitivity to isolate a SUSY signal over the SM background.
The essential observables, ordered by their relevance, are:\footnote{We note that, contrary to the SM studies
of \sct{sec:fiducial} and \sct{sec:distributions},
the cuts we consider here do not require to identify the lepton pair coming from a $Z$ boson.
A $Z$-boson identification is needed only for specific observables, namely
$\mtw$ and $\mll$. The identification is the same as used by the CMS SM analysis 
at $13$\,TeV, outlined in \sct{sec:fiducial}. The OSSF lepton pair with the invariant mass closest to $m_Z$ is associated with the $Z$ boson.}
\begin{itemize}
\item the missing transverse 
energy $\ptmiss$, which (in particular in its tail) is highly sensitive if unobserved 
SUSY particles, usually the lightest supersymmetric particle (LSP), are produced 
via chargino-neutralino pair production; 
\item the transverse mass of the $W$ boson $\mtw$, 
more precisely of the system of missing energy and the lepton not associated with the $Z$-boson decay, which is
to some extent complementary to $\ptmiss$;
\item the invariant mass of the lepton pair associated 
with the $Z$-boson decay $\mll$, which allows a discrimination between 
searches in the SUSY parameter space with a small ($\mll \ll m_Z$), intermediate ($\mll \sim m_Z$) and large ($\mll \gg m_Z$) mass difference of neutralino and LSP.
\end{itemize}

Based on these considerations, we choose four different 
categories, 
which are inspired by the categories considered in \citere{CMS:2016gvu}:

\renewcommand{\baselinestretch}{1.5}
\begin{table}[h]
\begin{tabular}{ll}
{\bf Category I}: &\quad\quad no additional cut\\
{\bf Category II}: &\quad\quad $\ptmiss>200$\,GeV\\
{\bf Category III}:&\quad\quad $\mtw>120$\,GeV\\
{\bf Category IV}:&\quad\quad $m_{ll}>105$\,GeV
\end{tabular}
\end{table}
\renewcommand{\baselinestretch}{1.0}

Our calculation is performed by using the setup discussed at the beginning of this section and employed in
\sct{sec:fiducial} and \sct{sec:distributions}.
However, since we are interested in studying the impact of QCD radiative corrections in a phase space region
which is characterized by relatively large transverse momenta (up to ${\cal O}(1\,{\rm TeV})$),
the fixed scale $\mu_0=\frac{1}{2}(m_Z+m_W)$ is not fully appropriate.
In the present study we use instead a dynamic scale defined as
\begin{align}
\label{eq:dynscale}
\mu_R=\mu_F=\mu_0\equiv \frac12\,\left(\sqrt{m_Z^2+p^2_{T,\lz\lz}}+\sqrt{m_W^2+p^2_{T,\lw\nu_{\lw}}}\right),
\end{align}
where $p_{T,\lz\lz}$ and $p_{T,\lw\nu_{\lw}}$ are the transverse-momenta of the identified $Z$ and $W$ bosons, 
respectively. In the limit of small transverse momenta 
\refeq{eq:dynscale} reduces to the fixed scale $\mu_0=\frac{1}{2}(m_Z+m_W)$ used in \sct{sec:fiducial} and \sct{sec:distributions}.

In \tab{tab:NP_rates} we report our results for the integrated cross sections in
the four categories. Four separate results are given in that table by dividing into $W^+Z$ and $W^-Z$ production as well as SF and DF channels:
$\elle^{'+} \elle^+\elle^-$, $\elle^{+} \elle^+\elle^-$, $\elle^{'-} \elle^+\elle^-$ and $\elle^{-} \elle^+\elle^-$. 
Throughout this section, flavour channels related by $e\leftrightarrow\mu$ exchange are summed over and the combination of individual channels is always done by summing them.
We start our discussion from Category\,I, for which the cross section is of the order of
the fiducial cross sections presented in \sct{sec:fiducial}
for the SM measurements at 13\,TeV, although with somewhat looser selection cuts. The 
relative radiative correction are large: They amount to about 94\% at NLO and 13\% at NNLO.
These relative corrections are slightly larger for $W^-Z$ production 
as compared to $W^+Z$ production as can be inferred from the separate rows in the table.
Results in the SF and DF channels are of the same size.

\renewcommand{\baselinestretch}{1.5}

\begin{table}[tp]
\begin{center}
\resizebox*{!}{0.9\textheight}{%
\begin{tabular}{c c c c c c}
\toprule

channel
& $\sigma_{\textrm{LO}}$ [fb]
& $\sigma_{\textrm{NLO}}$ [fb]
& $\sigma_{\textrm{NNLO}}$ [fb]
& $\sigma_{\textrm{NLO}}/\sigma_{\textrm{LO}}-1$
& $\sigma_{\textrm{NNLO}}/\sigma_{\textrm{NLO}}-1$\\
\bottomrule

\multicolumn{6}{c}{Category I}\\
\midrule
$\elle^{'+} \elle^+\elle^-$ & $49.45(0)_{-5.8\%}^{+4.9\%}$ & $94.12(2)_{-3.9\%}^{+4.8\%}$ & $105.9(1)_{-2.2\%}^{+2.3\%}$ & 90.3\% & 12.6\% \Bstrut\\
$\elle^{+} \elle^+\elle^-$ & $48.97(0)_{-5.8\%}^{+4.8\%}$ & $93.13(2)_{-3.9\%}^{+4.8\%}$ & $104.7(1)_{-2.1\%}^{+2.2\%}$ & 90.2\% & 12.4\% \Bstrut\\
$\elle^{'-} \elle^+\elle^-$ & $32.04(0)_{-6.3\%}^{+5.3\%}$ & $63.68(3)_{-4.1\%}^{+5.0\%}$ & $71.89(4)_{-2.2\%}^{+2.3\%}$ & 98.7\% & 12.9\% \Bstrut\\
$\elle^{-} \elle^+\elle^-$ & $31.74(0)_{-6.3\%}^{+5.3\%}$ & $63.00(2)_{-4.1\%}^{+5.0\%}$ & $71.13(4)_{-2.2\%}^{+2.2\%}$ & 98.5\% & 12.9\% \Bstrut\\
\midrule
combined & $162.2(0)_{-6.0\%}^{+5.0\%}$ & $313.9(1)_{-4.0\%}^{+4.9\%}$ & $353.7(3)_{-2.2\%}^{+2.2\%}$ & 93.5\% & 12.7\% \Bstrut\\
\bottomrule

\multicolumn{6}{c}{Category II}\\
\midrule
$\elle^{'+} \elle^+\elle^-$ & $0.3482(0)_{-2.8\%}^{+2.8\%}$ & $1.456(0)_{-11\%}^{+13\%}$ & $1.799(1)_{-5.4\%}^{+5.2\%}$ & 318\% & 23.6\% \Bstrut\\
$\elle^{+} \elle^+\elle^-$ & $0.3486(0)_{-2.8\%}^{+2.8\%}$ & $1.452(0)_{-11\%}^{+13\%}$ & $1.789(1)_{-5.4\%}^{+5.1\%}$ & 316\% & 23.2\% \Bstrut\\
$\elle^{'-} \elle^+\elle^-$ & $0.1644(0)_{-2.7\%}^{+2.6\%}$ & $0.5546(1)_{-9.9\%}^{+12\%}$ & $0.6631(4)_{-4.8\%}^{+4.3\%}$ & 237\% & 19.6\% \Bstrut\\
$\elle^{-} \elle^+\elle^-$ & $0.1645(0)_{-2.7\%}^{+2.6\%}$ & $0.5535(1)_{-9.9\%}^{+12\%}$ & $0.6600(3)_{-4.7\%}^{+4.2\%}$ & 237\% & 19.2\% \Bstrut\\
\midrule
combined & $1.026(0)_{-2.8\%}^{+2.7\%}$ & $4.015(1)_{-10\%}^{+13\%}$ & $4.911(3)_{-5.2\%}^{+4.9\%}$ & 292\% & 22.3\% \Bstrut\\
\bottomrule

\multicolumn{6}{c}{Category III}\\
\midrule
$\elle^{'+} \elle^+\elle^-$ & $0.3642(0)_{-2.2\%}^{+1.5\%}$ & $0.5909(1)_{-3.3\%}^{+4.3\%}$ & $0.6373(16)_{-1.6\%}^{+1.6\%}$ & 62.3\% & 7.86\% \Bstrut\\
$\elle^{+} \elle^+\elle^-$ & $1.090(0)_{-2.4\%}^{+1.7\%}$ & $1.904(0)_{-3.8\%}^{+4.8\%}$ & $2.071(2)_{-1.9\%}^{+1.9\%}$ & 74.7\% & 8.79\% \Bstrut\\
$\elle^{'-} \elle^+\elle^-$ & $0.2055(0)_{-2.8\%}^{+2.0\%}$ & $0.3447(1)_{-3.4\%}^{+4.5\%}$ & $0.3731(9)_{-1.7\%}^{+1.6\%}$ & 67.8\% & 8.22\% \Bstrut\\
$\elle^{-} \elle^+\elle^-$ & $0.6463(1)_{-2.9\%}^{+2.1\%}$ & $1.136(0)_{-3.7\%}^{+4.8\%}$ & $1.232(1)_{-1.7\%}^{+1.7\%}$ & 75.8\% & 8.42\% \Bstrut\\
\midrule
combined & $2.306(0)_{-2.5\%}^{+1.8\%}$ & $3.976(1)_{-3.7\%}^{+4.7\%}$ & $4.313(6)_{-1.8\%}^{+1.8\%}$ & 72.4\% & 8.50\% \Bstrut\\
\bottomrule

\multicolumn{6}{c}{Category IV}\\
\midrule
$\elle^{'+} \elle^+\elle^-$ & $2.500(0)_{-3.9\%}^{+3.1\%}$ & $4.299(1)_{-3.4\%}^{+4.1\%}$ & $4.682(2)_{-1.6\%}^{+1.7\%}$ & 72.0\% & 8.92\% \Bstrut\\
$\elle^{+} \elle^+\elle^-$ & $2.063(0)_{-4.2\%}^{+3.4\%}$ & $3.740(1)_{-3.6\%}^{+4.5\%}$ & $4.160(2)_{-2.0\%}^{+2.2\%}$ & 81.3\% & 11.2\% \Bstrut\\
$\elle^{'-} \elle^+\elle^-$ & $1.603(0)_{-4.4\%}^{+3.4\%}$ & $2.805(1)_{-3.5\%}^{+4.2\%}$ & $3.058(1)_{-1.6\%}^{+1.7\%}$ & 75.0\% & 9.01\% \Bstrut\\
$\elle^{-} \elle^+\elle^-$ & $1.373(0)_{-4.7\%}^{+3.8\%}$ & $2.591(1)_{-3.9\%}^{+4.7\%}$ & $2.904(1)_{-2.1\%}^{+2.2\%}$ & 88.7\% & 12.1\% \Bstrut\\
\midrule
combined & $7.540(1)_{-4.2\%}^{+3.4\%}$ & $13.44(0)_{-3.6\%}^{+4.4\%}$ & $14.80(1)_{-1.8\%}^{+1.9\%}$ & 78.2\% & 10.2\% \Bstrut\\
\bottomrule
\end{tabular}}
\end{center}
\renewcommand{\baselinestretch}{1.0}
\caption{\label{tab:NP_rates} Fiducial cross sections at LO, NLO and NNLO for all three categories split by SF ($\elle \elle \elle$) and DF ($\elle^{'} \elle \elle$) as well as $W^+Z$ and $W^-Z$ production. The last two columns contain the relative NLO and NNLO corrections.  ``Combined'' refers to the \textit{sum} sum of all separate contributions.}
\end{table}

\renewcommand{\baselinestretch}{1.0}

An additional and stringent cut on the missing transverse energy of $\ptmiss > 200$\,GeV (Category II) 
changes this picture dramatically: The cross section is reduced by roughly two orders of magnitude. 
The LO prediction vastly underestimates the cross section, with NLO corrections of several hundred 
percent. These corrections are significantly larger for the $W^+Z$ cross section ($\sim 320$\%) than 
for $W^-Z$ production ($\sim 240$\%). This is not unexpected:  A hard cut on \ptmiss\ enhances the relevance of
the high-$p_T$ region, where QCD corrections are more important.  Moreover, the $W^+Z$ final state is mainly produced through $u{\bar d}$ scattering, while $W^-Z$ originates from ${\bar u}d$ scattering. The $u$ quark carries on average more momentum than the $d$ quark, thus leading to harder $p_T$ spectra for the $W^+Z$ final states compared to $W^-Z$. Following similar arguments, also the NNLO contribution is sizeable. It is roughly $22$\%, which is in particular larger 
than in the more inclusive Category I. This clearly confirms the importance of NNLO corrections when scenarios with cuts on observables relevant to new-physics 
searches, such as $\ptmiss$, are under consideration.

In Category III (additional cut $\mtw>120$\,GeV), on the other hand, the cut 
has a rather mild effect on the NLO corrections, which are about $70\%$, i.e.\
even slightly lower than in Category I. NNLO corrections have an effect of 
about $8\%$. What turns out to be striking in this category 
is the difference between SF and DF channels, which are similarly large in the two 
previous categories. Here, the SF results are more than a factor 
of three higher than the corresponding DF cross section. We will discuss the origin 
and the implications of this observation in detail below.

QCD corrections are also very mildly affected by a high cut on \mll{} in Category IV ($\mll > 105$\,GeV)  
which forces the $Z$ boson to be off-shell. The difference between SF and DF results is smaller and has the opposite sign with respect to Category III, being, however, still of order $10\%-20$\% 
depending on the order.

Comparing the $W^+Z$ and $W^-Z$ ratios in the four categories, we see that,
due to the different contributing partonic channels,
they strongly depend on the applied phase-space cuts, with $\sigma_{W^+Z}/\sigma_{W^-Z}\approx 1.47$ in 
Category I, $\sigma_{W^+Z}/\sigma_{W^-Z}\approx 2.71$ in Category II,
$\sigma_{W^+Z}/\sigma_{W^-Z}\approx 1.69$ in Category III and $\sigma_{W^+Z}/\sigma_{W^-Z}\approx 1.48$ in Category IV at NNLO. We note that the precise
value of the ratio of 
$W^+Z$ and $W^-Z$ cross sections may be affected by the specific choice of the 
used PDFs.

\begin{figure}
\begin{center}
\begin{tabular}{ccc}
\hspace*{-0.2cm}
\includegraphics[trim = 7mm -7mm 0mm 0mm, width=.26\textheight]{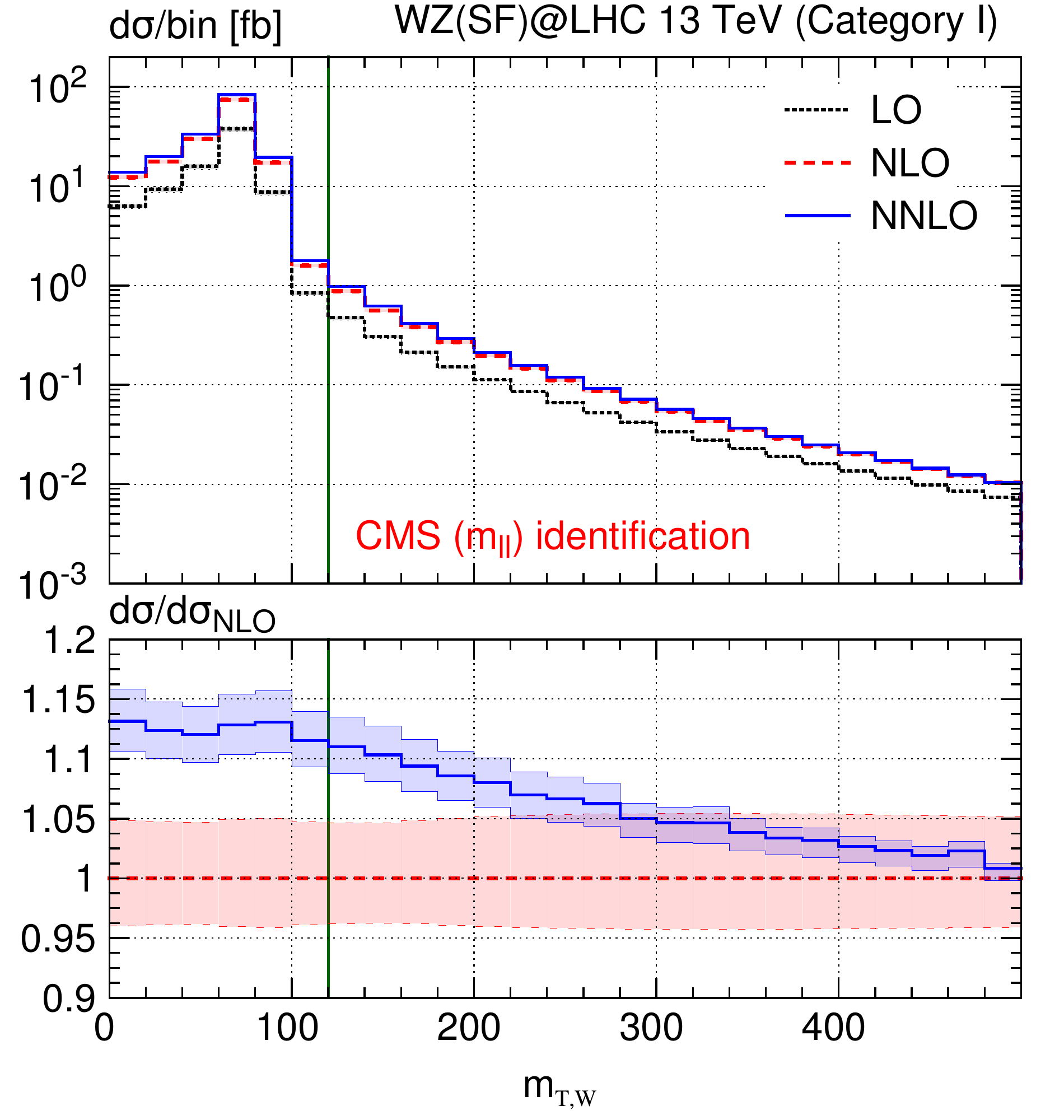} & \hspace{-1.17cm}
\includegraphics[trim = 7mm -7mm 0mm 0mm, width=.26\textheight]{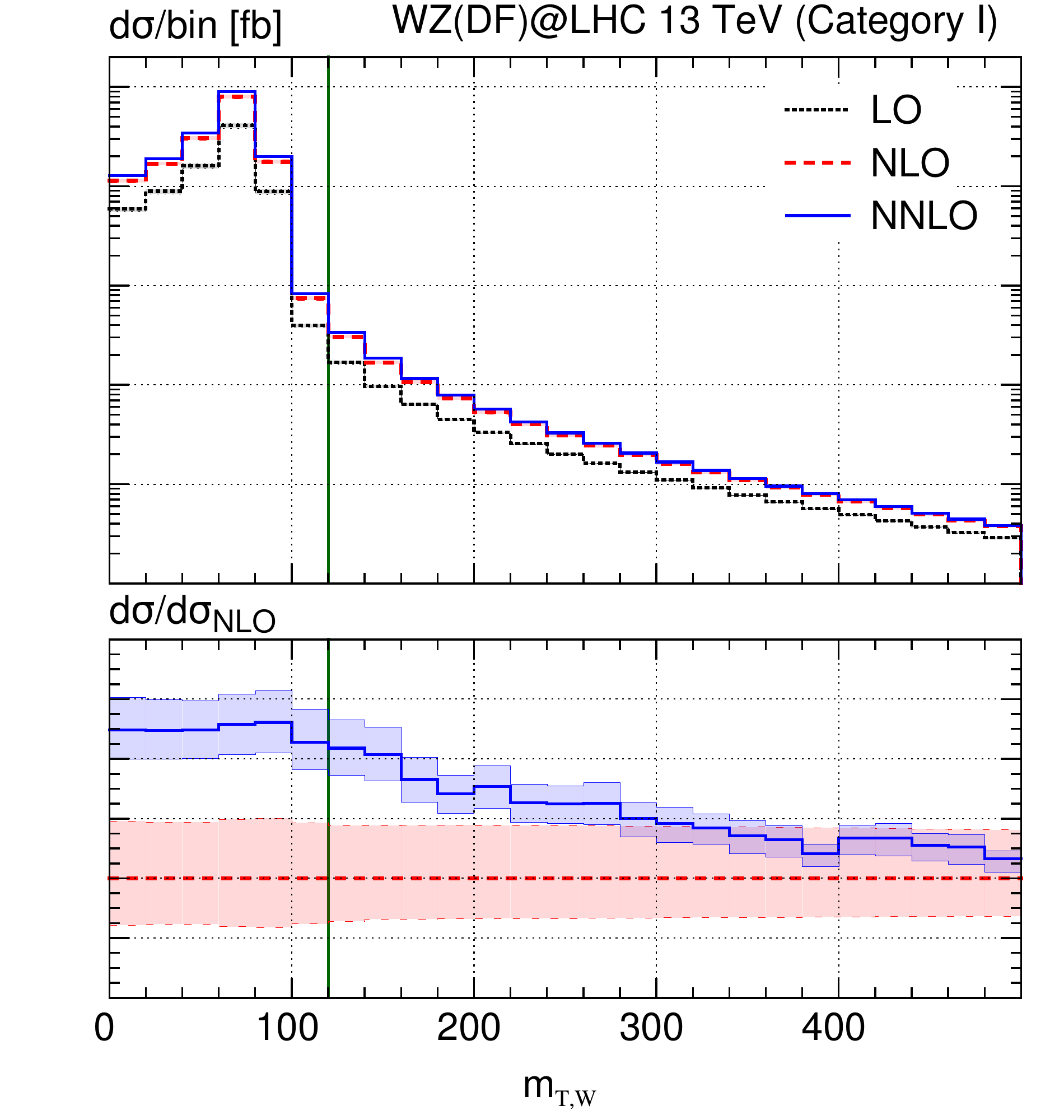} & \hspace{-1.17cm}
\includegraphics[trim = 7mm -7mm 0mm 0mm, width=.26\textheight]{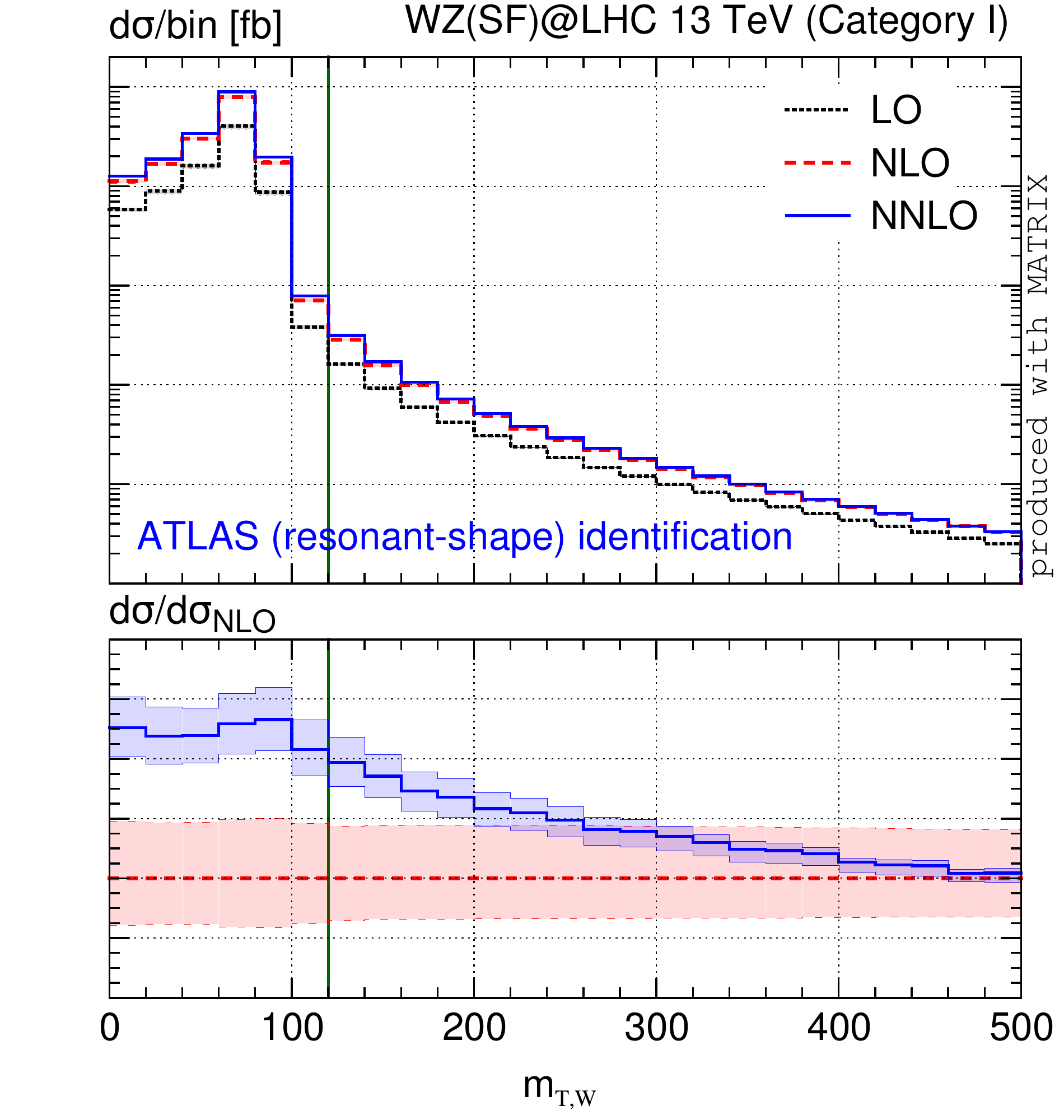}
\end{tabular}
\caption[]{\label{fig:mTWcomp}{Distributions with respect to $\mtw$ in the fiducial phase space without additional cuts (Category I); left: SF channel; centre: DF channel;  
right: SF channel, but using the resonant-shape identification of $W$ and $Z$ bosons as used by ATLAS. The green vertical line indicates the cut of $\mtw>120$\,GeV in Category III.}}
\end{center}
\end{figure}

Let us discuss in more detail the large difference between SF and DF cross sections in Category III. This seems 
surprising at first sight, since, as outlined in \sct{sec:calculation}, the SF and DF channels feature the same diagrams and have the same
generic resonant structures. Indeed, all SM results as well as BSM results in Category I and II show at most minor differences between SF and DF 
channels. This is true both for rates and distributions. Category III differs from Category I only by an additional cut on \mtw{}, whose distribution in 
Category I is shown separately for the SF and DF channels in the left and centre plots of \fig{fig:mTWcomp}. For reference we have added a 
green vertical line at $\mtw=120$\,GeV, which indicates the additional cut in Category III. Apparently, the \mtw{} tail, which is dominated by off-shell $W$ bosons, 
is considerably higher in the SF channel than in the DF channel. Thus, the origin of the different SF and DF rates is a different distribution of events, which 
are moved from the $W$-peak region to the tail.

This behaviour is not a particular feature of the SF channel, but a consequence of the $Z$ (and $W$) identification we are using,
which is entirely based on 
the invariant masses of the two possible combinations of OSSF pairs, by associating the $Z$ boson with the one closer to the $Z$ mass. 
We have repeated the computation of the \mtw{} distribution by replacing the CMS identification with the ATLAS resonant-shape identification (see \sct{sec:fiducial} and in particular \refeq{eq:pestimator}).
The ensuing distribution is shown in the right plot of \fig{fig:mTWcomp}. Indeed, by eye, no difference between right (SF channel with ATLAS 
identification) and centre (DF channel) plot is visible.
We stress that in the DF channel the $Z$ and $W$ bosons are unambiguously 
identified by the lepton flavours in the final state.
The resonant-shape identification takes into account information on both the 
$W$- and the $Z$-boson propagators in the dominant double-resonant topologies, 
which leads to a more accurate modelling of the $W$-boson peak in the $\mtw$ distribution.
This identification procedure distributes less events into the tail (similar to the DF channel) than the CMS identification. The resonant-shape identification is therefore much more effective 
in removing events from the peak region when cutting on $\mtw>120$\,GeV. This is also reflected by the ensuing total cross sections 
in Category III: At NNLO, for example, the SF cross section with the resonant-shape identification ($0.9265(7)_{-1.5\%}^{+1.5\%}$\,fb) is of similar size as the one in 
the DF channel ($1.010(2)_{-1.6\%}^{+1.6\%}$\,fb) as compared to $3.303(4)_{-1.8\%}^{+1.9\%}$\,fb in the SF channel when using the CMS identification. Thus, in more 
than two out of three events, 
in Category III the identification of the $Z$ and the $W$ boson is swapped in the case of CMS with respect to using the resonant-shape identification. 
Besides the potential risks that such different identification might have on shapes of certain distributions\footnote{We have checked explicitly several 
distributions in Category III and found quite substantial differences between SF with CMS identification and DF channels for, e.g.,
\dphill{}, \mll{}, \mlll{}, \mwz{}, \ptltwo{}, \ptlw{}. These differences are alleviated when using the resonant-shape identification, although 
some minor differences remain also in that case.}, a more effective identification would 
allow to suppress the SM background to new-physics searches in this category by more than a factor of three. Let us finally remark that also Category IV 
would benefit from a more effective identification, although the effects are much smaller and negative in that case.

\begin{figure}
\begin{center}
\begin{tabular}{ccc}
\hspace*{-0.2cm}
\includegraphics[trim = 7mm -7mm 0mm 0mm, width=.26\textheight]{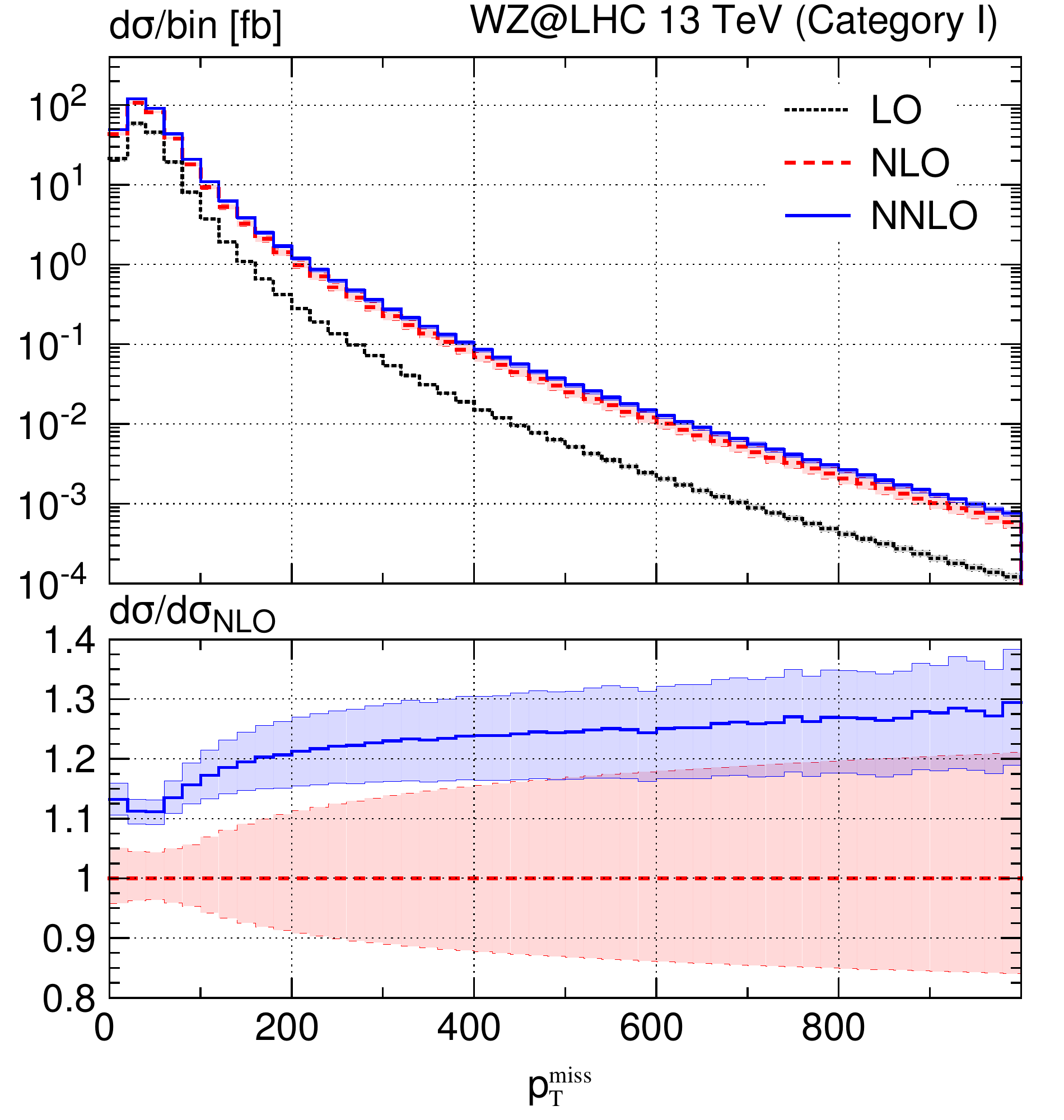} & \hspace{-1.17cm}
\includegraphics[trim = 7mm -7mm 0mm 0mm, width=.26\textheight]{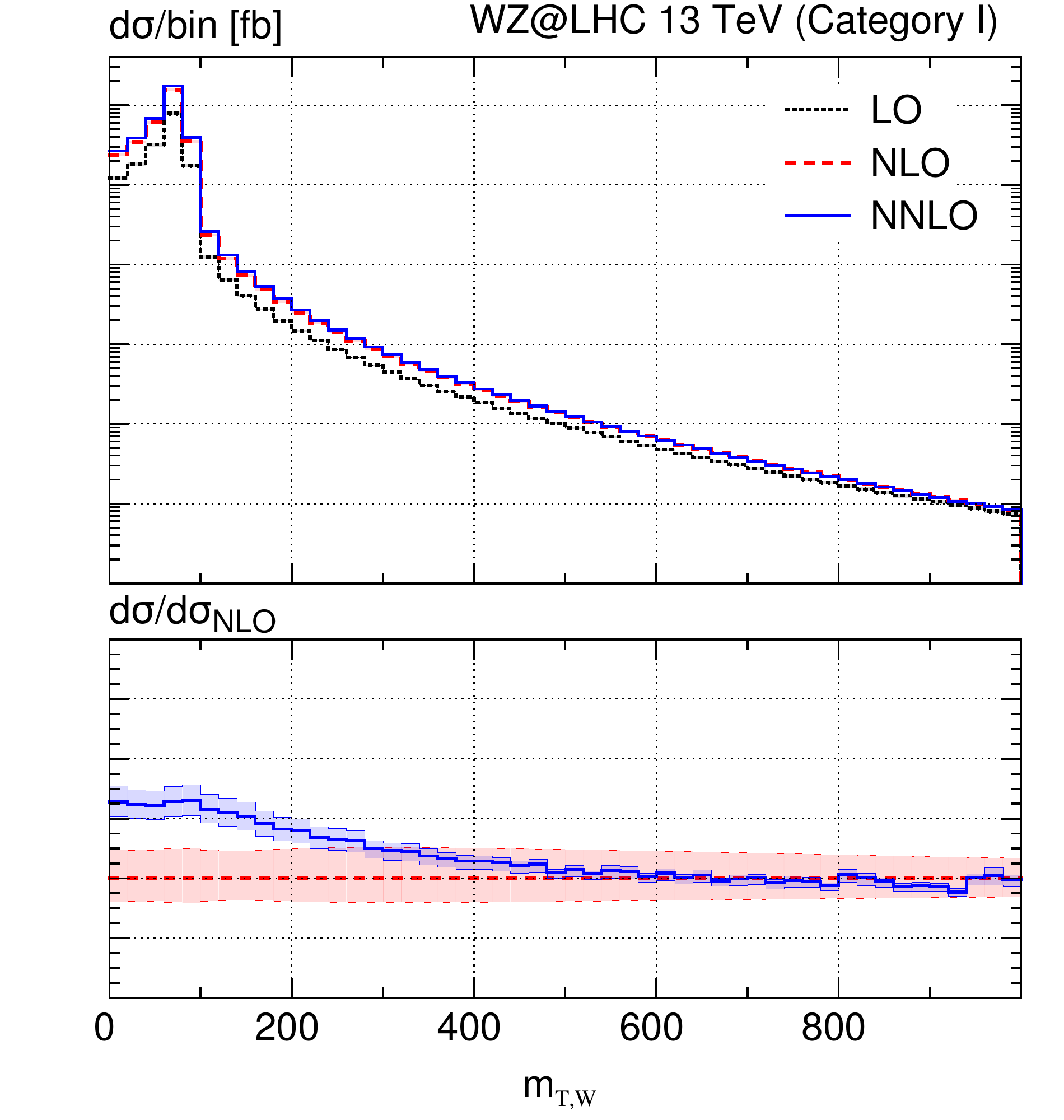} & \hspace{-1.17cm}
\includegraphics[trim = 7mm -7mm 0mm 0mm, width=.26\textheight]{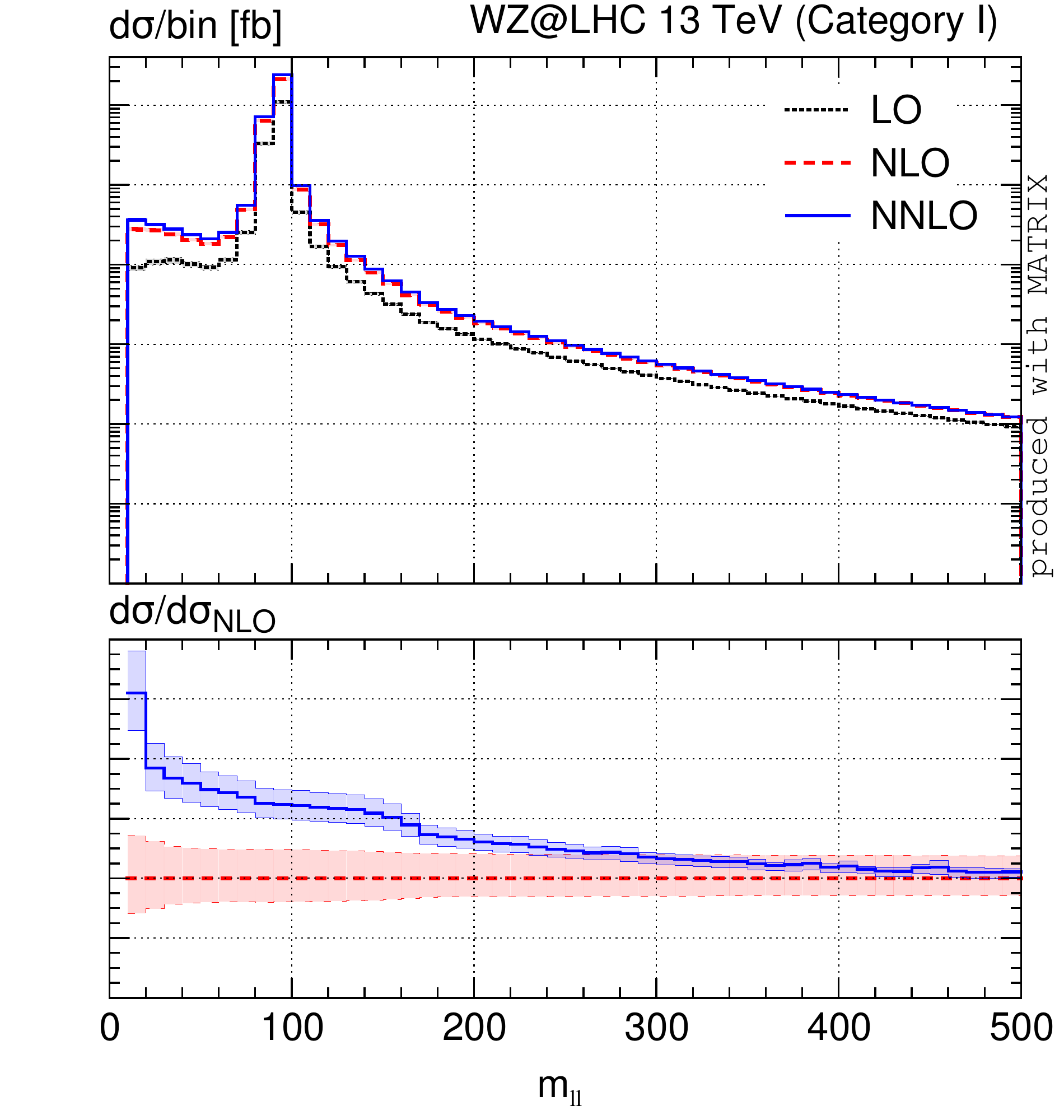}
\end{tabular}
\caption[]{\label{fig:catI}{Distributions with respect to $\ptmiss{}$ (left), $\mtw$ (centre) and 
$\mll$ (right) in the fiducial phase space without additional cuts (Category I).}}
\end{center}
\end{figure}

In terms of differential distributions, as previously pointed out, the most relevant observables for SUSY searches are $\ptmiss$, $\mtw$ and $\mll$. These distributions are shown 
in \fig{fig:catI} for the first category, i.e. without any additional restrictions on top of the default selection 
cuts of \tab{tab:SUSYcuts}.
The distribution in the missing transverse energy in the left panel of \fig{fig:catI} features large 
radiative corrections, ranging up to 30\% for the central curve, which, however, primarily affect the normalization.
Nevertheless, the shape of the distribution is affected by NNLO corrections at the 10\%-20\% level in 
the range up to $\ptmiss=1$\,TeV. We point out that the rather flat corrections at NNLO can only be 
achieved by using a dynamic scale (see \refeq{eq:dynscale}) that takes into account the effects of 
hard-parton emissions to properly model the tails of the distributions.  We have explicitly checked that the NLO \ptmiss{} distribution computed with a fixed 
scale is significantly harder in the tail with relatively large scale uncertainties, 
while the NNLO cross section --- as expected --- is quite stable with respect to 
the scale choice. As a consequence,
a fixed scale choice would lead to much larger, but negative NNLO corrections at high transverse momenta. 
Despite the considerable improvement in the  
perturbative stability achieved with the use of a dynamic scale,
a precise prediction of the fiducial cross 
section in Categories based on $\ptmiss$ still requires the inclusion of ${\cal O}(\as^2)$ terms,
since depending on the $\ptmiss$ cut the 
NNLO effects may still change by up to 20\%.

Similarly, also the $\mtw$ and $\mll$ distributions, in the centre and right
plots of \fig{fig:catI}, are subject to sizeable corrections due to the inclusion of 
${\cal O}(\as^2)$ terms. While in the tails of the spectra (for $\mtw\gtrsim300$\,GeV and 
$\mll\gtrsim200$\,GeV) the NLO and NNLO predictions  roughly agree within their respective 
uncertainties, at smaller \mtw{} and \mll{} values the shapes of the distributions are 
considerably modified, leading to NNLO corrections that are not covered by the lower-order 
uncertainty bands. These differences are alleviated to some extent by the fact that the low-\mtw{} 
and -\mll{} regions are usually less important to new-physics searches 
(where usually the phase-space region below
$\mtw \sim 120$\,GeV and $\mll \sim 100$\,GeV) is cut), 
but some region of phase space remains 
where NNLO corrections ought to be taken into account.

\begin{figure}[tp]
\begin{center}
\begin{tabular}{cc}
\hspace*{-0.17cm}
\includegraphics[trim = 7mm -7mm 0mm 0mm, width=.33\textheight]{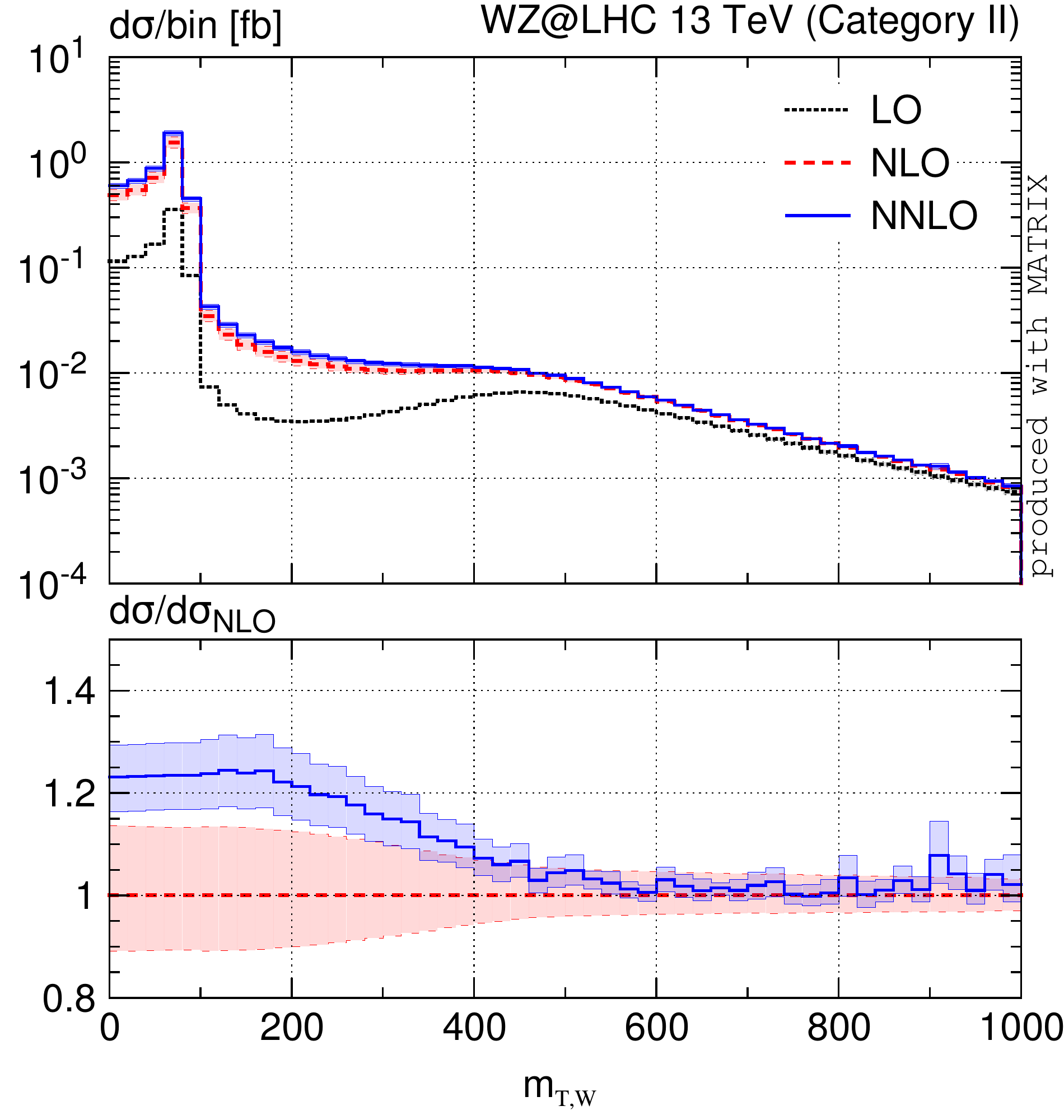} &
\includegraphics[trim = 7mm -7mm 0mm 0mm, width=.33\textheight]{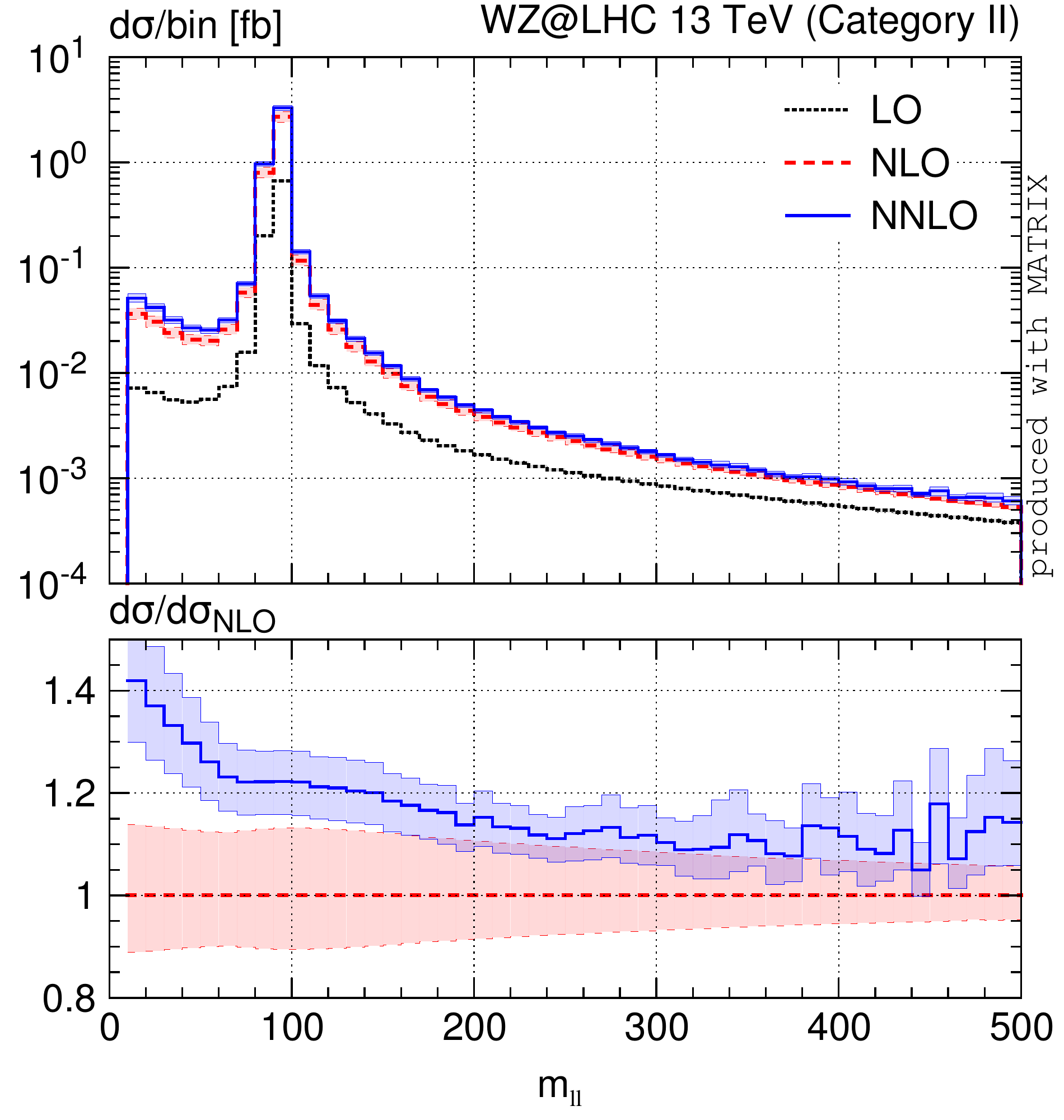} \\[-1em]
\hspace{0.6em} (a) & \hspace{1em}(b)
\end{tabular}
\caption[]{\label{fig:catII}{Distributions with respect to (a) $\mtw$ and 
(b) $\mll$ in the fiducial phase space with an additional $\ptmiss > 200$\,GeV cut (Category II).}}
\end{center}
\end{figure}

In \fig{fig:catII} we consider the $\mtw$ and $\mll$ spectra in Category II. Thus, these distributions include an 
additional cut of $\ptmiss > 200$\,GeV as compared to those in \fig{fig:catI}.
As pointed out before, such cut on \ptmiss{} requires NNLO accuracy on its own to ensure a proper 
modelling of the SM background. The specific value of $200$\,GeV, in fact, is incidentally in 
a region where the NNLO corrections start to become particularly large ($>20\%$), as can be inferred from 
the \ptmiss{} distribution in \fig{fig:catI}. Indeed, looking at \fig{fig:catII} both the distribution in \mtw{} and \mll{} feature NNLO and NLO cross sections without overlapping uncertainty 
bands in each peak region, with NNLO corrections of the order of 20\%.
For small \mll{} values
NNLO effects increase up to more than 40\%. This region, however, is less relevant
to new-physics searches. We note that, when going from NLO to NNLO 
scale uncertainties are reduced from about $15\%$ to at most $10\%$.
Overall, the results of the two distributions are very similar to the corresponding ones in 
\fig{fig:catI} for Category I. Although the NLO and NNLO scale uncertainties 
are generally larger, the ensuing bands do not overlap around the peak of the distributions.

\begin{figure}[tp]
\begin{center}
\begin{tabular}{cc}
\hspace*{-0.17cm}
\includegraphics[trim = 7mm -7mm 0mm 0mm, width=.33\textheight]{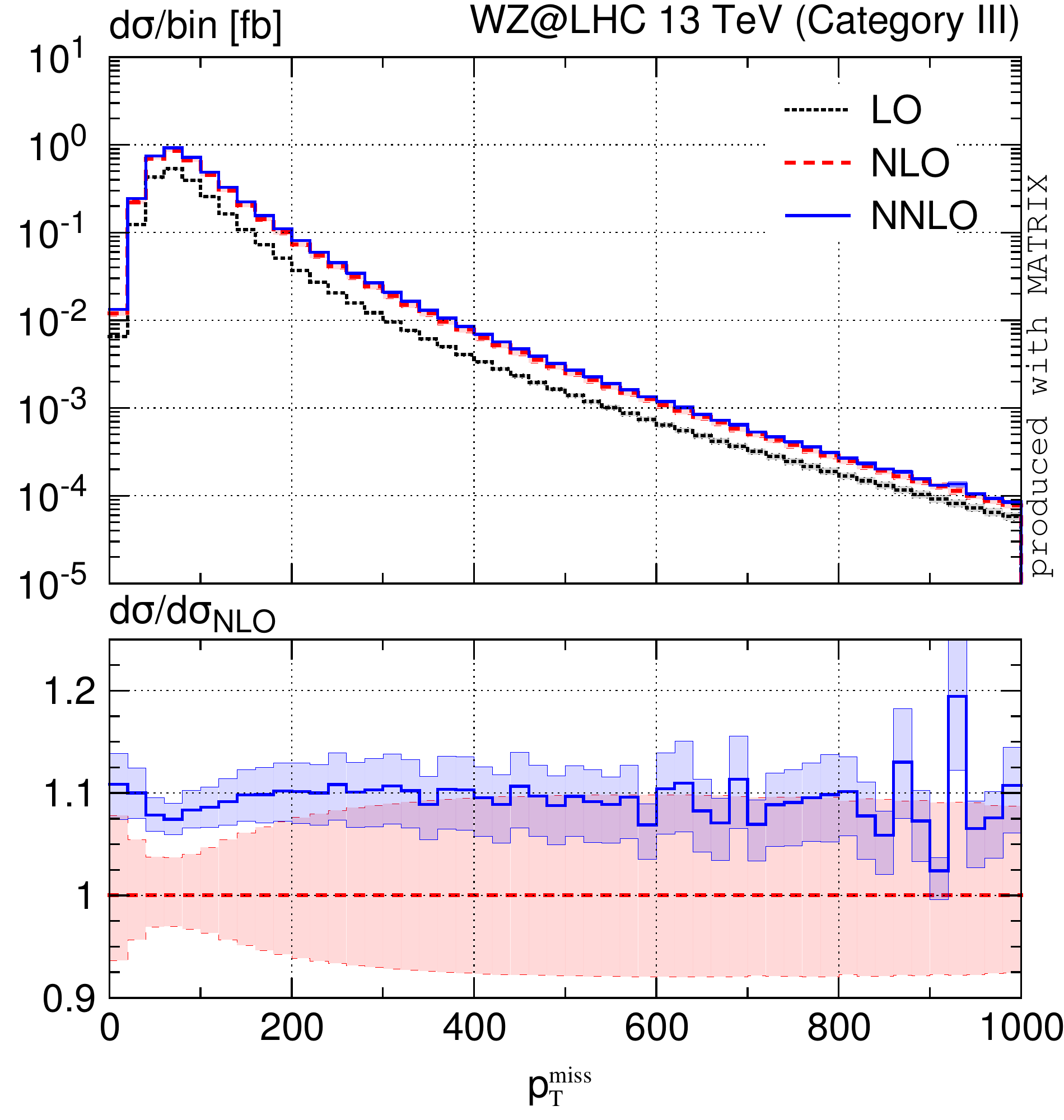} &
\includegraphics[trim = 7mm -7mm 0mm 0mm, width=.33\textheight]{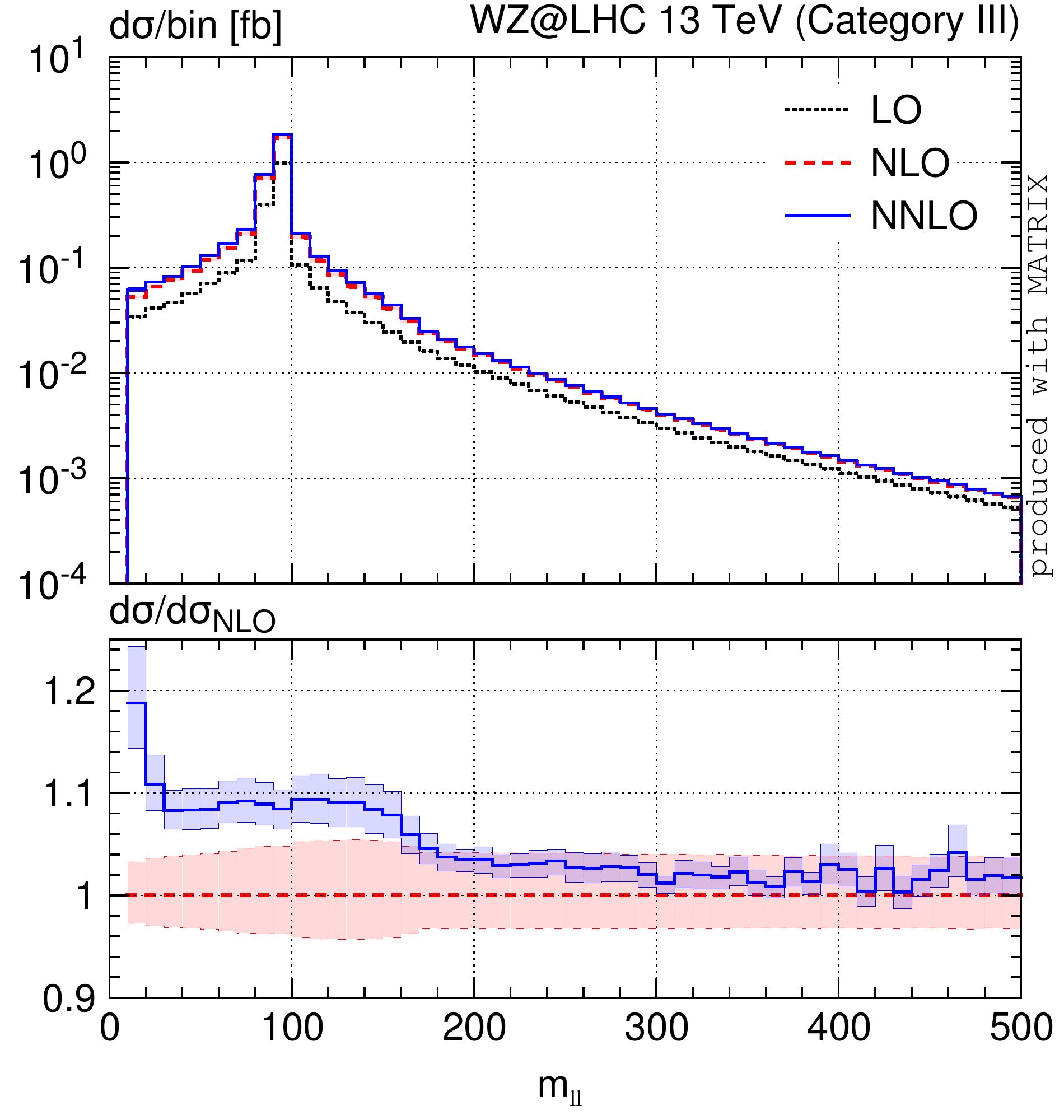} \\[-1em]
\hspace{0.6em} (a) & \hspace{1em}(b)
\end{tabular}
\caption[]{\label{fig:catIII}{Distributions with respect to (a) $\ptmiss$ and 
(b) $\mll$ in the fiducial phase space with an additional $\mtw > 120$\,GeV cut (Category III).}}
\end{center}
\end{figure}

\begin{figure}[tp]
\begin{center}
\begin{tabular}{cc}
\hspace*{-0.17cm}
\includegraphics[trim = 7mm -7mm 0mm 0mm, width=.33\textheight]{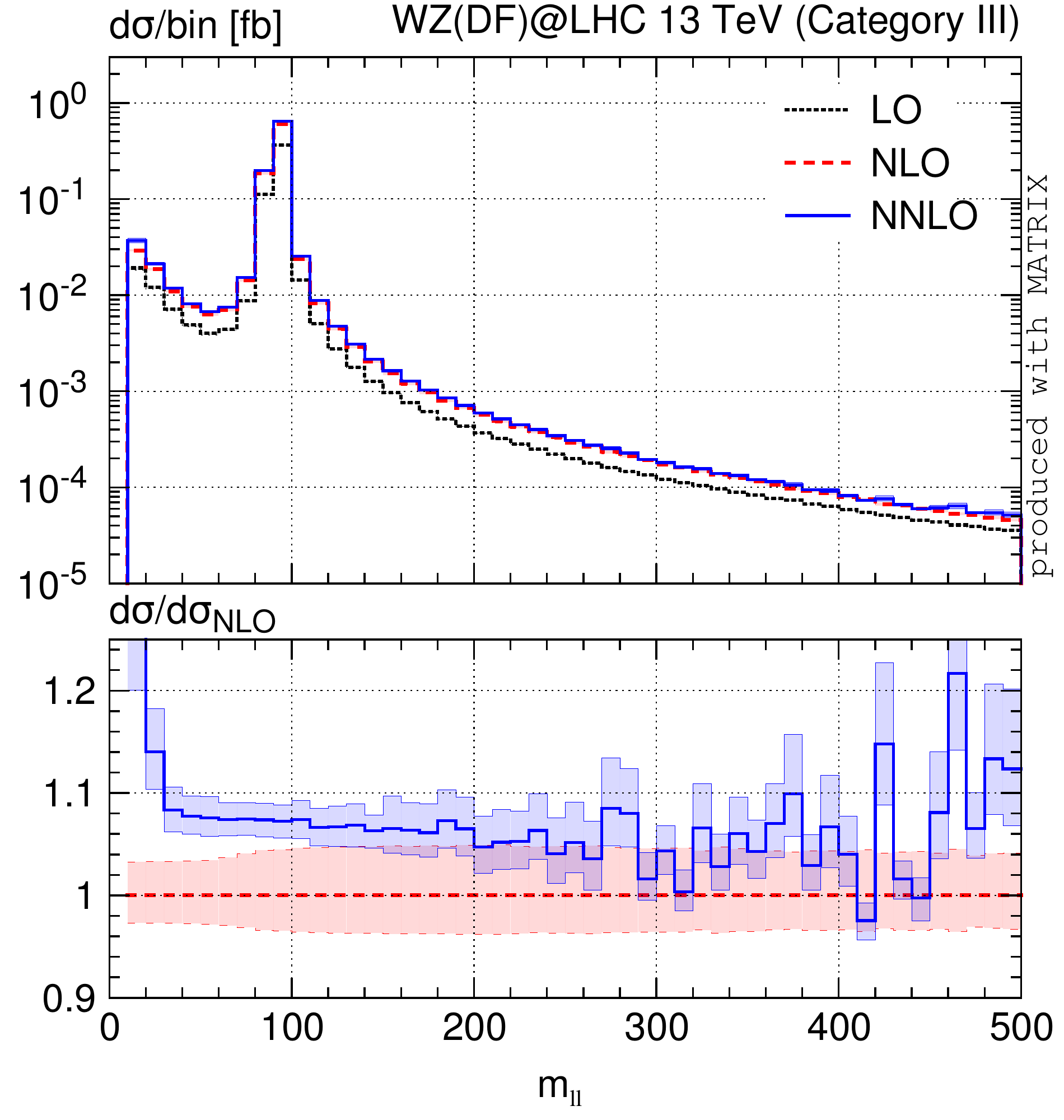} &
\includegraphics[trim = 7mm -7mm 0mm 0mm, width=.33\textheight]{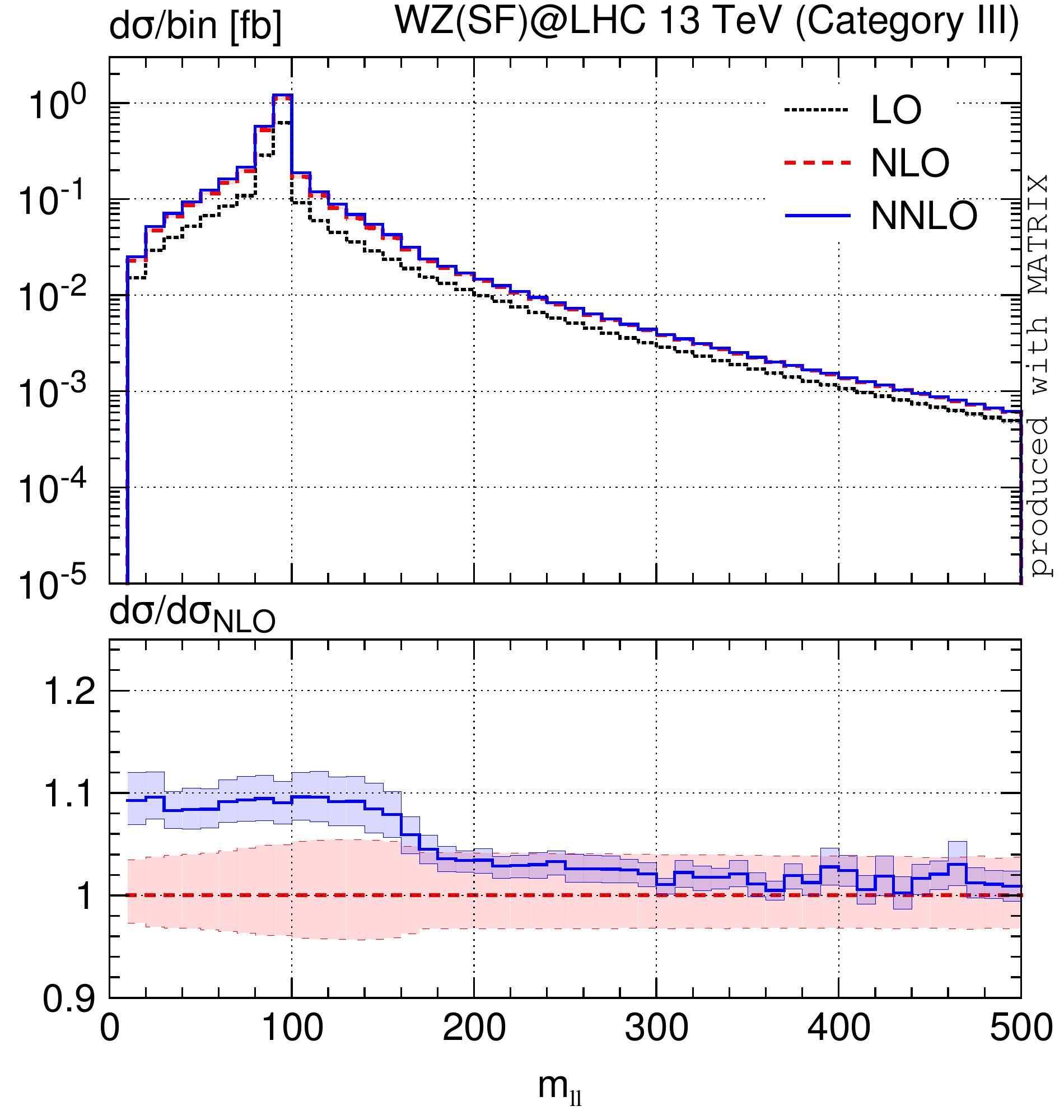} \\[-1em]
\hspace{0.6em} (a) & \hspace{1em}(b)
\end{tabular}
\caption[]{\label{fig:catIIImllsplit}{Distributions with respect to $\mll$ in the fiducial phase space with an additional $\mtw > 120$\,GeV cut (Category III), for (a) the SF 
and (b) the DF channel.}}
\end{center}
\end{figure}

\fig{fig:catIII} shows the $\ptmiss$ and $\mll$ spectra while including a
cut on $\mtw>120$\,GeV in addition to the standard selection cuts (Category III). 
Also in this case the general behaviour of these distributions is quite similar to 
those in Category I, however, the absolute size of the corrections at NNLO is reduced.
Thanks to the dynamic scale choice, the dependence of the NNLO correction 
on the value of $\ptmiss$ is quite flat. With a fixed scale we find a similarly 
strong $\ptmiss$ dependence in the tail of the distribution as pointed out for Category I.
NLO and NNLO uncertainty bands feature a satisfactory overlap starting from $\ptmiss\gtrsim200$\,GeV.
The $\mll$ distribution shows consistent NLO and NNLO predictions in the tail of the distribution. 
The NNLO corrections become larger ($\sim 10\%$) only at $\mll\lesssim150$\,GeV, where \wz{} 
production becomes less important as a SM background to new-physics searches. We point out
that, as shown in \fig{fig:catIIImllsplit}, the increase of the NNLO corrections 
at $\mll\lesssim150$\,GeV is only present in the SF channel, while the DF channel 
features a steep increase at $\mll\lesssim 50$\,GeV. It is clear from the main frame of that 
figure that the distributions in the two channels are modelled very differently, which can
again be traced back to the used identification procedure.

\begin{figure}[tp]
\begin{center}
\begin{tabular}{cc}
\hspace*{-0.17cm}
\includegraphics[trim = 7mm -7mm 0mm 0mm, width=.33\textheight]{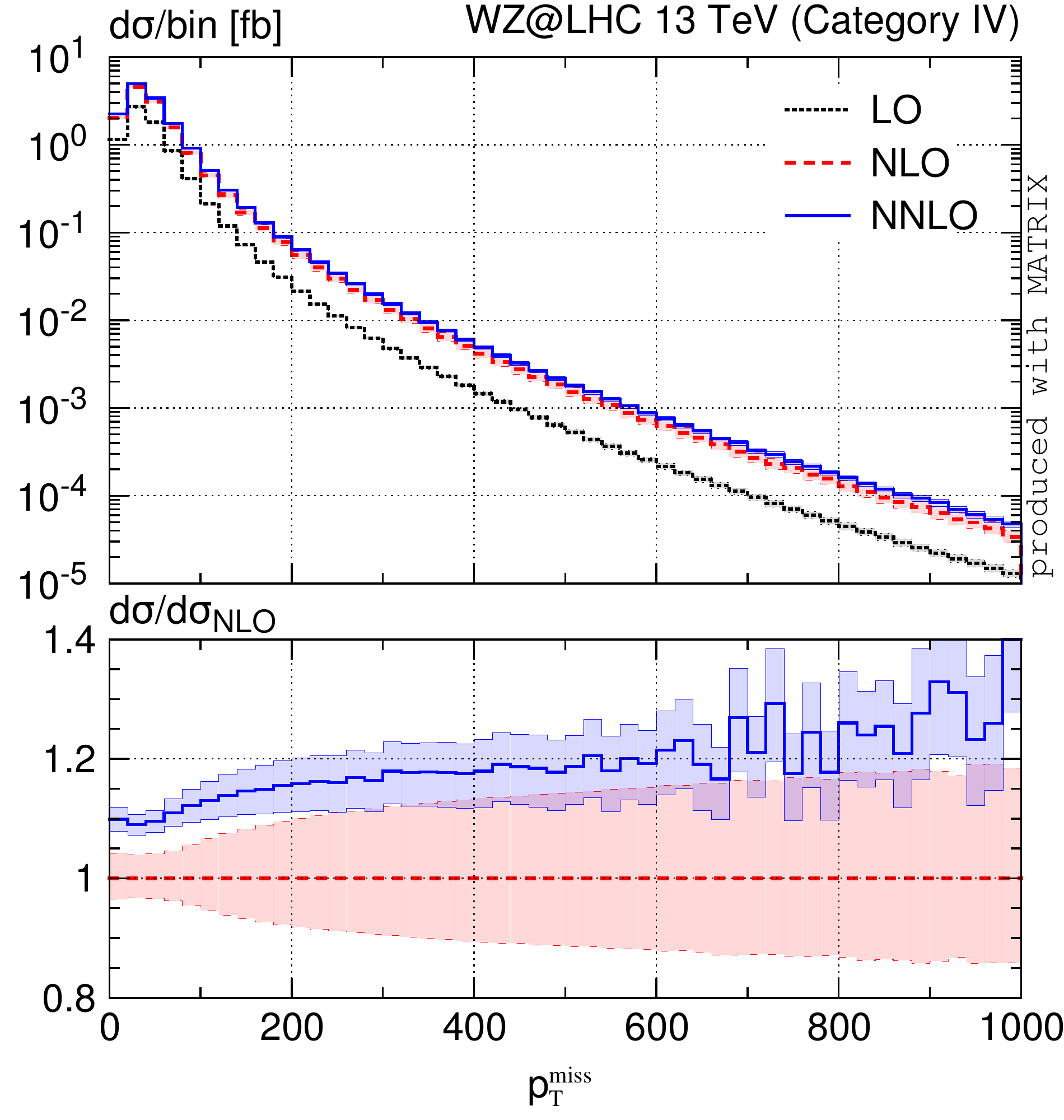} &
\includegraphics[trim = 7mm -7mm 0mm 0mm, width=.33\textheight]{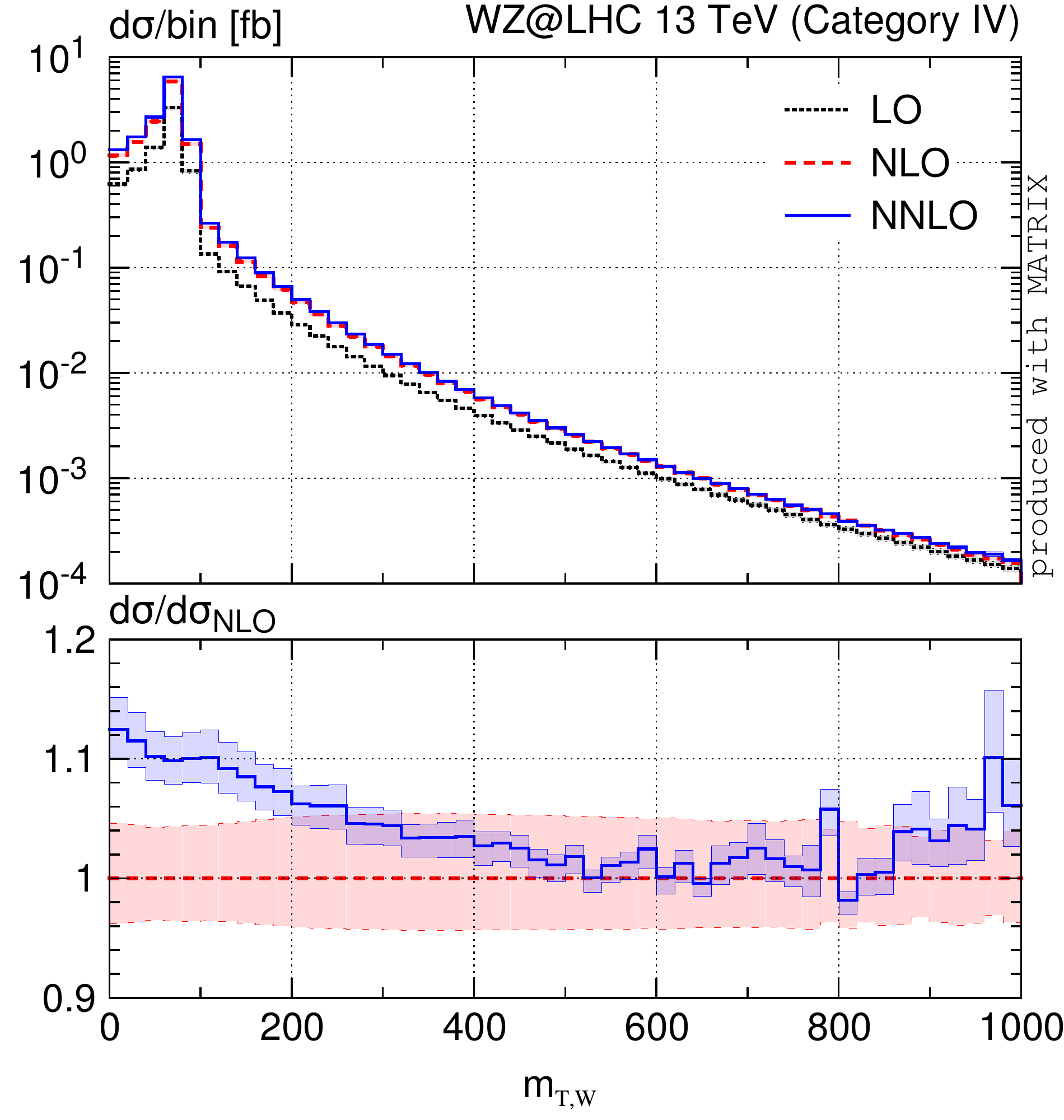} \\[-1em]
\hspace{0.6em} (a) & \hspace{1em}(b)
\end{tabular}
\caption[]{\label{fig:catIV}{Distributions with respect to (a) $\ptmiss$ and 
(b) $\mtw$ in the fiducial phase space with an additional $\mll > 105$\,GeV cut (Category IV).}}
\end{center}
\end{figure}

In \fig{fig:catIV} the \ptmiss{} and \mtw{} distributions in Category IV are shown.
We see that the $\mll >105$\,GeV cut has almost no impact on the shapes of the \ptmiss{} and \mtw{} 
spectra, apart from the general reduction of the absolute size of the 
NNLO corrections compared to Category\,I. Also in this category NNLO corrections are quantitatively relevant, 
and their impact on the tails of the distributions is reduced 
with the use of a dynamic scale.

In conclusion,
for the three observables relevant to new-physics searches that have been 
considered in this section,
the sizeable (10\%-30\%) NNLO corrections depend on the specific 
cut values. This demands NNLO accurate predictions for the \wz{} background 
when categories based on these observables are defined. Furthermore, a dynamic scale choice is crucial to properly
model the various distributions,
in particular the tail of the \ptmiss{} spectrum.
Moreover, NNLO corrections considerably reduce the perturbative uncertainties in all 
three distributions we investigated, regardless of the category under consideration.

\section{Summary}
\label{sec:summary}

In this paper, we have presented the first computation of fully differential cross sections for the production of a \wz{} pair at NNLO in QCD perturbation theory. 
Our computation consistently includes the leptonic decays of the weak 
bosons accounting for off-shell effects, spin correlations and interference 
contributions in all double-, single- and non-resonant configurations 
in the complex-mass scheme, i.e.\ we have performed a complete calculation for the process 
$pp\to \llln+X$ with $\ell,\ell'\in\{e,\mu\}$,
both in the SF and in the DF channel.
Our results are obtained with the numerical program MATRIX, which employs
the \qt{}-subtraction method to evaluate NNLO QCD corrections to a wide class of processes.
We have shown that the ensuing fiducial cross sections and distributions
depend very mildly on the technical cut-off 
parameter $r_{\rm cut}$, thereby allowing us to numerically control the predicted 
NNLO cross section at the one-permille level or better.

We have presented a comprehensive comparison of our numerical predictions 
with the available data from ATLAS and CMS at $\sqrt{s}=8$ and $13$\,TeV 
for both the fiducial cross sections and differential distributions in \wz{} production.
As in the case of the inclusive cross section \cite{Grazzini:2016swo}
QCD radiative corrections are essential to properly model the 
\wz{} cross section. They amount to up to $85\%$ at NLO, and NNLO corrections 
further increase the NLO result by about $10\%$. The inclusion of NNLO corrections 
significantly improves the agreement with the measured cross sections by ATLAS 
at both 8 and 13\,TeV centre-of-mass energies. The 13\,TeV CMS result is 
somewhat ($\sim 2.6\sigma$) lower than the theoretical prediction, which is about the same 
discrepancy that has been observed in the result extrapolated to the total inclusive 
cross section \cite{Grazzini:2016swo}. The full data set collected by the end of 
2016 ($\sim 40$\,fb$^{-1}$) will show whether this difference is a 
plain statistical effect of the small data set ($\sim 2.3$\,fb$^{-1}$) 
used for that measurement.

Distributions in the fiducial phase space of the \genllln{} final states are 
available only for the ATLAS $8$\,TeV data set. Our comparison 
reveals a remarkable agreement with the measured cross section in each 
bin upon inclusion of higher-order corrections, being typically 
within $1\sigma$ of the quoted experimental errors. Although this statement 
holds already at NLO, the NNLO cross sections display an 
improved description of the data not only in terms of normalization, but also 
regarding the shapes. 
Only the distribution in the missing transverse energy exhibits some tension 
between theory and data:
We observe deviations at the level of $1\sigma-2\sigma$ in some bins, leading to 
a more evident discrepancy in the shape of the distribution.
We have shown that this discrepancy is present only in $W^-Z$ production,
while our NNLO prediction nicely describes
the data in the case of $W^+Z$ production.

We have further shown that our computation of the ratio of
$W^+Z$ over $W^-Z$ distributions agrees well with the experimental data,
given the rather large experimental uncertainties.
Along with this study we have pointed out 
a number of distributions which signal significant differences between $W^+Z$ 
and $W^-Z$ production, and may be sensitive to disentangle genuine 
perturbative effects at NNLO.

We have completed our phenomenological study by considering a
scenario where \wz{} production is a background to new-physics searches 
in the three leptons plus missing energy channel. NNLO effects on the
background rates have been discussed in the relevant categories, together 
with the corresponding distributions.
Our findings can be summarized as follows:
\begin{itemize}
\item LO predictions cannot be used to model cross sections 
and distributions in a meaningful way: The size of NLO 
corrections can be, in some categories, of the order of several hundred percent.
\item NNLO corrections on the \wz{} rates range between roughly $8\%$ and $23\%$, 
while distributions are subject to considerable shape distortions when going from NLO to NNLO.
\item For cuts on the $\ptmiss$ observable, which is particularly important for 
categorization in new-physics scenarios, NNLO corrections turn out to be 
particularly important, as they may vary between 10\%-30\% depending on the specific 
value of the cut.
\item Only using a dynamic scale (see \refeq{eq:dynscale})
the shape of the relevant distributions is perturbatively stable.
This is in particular true for the 
\ptmiss{} distribution, which was found to be drastically impacted  
by NNLO corrections if a fixed scale was applied.
\item Finally, we have shown that in the SF channel an identification of the $Z$ boson
based solely on how close the dilepton-pair mass is to $m_Z$
may lead to some problems:
When a $\mtw>120$\,GeV cut is enforced,
in more than two out of 
three events the $Z$- and $W$-boson identification is swapped,
leading to a difference in the SF and DF rates by more than a factor of three.
We find that a resonant-shape
identification (see \refeq{eq:pestimator}) is much more efficient, thereby leading to a more effective background suppression.
\end{itemize}

We conclude by adding a few comments about the residual uncertainties of our calculation.
As is customary in perturbative QCD computations,
the uncertainties from missing higher-order contributions were estimated
by studying scale variations. We have seen that,
when going from NLO to NNLO scale uncertainties are generally reduced
both for fiducial cross sections and for kinematical distributions. 
It should be noted, however, that the uncertainties seem to underestimate the 
size of missing higher-order corrections at LO and NLO. This tendency decreases with
increasing perturbative order: While the LO uncertainty grossly underestimates 
the size of the NLO corrections (which, for this process, is in part due to the existence of an approximate radiation zero), the NLO and NNLO predictions are much closer, and almost consistent within uncertainties.
Considering that at NNLO all partonic channels are included and no regions of phase phase that are effectively only LO-accurate remain,
we conclude that the ${\cal O}(2-5\%)$ NNLO uncertainties on our fiducial cross sections (see Tables \ref{tab:ATLAS8}, \ref{tab:ATLAS13}, \ref{tab:CMS13} and \ref{tab:NP_rates})
are expected to provide the correct order of 
magnitude of yet uncalculated higher-order contributions.
EW corrections would affect the fiducial cross sections at the $1\%$ level or 
less \cite{Bierweiler:2013dja,Baglio:2013toa}, but are expected to become relevant
in the tails of the distributions, which will be potentially important for new-physics searches.
The inclusion of EW corrections is, however, left to future investigations. 
PDF uncertainties are expected to be at the $1\%-2\%$ level.

We believe that the calculation and the results presented in this paper will be
highly valuable both for experimental measurements of the \wz{} signal
and in new-physics searches involving the three lepton plus missing 
energy signature. The computation is available in 
the numerical program \Matrix{}, which is able to carry out fully-exclusive \nnlo{} 
computations for a wide class of processes at hadron colliders. We plan to
release a public version of our program in the near future.

\noindent {\bf Acknowledgements.}
We would like to thank Lucia Di Ciaccio, G{\"u}nther Dissertori, Thomas Gehrmann, Constantin Heidegger, Jan Hoss 
and Kenneth Long for useful discussions and comments on the manuscript.
This research was supported in part by the Swiss National Science Foundation (SNF) under 
contracts CRSII2-141847, 200020-169041, by 
the Research Executive Agency (REA) of the European Union under the Grant Agreement 
number PITN--GA--2012--316704 ({\it HiggsTools}), and by the National Science 
Foundation under Grant No. NSF PHY11-25915. MW has been partially supported by
ERC  Consolidator Grant 614577 HICCUP.

\clearpage

\begin{appendices}

\section{CMS cross sections at 8 TeV and 13 TeV}\label{app:rates_full}
    For completeness we quote below the cross-section predictions in the fiducial phase space for 
    CMS at 8 TeV and 13 TeV, separated by the individual leptonic channels in \tab{tab:CMS8_full}
    and  \tab{tab:CMS13_full}, respectively.\vspace{1cm}

\renewcommand{\baselinestretch}{1.5}
\begin{table}[h]
\begin{center}
\resizebox{\columnwidth}{!}{%
\begin{tabular}{c c c c c}
\toprule
channel
& $\sigma_{\textrm{LO}}$ [fb]
& $\sigma_{\textrm{NLO}}$ [fb]
& $\sigma_{\textrm{NNLO}}$ [fb]
& $\sigma_{\textrm{CMS}}$ [fb]\\
\bottomrule
$\mu^+ e^+e^-$ & \multirow{ 2}{*}{$14.72(0)_{-2.9\%}^{+2.1\%}$} & \multirow{ 2}{*}{$26.05(1)_{-4.1\%}^{+5.4\%}$} & \multirow{ 2}{*}{$28.16(1)_{-1.9\%}^{+1.8\%}$} \Bstrut\\
$e^+ \mu^+\mu^-$ &  &  &  &   \Bstrut\\
$e^+ e^+e^-$ & \multirow{ 2}{*}{$15.14(0)_{-3.0\%}^{+2.1\%}$} & \multirow{ 2}{*}{$26.97(1)_{-4.2\%}^{+5.5\%}$} & \multirow{ 2}{*}{$29.20(2)_{-1.9\%}^{+1.8\%}$} \Bstrut\\
$\mu^+ \mu^+\mu^-$ &  &  &  &   \Bstrut\\
\midrule
combined & $59.72(1)_{-3.0\%}^{+2.1\%}$ & $106.0(0)_{-4.1\%}^{+5.5\%}$ & $114.7(1)_{-1.9\%}^{+1.8\%}$ \Bstrut\\
\bottomrule
$\mu^- e^+e^-$ & \multirow{ 2}{*}{$8.432(1)_{-3.3\%}^{+2.4\%}$} & \multirow{ 2}{*}{$15.62(0)_{-4.5\%}^{+5.9\%}$} & \multirow{ 2}{*}{$16.98(1)_{-2.0\%}^{+1.9\%}$} \Bstrut\\
$e^- \mu^+\mu^-$ &  &  &  &   \Bstrut\\
$e^- e^+e^-$ & \multirow{ 2}{*}{$8.710(1)_{-3.4\%}^{+2.4\%}$} & \multirow{ 2}{*}{$16.24(0)_{-4.5\%}^{+5.9\%}$} & \multirow{ 2}{*}{$17.72(1)_{-2.0\%}^{+1.9\%}$} \Bstrut\\
$\mu^- \mu^+\mu^-$ &  &  &  &   \Bstrut\\
\midrule
combined & $34.28(0)_{-3.3\%}^{+2.4\%}$ & $63.72(2)_{-4.5\%}^{+5.9\%}$ & $69.39(4)_{-2.0\%}^{+1.9\%}$ \Bstrut\\
\bottomrule
$\mu^\pm e^+e^-$ & \multirow{ 2}{*}{$23.15(0)_{-3.1\%}^{+2.2\%}$} & \multirow{ 2}{*}{$41.67(1)_{-4.3\%}^{+5.6\%}$} & \multirow{ 2}{*}{$45.13(2)_{-1.9\%}^{+1.8\%}$} \Bstrut\\
$e^\pm \mu^+\mu^-$ &  &  &  &   \Bstrut\\
$e^\pm e^+e^-$ & \multirow{ 2}{*}{$23.86(0)_{-3.1\%}^{+2.2\%}$} & \multirow{ 2}{*}{$43.21(1)_{-4.3\%}^{+5.6\%}$} & \multirow{ 2}{*}{$46.91(3)_{-2.0\%}^{+1.9\%}$} \Bstrut\\
$\mu^\pm \mu^+\mu^-$ &  &  &  &   \phantom{\,$258\,\pm 8.1\%{\rm (stat)}^{+7.4\%}_{-7.7\%}{\rm (syst)}\pm 3.1{\rm (lumi)}$\quad\,}\Bstrut\\
\midrule
combined & $94.01(1)_{-3.1\%}^{+2.2\%}$ & $169.8(0)_{-4.3\%}^{+5.6\%}$ & $184.1(1)_{-1.9\%}^{+1.8\%}$ \Bstrut\\
\bottomrule
\end{tabular}}
\end{center}
\renewcommand{\baselinestretch}{1.0}
\caption{\label{tab:CMS8_full} Fiducial cross sections for CMS 8 TeV. Note that due to the flavour-unspecific lepton cuts the theoretical predictions are flavour-blind, which is why the results are symmetric under $e \leftrightarrow \mu$ exchange. No CMS data for the fiducial cross sections available at 8 TeV. ``Combined'' refers to the \textit{sum} of all separate contributions.}
\end{table}

\renewcommand{\baselinestretch}{1.0}
\renewcommand{\baselinestretch}{1.5}
\begin{table}[h]
\begin{center}
\resizebox{\columnwidth}{!}{%
\begin{tabular}{c c c c c}
\toprule
channel
& $\sigma_{\textrm{LO}}$ [fb]
& $\sigma_{\textrm{NLO}}$ [fb]
& $\sigma_{\textrm{NNLO}}$ [fb]
& $\sigma_{\textrm{CMS}}$ [fb]\\
\bottomrule
$\mu^+ e^+e^-$ & \multirow{ 2}{*}{$22.08(0)_{-6.2\%}^{+5.2\%}$} & \multirow{ 2}{*}{$43.91(1)_{-4.3\%}^{+5.4\%}$} & \multirow{ 2}{*}{$48.53(2)_{-2.0\%}^{+2.2\%}$} \Bstrut\\
$e^+ \mu^+\mu^-$ &  &  &  &   \Bstrut\\
$e^+ e^+e^-$ & \multirow{ 2}{*}{$22.73(0)_{-6.2\%}^{+5.2\%}$} & \multirow{ 2}{*}{$45.48(1)_{-4.4\%}^{+5.4\%}$} & \multirow{ 2}{*}{$50.39(3)_{-2.1\%}^{+2.3\%}$} \Bstrut\\
$\mu^+ \mu^+\mu^-$ &  &  &  &   \Bstrut\\
\midrule
combined & $89.62(1)_{-6.2\%}^{+5.2\%}$ & $178.8(0)_{-4.3\%}^{+5.4\%}$ & $197.8(1)_{-2.1\%}^{+2.3\%}$ \Bstrut\\
\bottomrule
$\mu^- e^+e^-$ & \multirow{ 2}{*}{$14.45(0)_{-6.7\%}^{+5.6\%}$} & \multirow{ 2}{*}{$30.04(1)_{-4.5\%}^{+5.6\%}$} & \multirow{ 2}{*}{$33.40(2)_{-2.1\%}^{+2.4\%}$} \Bstrut\\
$e^- \mu^+\mu^-$ &  &  &  &   \Bstrut\\
$e^- e^+e^-$ & \multirow{ 2}{*}{$14.92(0)_{-6.7\%}^{+5.6\%}$} & \multirow{ 2}{*}{$31.25(1)_{-4.6\%}^{+5.7\%}$} & \multirow{ 2}{*}{$34.83(2)_{-2.2\%}^{+2.4\%}$} \Bstrut\\
$\mu^- \mu^+\mu^-$ &  &  &  &   \Bstrut\\
\midrule
combined & $58.72(1)_{-6.7\%}^{+5.6\%}$ & $122.6(0)_{-4.6\%}^{+5.7\%}$ & $136.5(1)_{-2.2\%}^{+2.4\%}$ \Bstrut\\
\bottomrule
$\mu^\pm e^+e^-$ & \multirow{ 2}{*}{$36.52(0)_{-6.4\%}^{+5.3\%}$} & \multirow{ 2}{*}{$73.95(2)_{-4.4\%}^{+5.5\%}$} & \multirow{ 2}{*}{$81.93(4)_{-2.1\%}^{+2.3\%}$} \Bstrut\\
$e^\pm \mu^+\mu^-$ &  &  &  &   \Bstrut\\
$e^\pm e^+e^-$ & \multirow{ 2}{*}{$37.65(0)_{-6.4\%}^{+5.4\%}$} & \multirow{ 2}{*}{$76.74(2)_{-4.4\%}^{+5.5\%}$} & \multirow{ 2}{*}{$85.22(5)_{-2.1\%}^{+2.3\%}$} \Bstrut\\
$\mu^\pm \mu^+\mu^-$ &  &  &  &   \Bstrut\\
\midrule
combined & $148.3(0)_{-6.4\%}^{+5.4\%}$ & $301.4(1)_{-4.4\%}^{+5.5\%}$ & $334.3(2)_{-2.1\%}^{+2.3\%}$ & \quad\,$258\,\pm 8.1\%{\rm (stat)}^{+7.4\%}_{-7.7\%}{\rm (syst)}\pm 3.1{\rm (lumi)}$\quad\,\Bstrut\\
\bottomrule
\end{tabular}}
\end{center}
\renewcommand{\baselinestretch}{1.0}
\caption{\label{tab:CMS13_full} Fiducial cross sections for CMS 13 TeV. Note that due to the flavour-unspecific lepton cuts the theoretical predictions are flavour-blind, which is why the results are symmetric under $e \leftrightarrow \mu$ exchange. The available CMS data from \citeres{Khachatryan:2016tgp} are also shown. ``Combined'' refers to the \textit{sum} of all separate contributions.}
\end{table}

\renewcommand{\baselinestretch}{1.0}
\end{appendices}

\clearpage

\renewcommand{\em}{}
\bibliographystyle{UTPstyle}
\bibliography{diff_wznnlo}
\end{document}